\documentclass[aoas]{imsart}
\RequirePackage{natbib}
\usepackage{amsthm,amsmath,bbm,graphicx,listings,amssymb,mathrsfs}
\usepackage{amsmath}
\usepackage{amsfonts}
\usepackage{graphicx}
\usepackage{bbm}
\usepackage{amsthm}
\usepackage{booktabs}

\newtheorem*{remark}{Remark}
\usepackage{amssymb}
\usepackage{mathrsfs}
\usepackage{amsthm}
\usepackage{bbold}
\usepackage{dirtytalk}
\usepackage{float}
\usepackage{xr-hyper} 
\newcommand*\sref[1]{%
    \ref{#1}}
\usepackage{hyperref}
\hypersetup{
    urlcolor=black,
}

\usepackage{url}

\makeatletter
\newcommand*{\addFileDependency}[1]{
  \typeout{(#1)}
  \@addtofilelist{#1}
  \IfFileExists{#1}{}{\typeout{No file #1.}}
}
\makeatother

 \addFileDependency{supplement.aux}
\newcommand{\newedits}[1]{{#1}} 

\newcommand{\Var}{\textrm{Var}}

\begin{document}

\begin{frontmatter}
\title{A functional-data approach \\ to the Argo data}
\runtitle{A functional-data approach to the Argo data}
%
%
\begin{aug}
\author[A]{\fnms{Drew} \snm{Yarger}\ead[label=e1,mark]{dyarger@umich.edu}},
\author[A]{\fnms{Stilian} \snm{Stoev}\ead[label=e2,mark]{sstoev@umich.edu}}
\and
\author[A]{\fnms{Tailen} \snm{Hsing}\ead[label=e3,mark]{thsing@umich.edu}}
\address[A]{Department of Statistics,
University of Michigan
\printead{e1,e2,e3}}
\end{aug}


\begin{abstract}
The Argo data is a modern oceanography dataset that provides unprecedented global coverage of temperature and salinity measurements in the upper 2,000 meters of depth of the ocean.
We study the Argo data from the perspective of functional data analysis (FDA). 
We develop spatio-temporal \textit{functional kriging} methodology for mean and covariance estimation to predict temperature and salinity at a fixed location as a smooth function of depth. 
By combining tools from FDA and spatial statistics, including smoothing splines, local regression, and multivariate spatial modeling and prediction, our approach provides advantages over current methodology that consider pointwise estimation at fixed depths. 
Our approach naturally leverages the irregularly-sampled data in space, time, and depth to fit a space-time functional model for temperature and salinity. 
The developed framework provides new tools to address fundamental scientific problems involving the entire upper water column of the oceans such as the estimation of ocean heat content, stratification, and thermohaline oscillation.  For example, we show that our functional approach yields more accurate ocean heat content estimates than ones based on discrete integral approximations in pressure.
Further, using the derivative function estimates, we obtain a new product of a global map of the mixed layer depth, a key component in the study of heat absorption and nutrient circulation in the oceans. The derivative estimates also reveal evidence for density inversions in areas distinguished by mixing of particularly different water masses. 
\end{abstract}

\end{frontmatter}

\section{Introduction}

The development of technology has vastly increased the amount and complexity of data available that monitor the Earth's environment.
We focus on one type of such data collected by the Argo project, an international collaboration that oversees more than 3,800 devices called floats which measure the temperature and salinity of the oceans. 
Each float periodically ascends from 2 kilometers deep while collecting temperature and salinity measurements as a function of pressure -- a proxy for depth, with 1 decibar (dbar) roughly corresponding to 1 meter of depth. These data, referred to as profiles, are transmitted over satellite to data processing centers along with the float's coordinates and time stamps.
The drifting floats collect approximately 100,000 profiles each year, resulting in a large and complex space-time dataset, indexed by longitude, latitude, time, and pressure. See \cite{argo} for more information.

The global coverage of the Argo data and the depth of measurements provide previously unavailable richness of oceanography data (see Figure \ref{grid22}).
The data have begun to play a critical part in measuring sea level rise, currents, and the global distribution of temperature and salinity of the oceans. The oceans play a major role in the Earth's climate; for example, \cite{roemmich_unabated_2015} uses Argo data to study the warming oceans, which account for more than 90\% of the net planetary energy increase.
More than 1,500 papers that use Argo data have been published in the past five years; recently, the Argo data has begun to see research in the statistics community. For example, \cite{kuusela2017locally} is the first such publication, which enumerates some directions for future statistical research for the Argo data. To the best of our knowledge, none of the papers in this sizeable literature so far has fully taken into account the dependence of the Argo data across location, time, and pressure. For instance, the inference of the spatial dependence of temperature and salinity has thus far been conducted on a pressure-level by pressure-level approach.


We consider this problem of temperature and salinity estimation using data from all values of pressure simultaneously, under the framework of functional data analysis (FDA). The problem of spatial inference for functional data has only recently been addressed. 
See, for instance, \cite{baladandayuthapani_bayesian_2008}, \cite{gromenko_evaluation_2017}, \cite{zhang_CAR_2016}, \cite{zhang_spatial_fda}, \cite{zhou_reduced_2010}. A more thorough discussion of this area will be given in Section \ref{model_situation}.
One aspect of such inference is ``functional kriging,'' where the goal is to predict a function-valued variable at an unobserved location
based on spatially correlated function-valued covariates. Here, we develop a functional kriging methodology, tailored to the challenges and complexities of the Argo data, and aimed at producing maps or spatio-temporal predictions of temperature and salinity as functions of pressure along with functional uncertainties. 
In the context of the Argo data, each profile can be considered functional data, with measurements observed as a function of pressure for a fixed time and location. 
In this framework, we use nearby profiles in space and time to estimate temperature and salinity between the profile locations. This is done by using functional models for the mean and space-time covariance structure, which also yields uncertainties and confidence sets for the functional kriging estimates. 
The FDA approach provides computational, scientific, and methodological advantages over current approaches that consider models for one pressure level at a time by linearly interpolating temperature and salinity onto that pressure \cite{roemmich20092004, kuusela2017locally}. 
First, the FDA approach provides a principled way to share information in the irregularly-sampled measurements across pressure without perturbations (e.g., by linear interpolation).  
Second, the estimated functions capture the complex thermohaline structure in the oceans as a function of pressure that arises from the oceans' stratification and mixing. The FDA approach also naturally yields estimates of derivatives and integrals over the entire pressure dimension which can provide new insight into key scientific problems.

We directly compare our FDA approach with current ones that first linearly interpolate each profile onto fixed pressure levels.
While such an interpolation simplifies the data for the subsequent modeling compared to irregularly sampled pressures, it also introduces error or neglects data depending on whether the profiles observed are \textit{sparse} or \textit{dense} in pressure. Since Argo profiles typically range in number of observations from around 60 to 1,000 measurements, the Argo data present a combination of such heterogeneous data. When sparse functional data are observed, that is, there are just a few measurements per profile, interpolating or presmoothing each curve can decrease accuracy in comparison to pooling data from profiles \cite{hall_properties_2006, li_uniform_2010}.
When dense functional data are available, only some observations are used to interpolate onto pressure levels, and the smaller features of the temperature and salinity in the pressure dimension will be undetected. 
The FDA approach both avoids the interpolation error for the sparsely-observed profiles and leverages all measurements from each profile, and thus it
describes the pressure dimension in more intricate detail. Furthermore, when predicting at a large number of pressure levels \cite[e.g. the 58 pressure levels or more in][]{roemmich20092004}, the functional approach can considerably reduce computations by sharing information across pressure and providing functional predictions.
Perhaps most notably, estimating at fixed levels limits one's ability to predict derivative and integral functionals of the temperature and salinity, since these must be approximated from discrete predictions.
On the other hand, derivatives and integral estimates, along with their uncertainties, are readily available in our functional kriging approach and can be leveraged for fundamental scientific problems like the estimation of ocean heat content and mixed layer depth (see Section \ref{section_functional} below). 

We first introduce our notation for the data and our model:

\begin{itemize}
    \item \textbf{The data:} Denote the data for the $i$-th profile as $s_i, d_i, y_i, \left(p_{i,j}, Y_{i,j}\right)_{j=1}^{m_i}$ for $i=1, \dots, n$ where $j$ indexes the measurement, $s_i = (s_{i1}, s_{i2})$ is its location, $(d_i, y_i)$ is its day of year and year, respectively, and $(p_{i,j}, Y_{i,j})_{j=1}^{m_i}$ is the pressure and response measurements. Here, $Y_{i,j}$ denotes temperature or salinity, depending on the context; in actuality, both are observed for each $i$ and $j$. 
    In this analysis, different floats are treated identically, and the various float characteristics are not used. Data can be viewed using an \texttt{R} Shiny Application \citep{shinyapps}.
    \item \textbf{The model:} We assume that \begin{align}Y_{i,j} = \mu(s_i, d_i,y_i, p_{i,j}) +X(s_i, d_i, y_i, p_{i,j}) + \epsilon_{i,j}\end{align}where $\mu$ is a fixed mean function, $X$ is a
    zero-mean stochastic process that captures the dependence of the data, and $\epsilon_{i,j}$ is measurement error. We assume that the distribution of $X(\cdot,\cdot ,y,\cdot)$ is the same for all $y$ and that $X$ is weakly dependent in time, so that $X(\cdot,d_i,y_i,\cdot)$ and $X(\cdot, d_j,y_j, \cdot)$ are independent for $d_i$ near $d_j$ and $y_i \neq y_j$. The  $\epsilon_{i,j}$ are assumed to form a white noise process in space, time, and pressure, with mean zero and variance parameterized by $\kappa(s,d,p)$.
    
\end{itemize} 

Our new approach to the estimation of the functional mean $\mu$ combines two established approaches in nonparametric statistics: smoothing splines and local polynomial regression.
See \cite{green1994} and \cite{Fan:2012ht} for more information on these methodologies, respectively.
Specifically, we leverage irregularly sampled data in space and time using local regression to form a spline estimate of the mean function of pressure.
This approach can model the strong vertical stratification in the oceans where water masses at different depths can have drastically different characteristics. 
Our mean estimation reflects the advantages of both of these approaches: computations are reduced by using univariate B-splines along the pressure dimension while the nonlinear features of the oceans in space and time are estimated in a statistically efficient manner by local polynomial regression. As a byproduct, our approach extends that of \cite{Fan:2012ht} to the case of function-valued data and provides new functional estimates of derivatives of the mean with respect to space and time.

After subtracting the functional mean, we model the covariance structure of the residuals in space, time, and pressure. We first estimate the covariance between measurements in the same profile, decompose this estimate to form functional principal components (FPCs), and use the first $K$ FPCs to estimate a space-time covariance structure. 
As in \cite{kuusela2017locally}, locally-estimated space-time covariance models are used to perform kriging and obtain the uncertainty in the estimates. This entails a unified and computationally tractable functional modeling and prediction framework that takes into account the dependence in space, time and pressure.
Being able to fill data gaps in a principled manner to produce estimates of temperature and salinity continuously at all locations, times and pressures is of tremendous value to ocean research. Some examples and references of traditional ``mapping strategies,'' or interpolation approaches, in oceanography can be found in \cite{boyer1994quality} and \cite{ishii2009reevaluation}. As shown in Table 1 of \cite{cheng_uncertainties_2014}, the resolution of available ocean data continues to improve, the Argo Project being a contributing factor in the past ten years. Our functional data approach is motivated by fully leveraging the benefit of the high-resolution Argo data, but generalizations to other similar high-resolution data should be straightforward. These approaches will potentially play an important role in ocean and climate research in general.

We mention some applications to demonstrate the advantages of our approach.  
The first application is the estimation of the integrated ocean heat content at each location, which is related to the integral of the temperature curves. Traditionally, the integrated ocean heat content was studied through numerically interpolated data. 
The main focus of \cite{cheng_uncertainties_2014} is to address the bias in such estimates caused by data sparsity. 
In our opinion, however, this study and other traditional approaches do not always comprehensively consider the variability throughout the analysis.
Using our approach, distributional properties of the estimated heat content can be easily obtained from the overall analysis. In particular, if the heat content at a location is estimated with sparse data, then the model-based estimated error will reflect that. More generally, by pooling data across space and time, our model-based approach can be used to
identify statistically significant anomalies in the ocean heat content, which is one important and active area of research \citep[][]{roemmich_unabated_2015}.

For the second application, the functional predictions are used to estimate potential density, which provides valuable information about the vertical stratification of the oceans -- a key factor in their ability to absorb heat \citep[][]{li_increasing_2020}. For example, we use potential density to estimate the depth of the \textit{mixed layer} of the ocean, a region directly below the ocean surface where the ocean mixes uniformly and is characterized by near-constant ocean properties \citep[Sections 4.2 and 7.4 of][]{talley_chap4}. 
The mixed layer drives the ocean-atmosphere interactions and thus influences heat and carbon flux of the ocean, ocean circulation, and biological processes dependent on light \citep{holte_argo_2017}. Our functional estimates provide mixed layer depth estimates over all open oceans for each day-year combination that have minimal discretization error in pressure. 
Employing the bootstrap, we obtain distributions of the mixed layer depths and use them to assess the within and between-year variations in mixed layer depth. Our analysis shows, in particular, that summer mixed layer depths generally have smaller within-year variations and, at some locations, slightly larger between-year variations than do winter mixed layer depths, in proportion to their size. We also use the potential density estimates to evaluate the occurrence of non-monotone features of potential density that indicate vertical instability in the water column, and we find evidence of such features.  Our framework provides the means to identify such anomalies on a global scale, which can help oceanographers track the structural stability of the thermohaline oscillation -- a fundamental driver of Earth's climate \citep[][]{rahmstorf_exceptional_2015,li_increasing_2020}.

We outline the rest of the paper, which loosely follows the structure of the introduction. In Section \ref{ArgoData}, the Argo data is introduced in more detail. In Section \ref{meanest}, we develop our approach for mean estimation and its computational implementation over the Argo data. After subtracting the mean from the data, we estimate the covariance of the residuals, predict using the estimated covariance, and assess the quality of our predictions in Section \ref{covarianceest}. In Section \ref{section_functional}, a framework to use the functional estimates is developed, specifically applying the examples outlined above. Throughout our analysis, we provide the resulting estimates as data products to the community and introduce interactive {\tt R} Shiny web applications for visualizing the results \citep{shinyapps}. We conclude and identify future research directions in Section \ref{conclusion}. Throughout the paper, we refer to figures, tables, and text from the Supplementary Materials; such references are prefixed with ``S.'' 

\section{Argo Data and Existing Methodologies}\label{ArgoData}

In this section, we give a more detailed overview of the Argo data, give an introduction to mapping methods, and situate our approach within the spatial FDA literature. While there is a variety of measurements of the oceans, including sea surface temperature and ship-based measurements, we only use data from the Argo project because it provides a natural comparison to existing approaches. 

\subsection{Data from the Argo Project}

\begin{figure}[t]
\includegraphics[scale = .14]{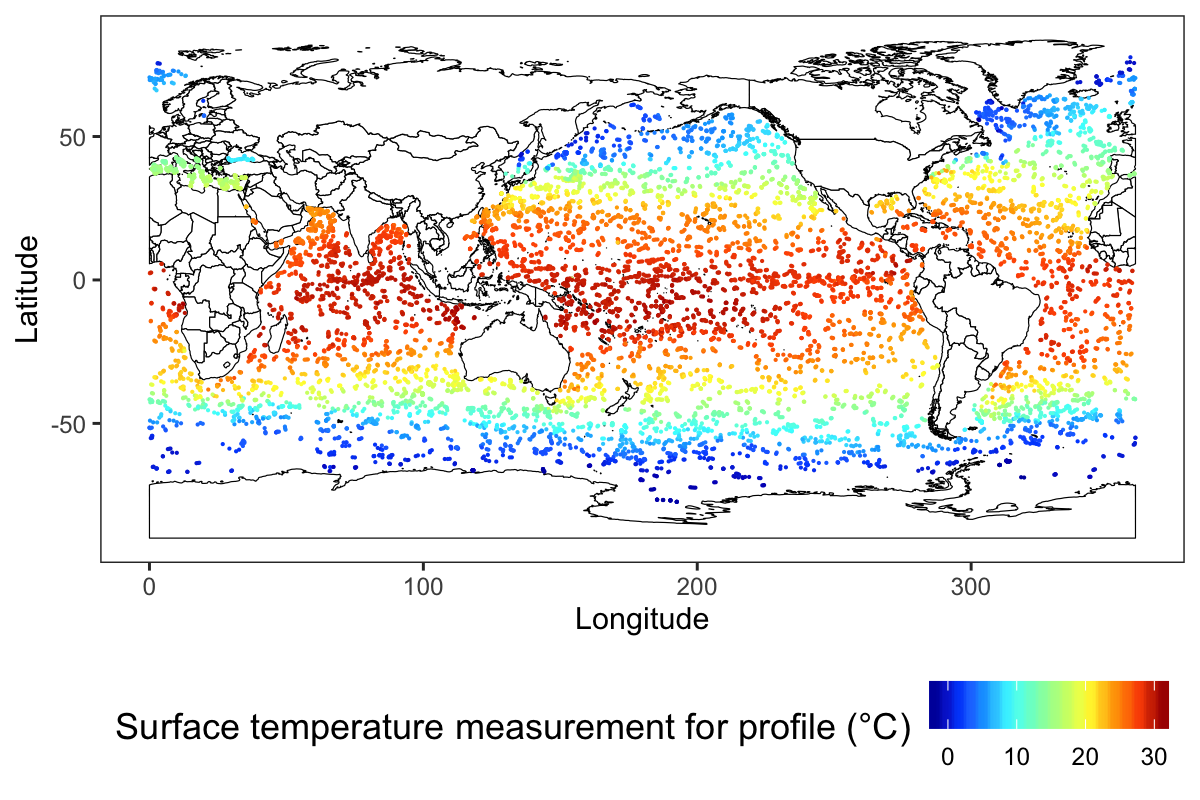}
\includegraphics[scale = .3]{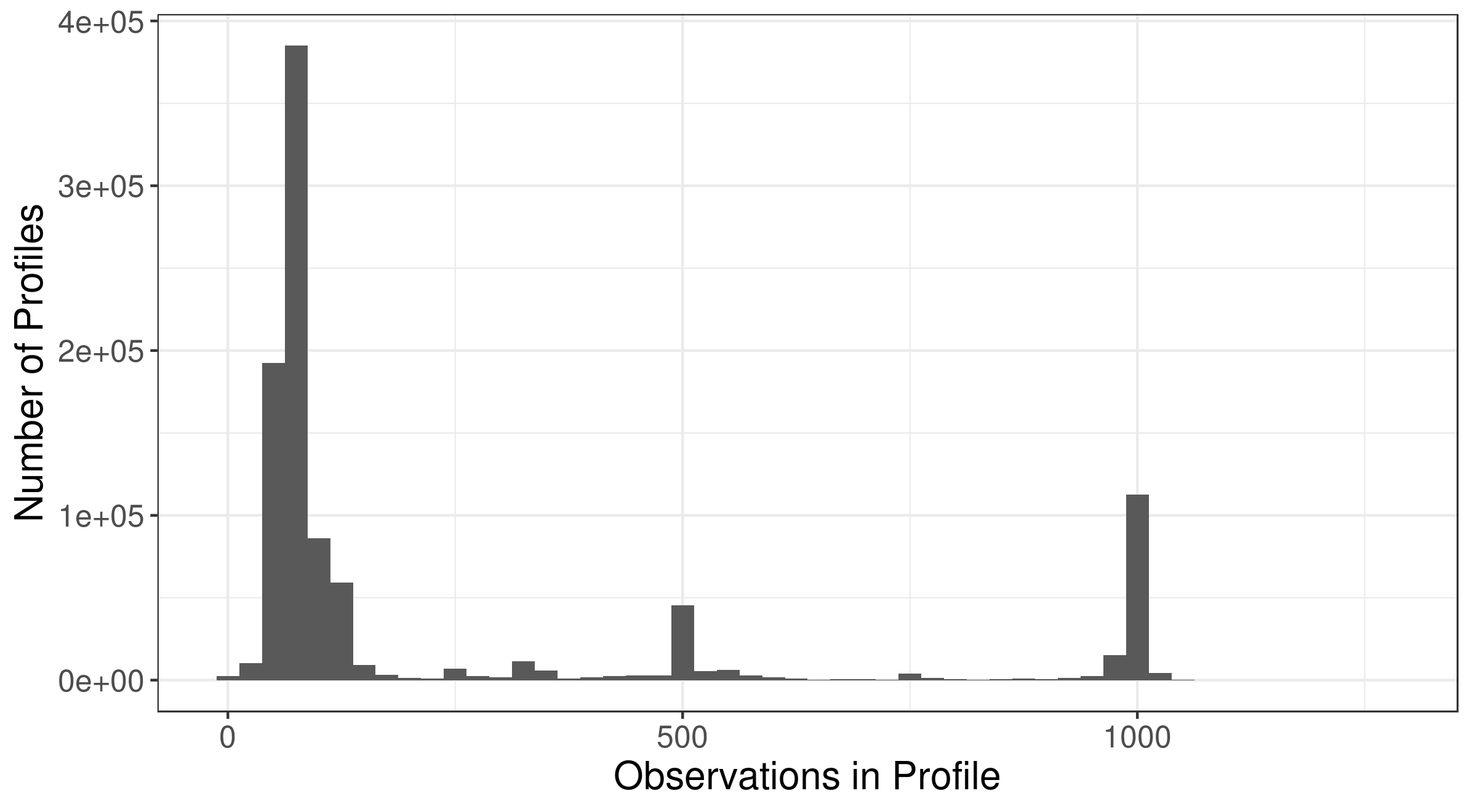}
   \caption{Argo data examples. (\textbf{Left}) Locations of profiles collected in February 2016, colored by the temperature of the measurement closest to the surface. (\textbf{Right}) Histogram of the number of measurements per profile.} \label{grid22}
\end{figure}

The Argo program is an international collaboration that develops and manages floats, mechanical devices that collect measurements on the world's oceans \citep{argo}. The Argo project reached a goal of global coverage in late 2007 and has continued to increase the number of floats to nearly 4,000 today; see, e.g., Figures \ref{grid22} and \ref{example_mean} or a Shiny application \citep{shinyapps}.
In ten-day cycles, each float descends from its parking depth at 1,000 dbar to a depth of 2,000 dbar, then rises over the course of six hours to the surface, collecting measurements of pressure, temperature, and salinity.
Upon surfacing, the float transmits the data via satellite. 
The pressure, temperature, and salinity data and its associated location and time for each cycle is called a profile.

The Argo program was designed to sample approximately one profile every 10 days in each 3 by 3 degree region of the open oceans. 
Before this relatively uniform sampling of the Argo program, sampling at greater depths was sparse and highly nonuniform in space and time, with fewer measurements in the Southern Hemisphere and during winter months \citep{roemmich20092004}. In terms of depth, the pressures at which each float samples can vary from float to float as well as from profile to profile due to varying data transmission technology, as seen in Figure \ref{grid22}. This heterogeneity in the sampling frequency is one important challenge addressed by our functional approach.


The Argo data is made publicly available after transmission through satellite and various data-quality control measures. For our analysis, we use a preprocessed version of the Argo data which was formed and used in \cite{kuusela2017locally}. The data spans the years 2007 to 2016 based on the May 2017 snapshot of the Argo data.
The data includes more than 245 million total point measurements from 994,709 profiles, of which 551,536 have extended data quality (delayed-mode) checks. 
Throughout our analysis, we generally use all profiles for temperature, while for salinity delayed-mode profiles are needed to ensure minimal drift or bias \cite{owens_improved_2009}.


\subsection{Argo Mapping Methodology}\label{rgsec}





The problem of mapping irregularly-sampled spatial data onto a grid or unobserved location is a common problem in spatial statistics, oceanography, and the geosciences in general. The main methods to address this problem are similar in the different fields, though they may be referred to with different names. In statistics, it is often called kriging or Gauss-Markov prediction, specifically referring to the conditional prediction of a Gaussian random vector based on a spatial covariance structure \citep[cf.][Section 4.1]{Cressie:2015uu}. In geology, this method is also referred to as kriging \citep[cf.][Chapter 3]{chiles_geostatistics_2012}, while in oceanography, this is usually called objective mapping or optimal interpolation (cf. Section 4.2 \citealp{thomson_chapter_2014}, \citealp{barth_introduction_nodate}) and focuses on constructing gridded predictions. Each, in essence, involves specifying a mean and covariance structure, then using these to form a prediction. If the true mean and covariance structure is specified, then the resulting prediction minimizes mean-squared error over the class of linear predictors. These approaches generally require inversion of the covariance matrix of size $n\times n$, where $n$ is the number of observed spatial locations. In optimal interpolation, the covariance structure is more often specified using subject-matter knowledge rather than being estimated from the data. 

In this framework, we review approaches for mapping that specifically use the Argo data. We focus on the important work of \cite{roemmich20092004}, who provide a methodology for mean estimation and analysis of anomalies using the Argo data, as well as \cite{kuusela2017locally}, who focus on covariance estimation and introduce maximum likelihood estimation for its model parameters in space and time. These are only two works in a wider array of temperature and salinity estimation works using Argo data. Other approaches used to form Argo data products include \cite{gray_method_2015}, who propose an iterative approach to estimating the covariance function, \cite{li_development_2017}, \cite{gaillard_isas-tool_2012}, \cite{hosoda_monthly_2008}, and \cite{udaya_bhaskar_operational_2007}. These focus on scalar data at a limited number of pressure levels, and each uses a Gaussian or exponential covariance function. 
We now turn to the Roemmich and Gilson product, which is available as the standard in global oceanography analysis using the Argo data. This product provides estimates of the mean temperature and salinity separately, as well as monthly anomalies from the mean over grids of different resolutions in space and fixed pressure levels.
Before estimation, the temperature and salinity for each profile is interpolated onto 58 fixed nonuniformly-spaced pressure levels. 
Throughout, they use a distance based on latitude, longitude, and the depth of the ocean floor at each location. 
The inclusion of the depth of the ocean floor better handles areas where ocean currents run along the shores of continents like the Western boundary currents \citep[see, for example, Section 7.8 of ][]{talley_chap4}.
To estimate the mean, for each pressure level and grid point of space, they combine data from the years 2004-2016, using the 100 nearest profiles from each of the twelve months of the year. 
In addition, they only use the interpolated values at a pressure level as well as the two adjacent pressure levels.  
A weighted least squares approach based on distance from the grid point is used to fit a model of the form: \begin{align}\begin{split}&\beta_0 + \beta_1(s_{i1} - s_{01}) + \beta_2(s_{i2} - s_{02}) +   \beta_3(s_{i1} - s_{01})^2 + \beta_4(s_{i2} - s_{02})^2   \label{RG}\\
& \ \ \ \ \ \ \   +\beta_5(p_i - p_0) +  \beta_6(p_i - p_0)^2 +\sum_{k=1}^6 \gamma_k \sin\left( \frac{d_i2\pi k}{365.25}\right) + \sum_{k=1}^6 \delta_k \cos\left(\frac{d_i2\pi k}{365.25}\right)\end{split} \end{align}where $s_{i1}$ and $s_{i2}$ give the location of profile $i$, $p_i$ the pressure level of the interpolated measurement, $d_i$ is the day of the year profile $i$ was observed, and $\beta_k$, $\gamma_k$, and $\delta_k$ are scalar coefficients. The coefficient $\beta_0$ represents the time-averaged mean, while the $\gamma_k$, and $\delta_k$ give the deviations from this mean at different times of the year. Overall, this approach is a form of local regression, where the time dimension is estimated using a fixed Fourier basis. 

After subtracting the mean, Roemmich and Gilson then provide a field of anomalies for each month of each year that describes the variation away from the mean at a particular location. These are formed by computing the conditional mean at each grid point in space and pressure level assuming Gaussianity and using a covariance of the form \begin{align}
\begin{split}
    C_{RG}(\Delta_{RG}) &= 0.77\cdot \textrm{exp}\left( - (a^\top \Delta_{RG}/140)^2\right) +0.23 \cdot \textrm{exp}\left( - | a^\top\Delta_{RG} | /1111\right)\label{RG_cov}
    \end{split}
\end{align}Here, $\Delta_{RG}= (\Delta_{s_1}, \Delta_{s_2}, \Delta_{dep})^\top$ denotes a vector of distances between two locations $s$ and $s^\prime$ for the zonal direction (East-West), meridional direction (North-South), and the distance penalty for ocean depth described above. The vector $a$ scales the relative directions and is $(1,1,1)$ above 20 degrees North and below 20 degrees South, but changes linearly to $(.25, 1, 1)$ at the equator, which increases the covariance in the zonal direction in the tropics. This choice is supported by empirical estimates near the surface. The covariance in \ref{RG_cov} is nonstationary due to its dependence on $a$, though the covariance does not depend on time or the pressure level. 
To form the final product, the anomalies over all months and years are averaged and added to the mean.

\cite{kuusela2017locally} employ the Roemmich and Gilson mean and study the covariance structure in more detail by proposing a space-time covariance model and fitting it using maximum likelihood. To address the nonstationarity of the data, they use the \textit{locally stationary} assumption; that is, at each location, parameters of a stationary covariance are estimated using data nearby, and the local covariance estimates are used for prediction at that location. Data from different years are assumed independent, and one stationary covariance function for data observed in the same year they consider is 
\begin{align*}
    C_{KS}(\Delta_{KS}) &= \phi \cdot \textrm{exp}\left(-\sqrt{ \frac{\Delta_{s_1}^2}{\theta_{s_1}^2} + \frac{\Delta_{s_2}^2}{\theta_{s_2}^2} + 
    \frac{\Delta_{d}^2}{\theta_{d}^2}} \right) + \sigma^2\cdot 1(\Delta_{KS} = 0)
\end{align*}where $\Delta_{KS} = (\Delta_{s_1}, \Delta_{s_2}, \Delta_d)^\top$ is the relevant distance between two locations and times in longitude, latitude, and day of the year, respectively. The estimated parameters are the process variance $\phi$, nugget variance $\sigma^2$, and three scale parameters $\theta$ subscripted by their direction. Thus, since the model is estimated at each pressure level, it can adapt to the large differences in the covariance structure at different depths. Furthermore, the model provides uncertainty for the estimates, which are validated using cross validation for both Gaussian and t-distributed measurement errors. At many depths, the residuals may have non-Gaussian features as noted in \cite{kuusela2017locally}. To further address this issue, \cite{bolin_multivariate_nodate} explore a class of multivariate non-Gaussian spatial models that offer some improvements in prediction on a limited analysis of Argo data.

To conclude this section, we recognize that some aspects of FDA are not altogether new to oceanography. For example, splines have been used as a smoothing approach to interpolate sparse observations in a profile; principal component analysis (PCA), known as empirical orthogonal functions (EOF) analysis, is a common dimension-reduction approach \citep{thomson_chapter_2014}. However, these are applied in somewhat limited ways that include little to no statistical considerations.




\subsection{Spatial FDA Literature}\label{model_situation}

The extension of spatial prediction for scalar data to functional data has primarily been developed recently in the statistics discipline. For independent and identically distributed functional data, the literature has been well developed and presented, for example, in the books of \cite{Ramsay:2013hf}, \cite{Hsing2015}, and \cite{Kokoszka}. For spatially-dependent functional data, most of the literature has focused on the idealized regime where entire functions are observed. In particular, there are detailed reviews in \cite{delicado2010}, \cite{aguilera-morillo_prediction_2017}, \cite{koko2019}, and \cite{martinez-hernadez_recent_2020}. We outline some of the work in this area. 

Recent developments in spatial FDA have provided increasingly comprehensive approaches for complex spatio-temporal data. Most of the literature focuses on geostatistical (point-referenced) data, though approaches for areal data and point processes have been considered \citep{delicado2010, zhang_CAR_2016, cronie_functional_2019}. 
In addition, methods for hierarchical spatial functional data have been developed through the work of \cite{baladandayuthapani_bayesian_2008}, \cite{staicu_fast_2010}, and \cite{zhou_reduced_2010}. \cite{ruiz-medina_spatial_2011} and \cite{zhang_CAR_2016} extend spatial autoregressive or moving-average processes to functional data. \cite{staicu_modeling_2012} develop copula-based methods for skewed spatial functional data. Methods for clustering spatio-temporal functional data have been proposed in \cite{jiang_clustering_2012} and \cite{romano_spatial_2017}, among others.
Theory and methodology for spatial FDA has been explored in \cite{zhang_CAR_2016}, and \cite{gromenko_evaluation_2017} and \cite{zhang_spatial_fda}. In particular,  \cite{gromenko_evaluation_2017} propose an iterative approach for modeling the mean and covariance structure, while also addressing inference on the mean function. Bayesian approaches to spatial FDA are proposed in \cite{baladandayuthapani_bayesian_2008} and \cite{song_hierarchical_2019}.
The spatial FDA perspective has mostly been considered in applications to environmental data \citep{monestiez_cokriging_2008, rodriguez_bayesian_2009, 
king_functional_2018,Pauthenet} and medical applications including neuroscience \citep{lynch_test_2018} and a cancer study \citep{baladandayuthapani_bayesian_2008, staicu_fast_2010, zhou_reduced_2010}. 
Here, we directly situate our approach within this rich literature. Most of the mentioned approaches propose basis expansions of mean and principal component functions and model the principal component scores as a spatial process (e.g. Mat\'ern); our approach does as well. At the broad level, our approach is similar to \cite{gromenko_evaluation_2017}. Specifically, our two-stage approach, where we estimate the covariance only after estimating the mean and forming residuals, is similar to steps 1-3 of their Algorithm 3.1. Since there are a large number of parameters, this two-stage approach helps reduce the parameter space. This approach is common in FDA and is supported by the theoretical work in, e.g.,  \cite{li_uniform_2010} or \cite{yao_functional_2005}. Ideally, we would employ the iterative algorithm in \cite{gromenko_evaluation_2017} in the spirit of iteratively reweighted least squares,  but we are limited by the computation. We justify this two-stage approach with an appeal to profile likelihood, where the mean is estimated assuming a fixed within-profile covariance, after which the covariance is estimated assuming a fixed mean.
We also extend their methodology by proposing a nonseparable covariance structure, as discussed below. Some of the methodology and motivation is similar to \cite{king_functional_2018}.


The Argo data calls for more involved modeling than in the existing spatial FDA literature in a number of respects. The challenges include: the addition of another dimension (pressure) to the space and time dimensions, irregularly-spaced data in each of these dimensions, the varying number of measurements per profile, the sparsity of data in space and time, and the large size of the data. Here, we detail a few aspects of our approach that address these complexities. 
First, the referenced approaches assume a constant mean in space, i.e.  $\mu(s,d,y,p) = \mu(p)$ for some function $\mu(p)$, as well as a stationary covariance. Due to the nonstationary nature of the Argo data, a constant mean would not be physically adequate. We allow the mean and covariance structure to change in space, providing a way to model nonstationary functional data in space and time, extending the local stationarity assumption of \cite{kuusela2017locally} for their setting of a fixed pressure level. This local approach for the mean and covariance also helps address the computational challenges with respect to the size of the data. 

Next, most approaches in the literature depend on a basis representation of profiles. By projecting each profile onto a suitable basis before modeling, this simplifies the subsequent analysis, but such a step introduces systematic error. For such an interpolation approach to be justifiable, all profiles should be densely sampled \citep[cf.][]{hall_properties_2006, li_uniform_2010}. For many Argo profiles, in particular ones sampled 2007-2010 that have fewer measurements, an appropriate basis representation cannot be obtained. \newedits{Our estimation and prediction methodology avoids this issue and naturally accommodates both sparse and dense data at their measured pressure.}

Finally, in Section \sref{discussseparability}, we compare the non-separability of our covariance model described in Section \ref{covarianceest} and compare the model with existing literature. \newedits{Our covariance model, by assuming separability for each principal component direction, allows for a varying space-time covariance structure as a function of depth. Such flexibility is necessary for the Argo data, since processes at the surface can be much different than those at greater depth.}

\section{Functional Mean Estimation for Argo Data} \label{meanest}

In this section, we introduce our functional approach for mean estimation, in which we estimate a smooth function $\mu(s_0, d_0, y, p_{i,j})$ for a location $s_0$. We assume that the mean function $\mu$ is smooth in terms of space, time, and pressure. Due to the functional nature of the data, we focus on and formalize the smoothness in pressure here.
In particular, consider the class of functions \begin{align*}\mathbb{W}_2 &= \left\{f \ | f^{(2)} \textrm{ exists almost everywhere,} \textrm{ and }  \int_0^{2000} (f^{(2)}(p))^2 dp < \infty\right\}\end{align*} where $f^{(k)}$ is the $k$th derivative of $f$. The space $\mathbb{W}_2$ is a Sobolev space of functions widely used for nonparametric inference including problems in FDA \citep{Hsing2015,wahba1990spline}. The size of $\left\lVert f^{(2)}\right\rVert_{\mathbb{L}_2}^2 = \int (f^{(2)}(p))^2dp$ quantifies the smoothness of $f$, i.e., if $\int (f^{(2)}(p))^2dp=0$, then $f$ takes the form of a line.  

\subsection{A Functional Approach to Mean Estimation}

We consider a mean estimated locally in space and day of the year which can be evaluated at any pressure in $[0,2000]$. Our novel approach combines local regression (to smooth space and time) and smoothing splines (to smooth pressure) by estimating the function:
\begin{align}\begin{split}f_{\beta, s_0, d_0}(s_i, d_i, y_i, p) &= \sum_{y=2007}^{2016}\beta_{0, y}(p)1(y_i = y) + (s_{1i} - s_{10}) \beta_1(p) + (s_{2i} - s_{20}) \beta_2(p) + \\& \  \ \ \ \ (s_{1i} - s_{10})^2 \beta_3(p) + (s_{2i} - s_{20})^2 \beta_4(p) +\\ &  \ \ \ \  (s_{1i} - s_{10})(s_{2i} - s_{20}) \beta_5(p)+(d_i - d_0)\beta_6(p) + (d_i - d_0)^2 \beta_7(p)\end{split}\label{functional_form_mean}\end{align}
where $s_0 = (s_{10}, s_{20})$ is a fixed location and $d_0$ is a fixed day of the year. Here, the functions $\beta_{0,y}$ and $\beta_k$ are specific to $s_0$ and $d_0$, though we omit this notation for ease of writing. Assuming each function denoted with $\beta_{0,y}$ or $\beta_k$ falls in the class $\mathbb{W}_2$, we include the standard smoothing spline penalty on the second derivative of each function:
\begin{align*}
\textrm{Pen}(\lambda) &=\lambda_0 \sum_{y=2007}^{2016} \left\lVert \beta_{0,y}^{(2)}\right\rVert_{\mathbb{L}_2}^2 + \sum_{k = 1}^7 \lambda_k \left\lVert \beta_k^{(2)}\right\rVert_{\mathbb{L}_2}^2
\end{align*}where the $\lambda_j$ are nonnegative smoothing parameters. This penalty controls the smoothness of the estimated functions. With this notation, for a fixed location $s_0$ we solve the optimization problem:
\begin{equation}
    \min_{\beta_k \in \mathbb{W}_2}  \left(\ell_{s_0,d_0}(\beta) + \textrm{Pen}(\lambda)\right),\label{meanoptim}
\end{equation}where \begin{equation} 
\ell_{s_0, d_0}(\beta) = \frac{1}{n} \sum_{i=1}^n \frac{K_{h_s, h_d}(s_i - s_0, d_i - d_0)}{m_i} \left\lVert \Sigma_i^{-\frac{1}{2}} \left(Y_i - f_{\beta, s_0,d_0, i}\right)\right\rVert^2_2,
\end{equation}and $Y_i$ and $f_{\beta, s_0, d_0, i}$ are vectors with entries $\{Y_{i,j}\}_{j=1}^{m_i}$ and $\{f_{\beta, s_0, d_0}(s_i, d_i,  y_i, p_{i,j})\}_{j=1}^{m_i}$, respectively. Here, $K_{h_s, h_d}$ is a product of Epanechnikov kernels, the first based on the great-circle distance between $s_i$ and $s_0$ with bandwidth $h_s$ and the second based on the difference in day of the year between $d_i$ and $d_0$ with bandwidth $h_d$. This type of kernel is commonly used for local regression \citep{Fan:2012ht}. Also, $\Sigma_i$ is a matrix that specifies the working correlation between measurements in the same profile; we address choosing its form in the next section. Dividing by $m_i$ in (\ref{meanoptim}) ensures that profiles with more measurements do not contribute in greater proportion to the loss function compared to profiles with fewer measurements. Once again, the resulting functions $\beta_{0,y}$ and $\beta_k$ are estimated for each fixed location $s_0$ and time $d_0$, omitted for simplicity in the notation.
We propose this new general nonparametric approach of combining local regression and spline smoothing for estimating a spatially-varying functional mean. 
\begin{figure}[t]
\includegraphics[scale = .28]{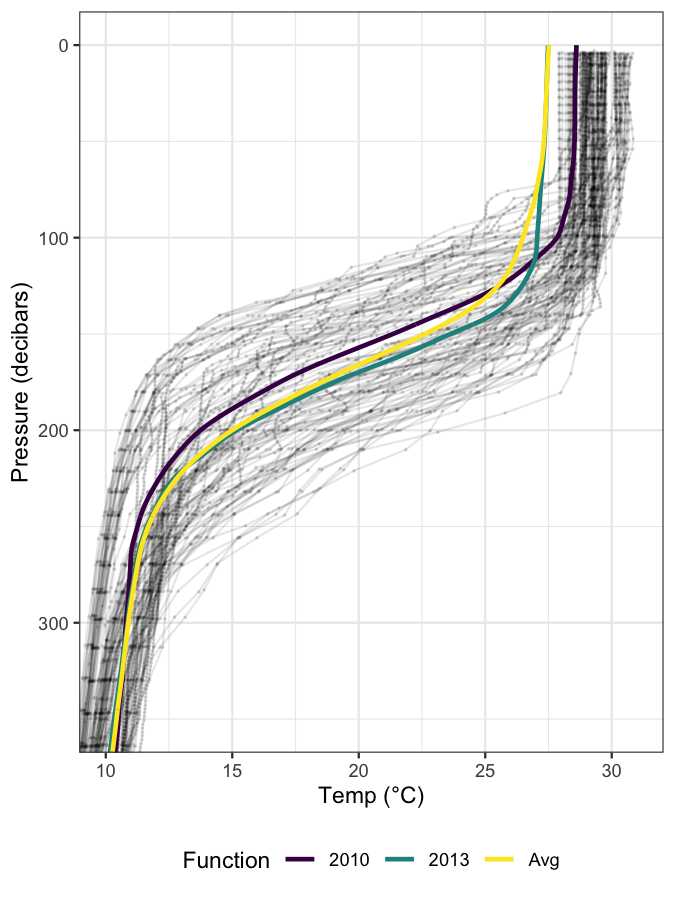}
\includegraphics[scale = .28]{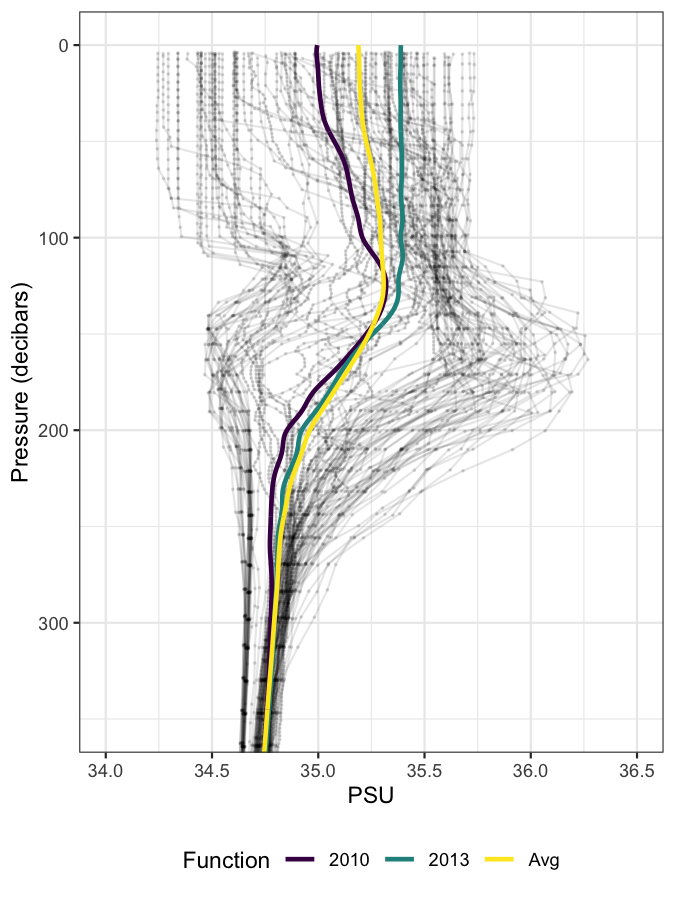}
\caption{Data used from 2010 for first 350 dbar for $-170.5^\circ$ W $0.5^\circ$ N, $d_0 = 45.25$ for (\textbf{Left}) temperature ($^\circ$C) and (\textbf{Right}) salinity (practical salinity units, PSU). We plot the estimated mean functions $\beta_{2010}(p)$ and $\beta_{2013}(p)$ for two of the years and the year-averaged estimate $\overline{\beta}(p)$. In this figure as well as Figures \ref{pcexample} and \sref{cv_example} we plot according to the oceanography convention with pressure on the $y$ axis in reference to depth in the ocean.}\label{example_mean}
\end{figure}

The optimization problem (\ref{meanoptim}) is solved for temperature and salinity separately. The functions $\beta_{0,y}$ for $y=2007$, $\dots$, $2016$ give a mean function estimated from each year. The function $\overline{\beta}(p)= \frac{1}{10}\sum_{y=2007}^{2016} \beta_{0,y}(p)$ is the year-averaged mean at $s_0$ and $d_0$. 
The additional functions $\beta_1$ through $\beta_7$ are used to estimate the derivatives of the mean with respect to space and time, for each pressure.
Figure \ref{example_mean} gives results at one location in the Pacific Ocean for the first 350 dbar with $d_0 = 45.25$, corresponding the mid-February. The mean functions are able to capture the water column with constant temperature near the surface known as the mixed layer, which we address in more detail in Section \ref{section_functional}.
The reader can compute (\ref{meanoptim}) for fixed smoothing parameters using an {\tt R} Shiny application \citep{shinyapps}.

We motivate our functional approach by qualitatively comparing it to a multivariate local regression approach with respect to pressure, space, and time. Both approaches are nonparametric and should behave relatively similarly given appropriate bandwidths and smoothing parameters. However, some key advantages of the functional approach are as follows. First, multivariate local regression can be challenged by the curse of dimensionality since no points are truly ``local'' \citep{Fan:2012ht}, while our approach reduces this problem by using all data in pressure simultaneously. This provides a ``middle-ground'' nonparametric technique between local estimation (computationally manageable, using a limited amount of data) and multivariate/thin-plate splines (computationally intractable, using all of the data).
Also, applying local regression in pressure can introduce new challenges of bandwidth selection, which we avoid. Due to the differences in variability and sampling in pressure, a constant bandwidth in pressure would not be appropriate. Our approach can also save computation, since re-estimation is not needed for any additional pressure measurement. Relatedly, derivatives and integrals of pressure are immediately available for the entire pressure dimension, while, for local regression, only derivatives are available at the points of computation. Our approach {is tailored to and reflects the functional nature of the data}, so that profiles and estimated mean functions can be easily compared.  

 \subsection{Computation and Cross Validation} \label{CV}

In this section, we give an overview of our approach for computation and how to choose smoothing parameters. More details are shown in Section \sref{leverage_score_computation}. The solution to (\ref{meanoptim}) must be computed for each location of interest $s_0$; however, calculations for different $s_0$ do not rely on each other, so they can be easily made in parallel over multiple computer cores.
By applying Theorem 6.6.9 of \cite{Hsing2015} to losses that include multiple functions in $\mathbb{W}_2$, we obtain that each function of the resulting solution to the infinite dimensional optimization problem (\ref{meanoptim}) is a natural cubic spline with knots at each uniquely observed $p_{i,j}$ for each $i$ such that $K_{h_s, h_d}(s_i - s_0, d_i - d_0) > 0$. Since the smoothness of each function is penalized in the objective function using GCV, having a large number of knots does not contribute to overfitting \citep{ruppert_selecting_2002}. On the other hand, placing a knot at each observed pressure value is prohibitively costly when a large number of profiles are included in each fit. Commonly, in nonparametric regression, reducing the number of knots is done using the quantiles of $p$ or equispaced knots \citep{ruppert_selecting_2002}. We adopt a similar strategy 
in this functional setting by employing penalized cubic B-splines bases with 200 equispaced knots in $[0, 2000]$. Our experiments indicated that the difference with the exact solution involving knots at all relevant pressures is small, and this basis provides knots at intervals near the size of Argo pressure uncertainties of $\pm 2.4$ dbar. Due to the local nature of the B-splines, the relevant matrices needed to compute the solutions are sparse and banded, which leads to further computational gains. In particular, the Cholesky decomposition of matrices is numerically efficient. To compute the B-spline basis functions and penalty, we use the \texttt{fda} package.

In addition to computing the solution, we also need to choose the smoothing parameters $\lambda_j$ and bandwidths $h_s$ and $h_d$. Smoothing parameters $\lambda_j$ are currently chosen assuming $h_s$ and $h_d$ fixed. We set $h_s= 900$ kilometers for both temperature and salinity and $h_d = 45.25$ days. This provides nearly enough profiles for each grid point and year and uses data from three months of the year. Also, if fewer than 10 profiles were used for each year, $h_s$ is increased so that there are at least 10 profiles used for each year. To choose $\lambda_j$, we use generalized cross validation (GCV) for its favorable properties, ease of calculation, and ability to include a correlation structure in the observations \citep{wahba1990spline}. The GCV score in the context of our problem is \begin{align*}
\textrm{GCV}(\lambda) = \frac{(Y - \hat{Y})^\top \Sigma^{-1} (Y - \hat{Y})}{(1- \textrm{tr}(A(\lambda))/n_{s_0})^2}
\end{align*}where $n_{s_0}  = \sum_{i=1}^n 1\left(K_{h_s, h_d}(s_i - s_0, d_i - d_0)  > 0\right)m_i$, $A(\lambda)$ is the \say{hat} matrix defined in the Section \sref{leverage_score_computation}, $Y$ are the observations for temperature or salinity, $\hat{Y}$ are predictions using smoothing parameters $\lambda$, and $\Sigma^{-1}$ is the block-diagonal matrix $$\Sigma^{-1} = \textrm{diag}\left[ \frac{K_{h_s, h_d}(s_i - s_0, d_i - d_0)}{n m_i}\Sigma_i^{-1}, i=1, \dots, n\right].$$
Considering approaches like variable bandwidth selection, jointly choosing bandwidths and smoothing parameters, and leave-one-profile-out cross validation are methodological and computational challenges that can motivate further research. Computing the leverage scores for the calculation of GCV is the largest computational cost in the selection of $\lambda$, and we detail how to compute them in Section \sref{leverage_score_computation}. For choosing multiple smoothing parameters, computing the GCV function on a two-dimensional or larger grid becomes prohibitively expensive. We have taken the approach of finding suitable fixed ratios $\eta_\ell = \lambda_\ell/\lambda_0$ for each $\ell$, then using the smoothing parameters $a\eta$ and cross validating on the single parameter $a>0$. These ratios are chosen to balance the units of each of the covariates, and the quadratic terms require larger amounts of smoothing. In particular, we let
$\eta = (1, 10^8, 10^8, 10^{13}, 10^{13}, 10^{13},10^{9}, 10^{13})$ and conduct standard 1-d optimization using \texttt{optimize} in {\tt R} to search for $a \in (10^{-3}, 10^7)$.

The irregular sampling of Argo profiles over pressure can present challenges for naive spline estimation. This issue can be addressed using a working correlation structure in pressure. A simple choice employed here is Markovian-type dependence in continuous pressure. Specifically, we consider $(\Sigma_i)_{j,k} = \textrm{exp}(-\tau  |p_{i,j}-p_{i,k}|)$ 
with $\tau \in (0,\infty)$, resulting in a tridiagonal precision matrix $\Sigma_i^{-1}$. 
In practice, we have found that using this within-profile correlation with $\tau = 0.001$, which corresponds to a correlation of about $0.9512$ for measurements 50 dbar apart, helps both the selection of $\lambda$ as well as the quality of solution. 
In Section \ref{covarianceest}, the within-profile covariance is estimated, and in Section \sref{comparecor} it is shown that the empirical covariance estimates generally match well with this choice.  
One could include a non-constant working variance as well, though such benefit may be marginal. 

We compute the solution to (\ref{meanoptim}) in {\tt R} on a 1 degree by 1 degree grid in space for mid-February ($d_0 = 45.25$) between $-80^\circ $ S and $80^\circ $ N. This results in 47,938 and 46,023 grid points computed for temperature and salinity, respectively. For salinity, we use only delayed-mode data.
For each profile $i$ from the first three months of the year, residuals were computed by using the mean estimate at the nearest grid point to profile $i$ as $Y_{i,j} - \hat{f}_{\beta}(p_{i,j})$. The implicit assumption of computing these residuals is that the mean is represented well by a locally quadratic function of day of the year for these three months as in (\ref{functional_form_mean}). In Section  \sref{RG_comparison}, we compare with the February mean field estimates of \cite{roemmich20092004}.

\subsection{Functional Derivatives}

\begin{figure}[t]
\begin{minipage}{.35\textwidth}
        \centering
        \includegraphics[width=0.95\textwidth]{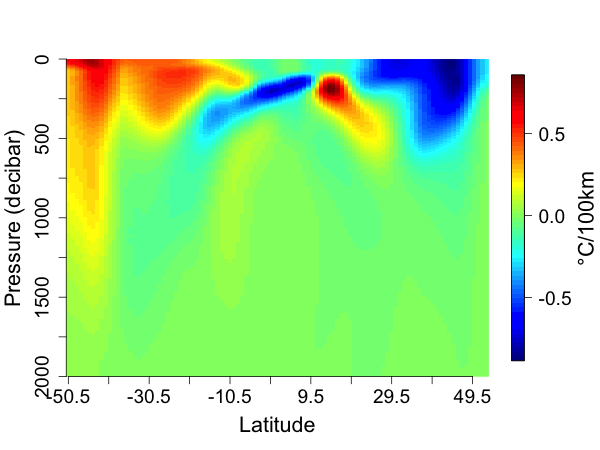}
        
        \includegraphics[width=0.95\textwidth]{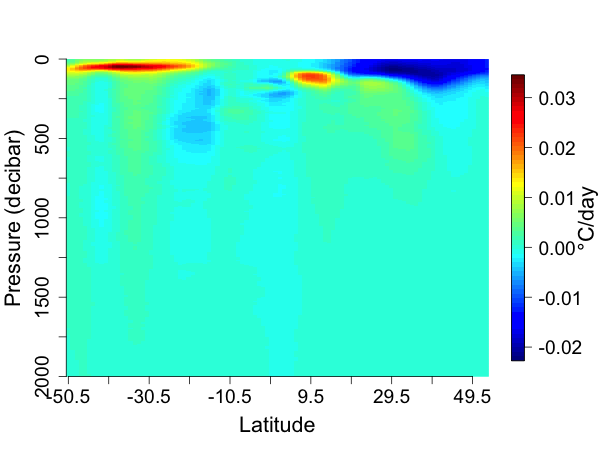}
    \end{minipage}
    \begin{minipage}{0.55\textwidth}
        \centering
       \includegraphics[width=0.95\textwidth]{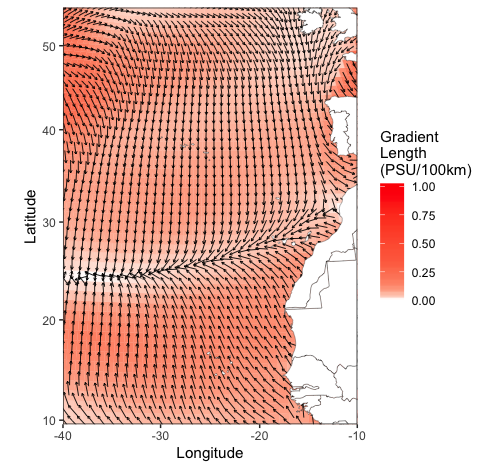}
    \end{minipage}
\caption{Estimates of derivatives of temperature in latitude (\textbf{Top Left}) and time (\textbf{Bottom Left}) for a fixed longitude 179.5 West. The derivative with respect to latitude reflects that in the middle latitudes (between (30, 50) and (-50, -30)) the temperature increases as one moves towards the equator near the surface. The derivatives also identify two separate areas of high temperature on either side of the equator near 250 dbar. The derivative with respect to time demonstrates that in mid-February, the temperature is mostly increasing in the Southern Hemisphere and mostly decreasing in the Northern Hemisphere as the time of year suggests. On the right, we give the gradient in space in PSU per 100 km for salinity at 10 dbar in mid-February (\textbf{Right}).
The gradient points toward the salinity maximum in the Central North Atlantic, identifies the flow of salty water from the Mediterranean Sea, and shows accordance with \cite{talley_chap4} Figure 4.15 that gives the distribution of sea-surface salinity in the oceans in Jan-March.}\label{functional_derivs_plot}
\end{figure}

One novelty in our approach of combining local regression and spline smoothing is its estimation of functional derivatives. 
Namely, writing the mean averaged over years as $\overline{\mu}(s,d, p)$, the functions $\left(\hat{\beta}_1(p), \hat{\beta}_2(p)\right)$ estimate $$ \left(\frac{\partial \overline{\mu}}{\partial s_1}(s, d, p),\frac{\partial \overline{\mu}}{\partial s_2}(s, d, p)\right),$$ the gradient consisting of the partial derivatives at $s = s_0$ and $d = d_0$ of the response with respect to zonal distance and meridional distance, respectively. Likewise, $(2\cdot\hat{\beta}_3 ,2\cdot \hat{\beta}_4, \hat{\beta}_5)$ estimate the second-order derivatives \begin{align*}
   \left( \frac{\partial^2 \overline{\mu}}{\partial s_1^2}(s, d, p), \  \frac{\partial^2 \overline{\mu}}{\partial s_2^2}(s, d, p),   
  \ \frac{\partial^2 \overline{\mu}}{\partial s_1\partial s_2}(s, d, p)\right),
\end{align*}and ($\hat{\beta}_6, 2\cdot \hat{\beta}_7$) estimate \begin{align*}
\left(\frac{\partial \overline{\mu}}{\partial d}(s, d, p), \  \frac{\partial^2 \overline{\mu}}{\partial d^2}(s, d, p)\right)
\end{align*}at the location and time $s = s_0$ and $d =d_0$. These functions collectively describe the local quadratic behavior of the mean near $s_0$ and $d_0$. In Figure \ref{functional_derivs_plot}, the derivatives in latitude and time for temperature are given for a cross-section of the ocean for a fixed longitude. Also, the figure includes the direction and strength of the spatial gradient at a fixed pressure of 10 dbar for salinity. These derivatives can identify the direction of warming and cooling for each location and pressure, as well as physical properties including the exchange of salty and fresh waters near the Strait of Gibraltar. Our functional approach facilitates this detailed description of the ocean properties at any pressure.

\section{Covariance Estimation} \label{covarianceest}

After subtracting the mean from the data, the spatial dependence structure of the residuals can be modeled to provide predictions and estimate uncertainties. Modeling the covariance in space, time, and pressure is a challenging task.
For example, there are considerable differences in the spatial dependence structure and residual variances at different pressures and locations.

Our covariance estimation can be described in three steps. First, we estimate the functional principal components (FPCs), which explain the first few dimensions of variability in pressure \citep[cf.][Chapter 9]{Hsing2015}. Next, each profile is summarized by these principal components, and the resulting scores are modeled. Lastly, we estimate the remaining variability not accounted for by the principal components. The implicit assumptions in this approach are that the covariance structure of temperature and salinity changes smoothly as a function of pressure and only a small number of FPCs are needed to approximate the spatial and temporal structure in pressure. Since modeling the dependence between the raw measurements in space, time, and pressure simultaneously is not practical or appropriate due to the number of observations and flexibility of covariance models, our functional approach facilitates a dimension reduction strategy that shares information across pressure through the FPCs.

We develop this approach in mathematical notation first by assuming \begin{align}
    Y_{i,j}^0 &=  X(s_i,d_i,y_i,p_{i,j}) + \epsilon_{i,j}\label{dgp}
\end{align}where $\{Y_{i,j}^0\}_{j=1}^{m_i}$ are the residuals for profile $i$ formed by subtracting the mean estimate from the data, $X(s, d, y_i, \cdot)$ for $y_i = 2007, \dots, 2016$ are identically-distributed realizations of a functional random field with mean $0$, and $\epsilon_{i,j}\stackrel{ind}{\sim} N(0,\kappa(p,s,d))$ is an independent measurement error noise with mean $0$ and finite variance 
that may depend on pressure, location, and day of the year (see Section \ref{me_var}, below).
If $X(s_i, d_i, y_i, \cdot) \in \mathbb{L}_2$ for each $s_i$ and $d_i$, one can write \begin{align*}
    X(s_i,d_i,y_i, p) &= \sum_{k=1}^\infty Z_{k}(s_i, d_i, y_i) \phi_k(p)
\end{align*}where $\phi_k$ are fixed orthonormal functions, and the $Z_{k}(s, d, y)$ are scalar random fields that are weakly dependent in time that we refer to as \textit{scores}. This is similar to the Karhunen-Lo\'eve expansion for zero-mean square-integrable stochastic processes, though the scores may be correlated across $k$ due to their spatial dependence. In the subsequent development, we simplify the notation by defining $Z_{i,k} = Z_{k}(s_i, d_i, y_i)$ where it does not cause confusion.
For an adequate choice of $\phi_k$, we would expect that $X(s_i, d_i,  y_i, \cdot)$ can be approximated as 
\begin{align}
   X(s_i,d_i,y_i,p) = \sum_{k=1}^{K_1} Z_{i,k}\phi_k(p)\label{pca_approx}
\end{align}
for some small number $K_1$. This effectively reduces the dimension of our problem. Here, each $\phi_k$ is a fixed function that has been estimated through some form of functional principal component analysis, with one such approach given in Section \ref{marginalest}. For a choice of $\phi_k$ and a profile $i$, the scores are estimated by the least squares solution \begin{align}Z_{i, \cdot} = \left(\Phi_i^\top \Phi_i\right)^{-1} \Phi_i^\top Y_i^0\label{estimate_scores}\end{align}where $Y_i^0$ is are the residuals for profile $i$ and $\Phi_i \in \mathbb{R}^{m_i \times K_1}$ is the matrix with $j,\ell$ entry $\phi_{\ell}(p_{i,j})$. The principal component functions $\phi_k$ and the scores $Z_{i, \cdot}$ are only estimates and not the truth, though we use the same notation for convenience. We found that the alternative approach to estimating the scores proposed by \cite{yao_functional_2005} to give similar results, though this approach is computationally expensive to implement on a large scale.

We assume that the decomposition (\ref{pca_approx}) of $X(s_i, d_i, y_i,p)$ holds locally with respect to both $Z_{i,k}$ and $\phi_k$, similar to the locally stationary assumption of \cite{kuusela2017locally}. That is, for a fixed location $s_0$ and time $d_0$, the functions $\phi_k$ are estimated and used to form estimates of the $\{Z_{i,k}\}_{k=1}^{K_1}$ and the measurement error variance $\kappa(p) := \kappa(s_0,d_0,p)$ for all nearby profiles. Next, the joint distribution of the nearby scores is modeled. For different choices of $(s_0,d_0)$, the functions $\phi_k$ and resulting scores $\{Z_{i,k}\}_{k=1}^{K_1}$ and measurement error variance $\kappa(p)$ are different.

The model gives a clear approach to address the fundamental problem of functional kriging, i.e. spatial prediction of functional data, using the conditional distribution at an unobserved location given the data observed. For any set of data $Y^0$, to provide a prediction for the function-valued random field $X(s_*, d_*, y,\cdot)$ for an unobserved location $s_*$ at time $d_*$, one has \begin{align}
\mathbb{E}\left\{X(s_*, d_*, y,p) | Y^0\right\} &= \phi(p)^\top \mathbb{E}\left\{Z_{\cdot}(s_*, d_*, y) | Y^0 \right\} \label{predictivemean_prelim}\\
\Var\left\{X(s_*, d_*,y,p) | Y^0\right\} &= \phi(p)^\top \Var\left\{Z_{\cdot}(s_*, d_*,y) | Y^0\right\}\phi(p)\label{predictive_prelim}
\end{align}where $\phi(p) = \left( \phi_1(p), \  \phi_2(p), \ \dots, \ \phi_{K_1}(p)\right)^\top$ are the principal components and $Z_{ \cdot}(s_*, d_*,y) = \begin{pmatrix} Z_{1}(s_*, d_*, y), \ \dots,\   Z_{K_1}(s_*, d_*,y)\end{pmatrix}^\top$ are the scores of $X(s_*, d_*, y,\cdot)$. Furthermore, for each residual point $Y_{i,j}^0$, \begin{align}
    \mathbb{E}\left\{Y_{i,j}^0 | Y^0\right\} &= \mathbb{E}\left\{X(s_i, d_i,y_i, p_{i,j}) | Y^0\right\}\label{predictivemean}\\
\Var\left\{Y_{i,j}^0 | Y^0\right\} &= \Var\left\{X(s_i,d_i,y_i,p_{i,j}) | Y^0\right\} + \kappa(p_{i,j})\label{predictive}
\end{align} Thus, if one assumes that the field of $\{Z_{k}(s,d,y); k=1, \dots, K_1; (s,d) \in \mathbb{R}^3\}$ is Gaussian, one only needs a spatio-temporal model of the scores $Z_{k}(s,d,y)$ for $k=1, \dots, K_1$ using the conditional mean and variance, as well as estimate $\kappa(p)$.
We address the estimation of $\phi_k(p)$ in Section \ref{marginalest}, the modeling of the scores $Z_{i,k}$ in Section \ref{scoremodelling}, and the estimation of $\kappa(p)$ in Section \sref{me_var}.

\subsection{Marginal Covariance Estimation in Pressure}\label{marginalest}

In this section, we focus on the estimation of $\phi_k$ in (\ref{pca_approx}), which amounts to performing local functional principal component analysis (FPCA). 
 A fixed set of basis functions may not be suitable for different locations or seasons, and the resulting decomposition would be suboptimal at most locations. We thus estimate $\phi_k$ locally in space and time as done with the mean to provide an optimal decomposition. At each location, a local version of the approach given in Section 8.3 of \cite{Hsing2015} is used to estimate the entire within-profile covariance. Then, the covariance is decomposed to obtain the functional principal components. This approach uses data from both sparse and dense profiles and avoids needing a basis representation of each profile as in \cite{Ramsay:2013hf}. Also, it resembles our approach for mean estimation by treating the covariance as an expectation, and it provides advantages over other approaches like thin plate splines by using B-splines that greatly reduce computations \citep{wahba1990spline}. 

For fixed $s_0$ and $d_0$, we solve the optimization problem:
\begin{equation}
    \min_{f_{s_0, d_0} \in \mathbb{W}_2 \otimes \mathbb{W}_2}  \left(\ell_{s_0, d_0}(f_{s_0, d_0}) + \textrm{Pen}_{f_{s_0, d_0}}(\lambda)\right)
\end{equation}where \begin{equation}
\ell_{s_0, d_0}(f_{s_0, d_0}) = \frac{1}{n} \sum_{i=1}^n \frac{K_{h_s, h_d}(s_i - s_0, d_i - d_0)}{m_i(m_i - 1)} 
\mathop{\sum\sum}_{1 \leq j\neq k \leq m_i}
\left(Y_{i,j}^0Y_{i,k}^0 - f_{s_0, d_0}(p_{i,j}, p_{i,k})\right)^2.\label{margcovopt}
\end{equation}
In particular, $f_{s_0, d_0}$ is restricted to be of the form $$f_{s_0, d_0}(p_1, p_2) = \sum_{k_1=1}^M\sum_{k_2 = 1}^M \alpha_{k_1, k_2}\chi_{k_1}(p_1)\chi_{k_2}(p_2)$$where $\{\alpha_{k_1, k_2}\}_{k_1, k_2 = 1}^M$ are scalar coefficients and $\{\chi_k(p)\}_{k=1}^M$ is a univariate B-spline basis over a fixed set of knots. 
As suggested in \cite{wood_low-rank_2006}, the penalty used is $$ \textrm{Pen}_{f_{s_0, d_0}}(\lambda) =\lambda \textrm{vec}(\alpha)^\top(\Omega \otimes I_M  + I_M \otimes \Omega) \textrm{vec}(\alpha)$$where $\otimes$ is the standard Kronecker product, $\Omega$ is the univariate smoothing matrix for the B-splines used with $k_1,k_2$ entry $\int_{0}^{2000} \chi_{k_1}^{(2)}(p)\chi_{k_2}^{(2)}(p) dp$, and $I_M$ is the $M\times M$ identity matrix. 
This penalty approximates $$\lambda \int_0^{2000}\int_0^{2000}\left[ \left(\frac{\partial^2 f_{s_0, d_0}}{\partial p_1^2}\right)^2 + \left(\frac{\partial^2 f_{s_0, d_0}}{\partial p_2^2}\right)^2 \right]dp_1dp_2$$as given in \cite{wood_low-rank_2006}.
The computation is similar to the approach for mean estimation, with $\lambda$ chosen by cross validation and using a product kernel with $h_s=550$ kilometers and $h_d = 45.25$; this smaller spatial bandwidth is possible since we pool together data from all years.
We use $M=102$ with equally spaced knots over $[0, 2000]$ for the basis $\chi_k$. 
The overall size of the problem is $M^2$, whose computational cost increases much faster compared to the mean estimation. This choice of knots is able to approximate the covariance operator reasonably well while ensuring the calculations are computationally manageable. The exclusion of points with $j=k$ ensures that the measurement error $\epsilon_{i,j}$ is not included in the estimates along the diagonal. This allows us to 
formally identify the variance parameter of the measurement error in \eqref{dgp}, similar to \cite{yao_functional_2005}. 

The main goal of the covariance estimation is to obtain a basis of functional principal components for the space-time modeling; we detail how to obtain orthonormal principal component functions in Section \sref{pc_ortho}.
At this point, we estimate the FPCs for temperature and salinity separately. By working in the corresponding FPC bases for future modeling, we optimally reduce the infinite-dimensional kriging problem in pressure to a finite-dimensional one using principal components.

\begin{figure}[t]
\includegraphics[scale = .07]{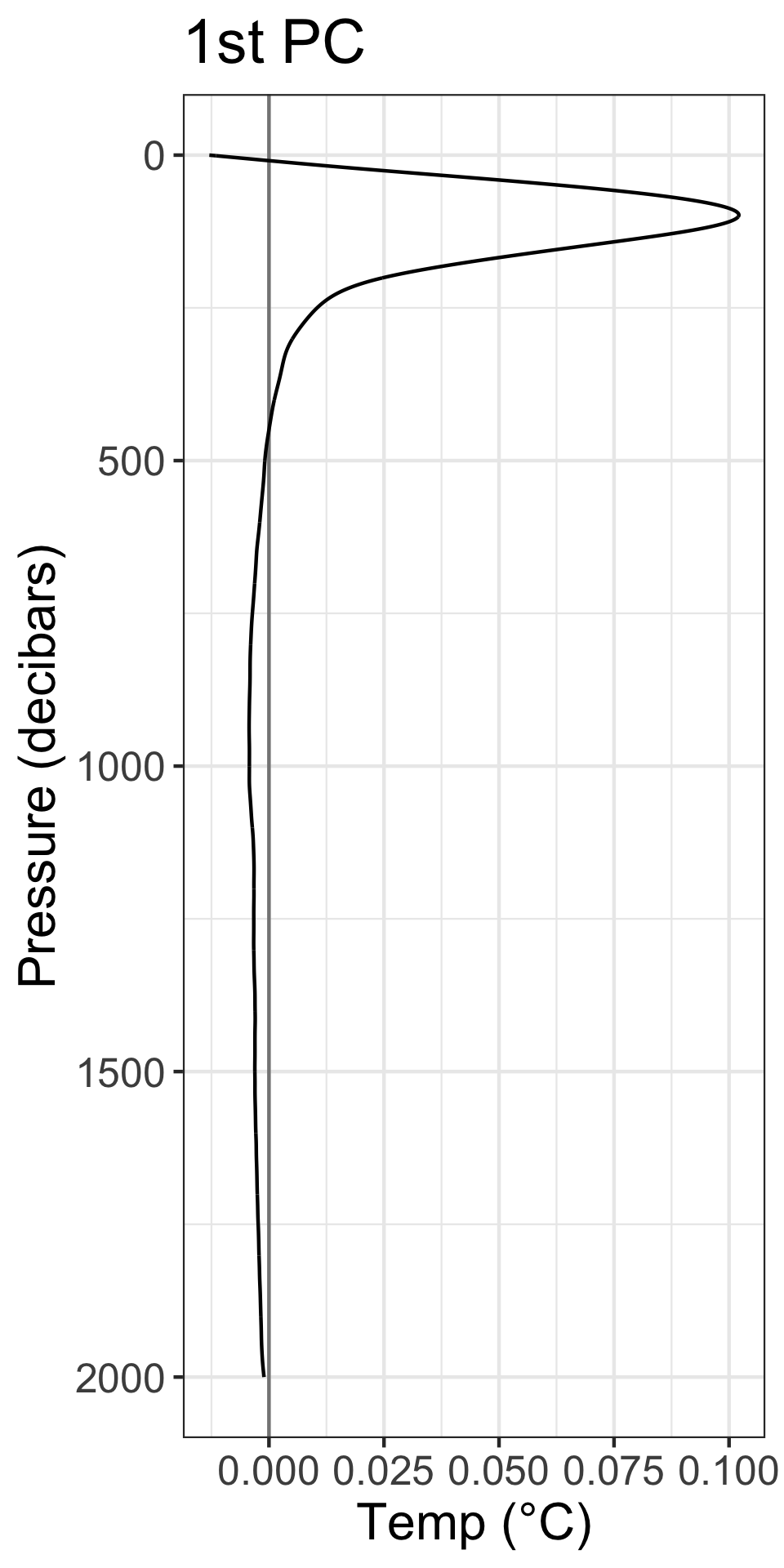}\includegraphics[scale = .07]{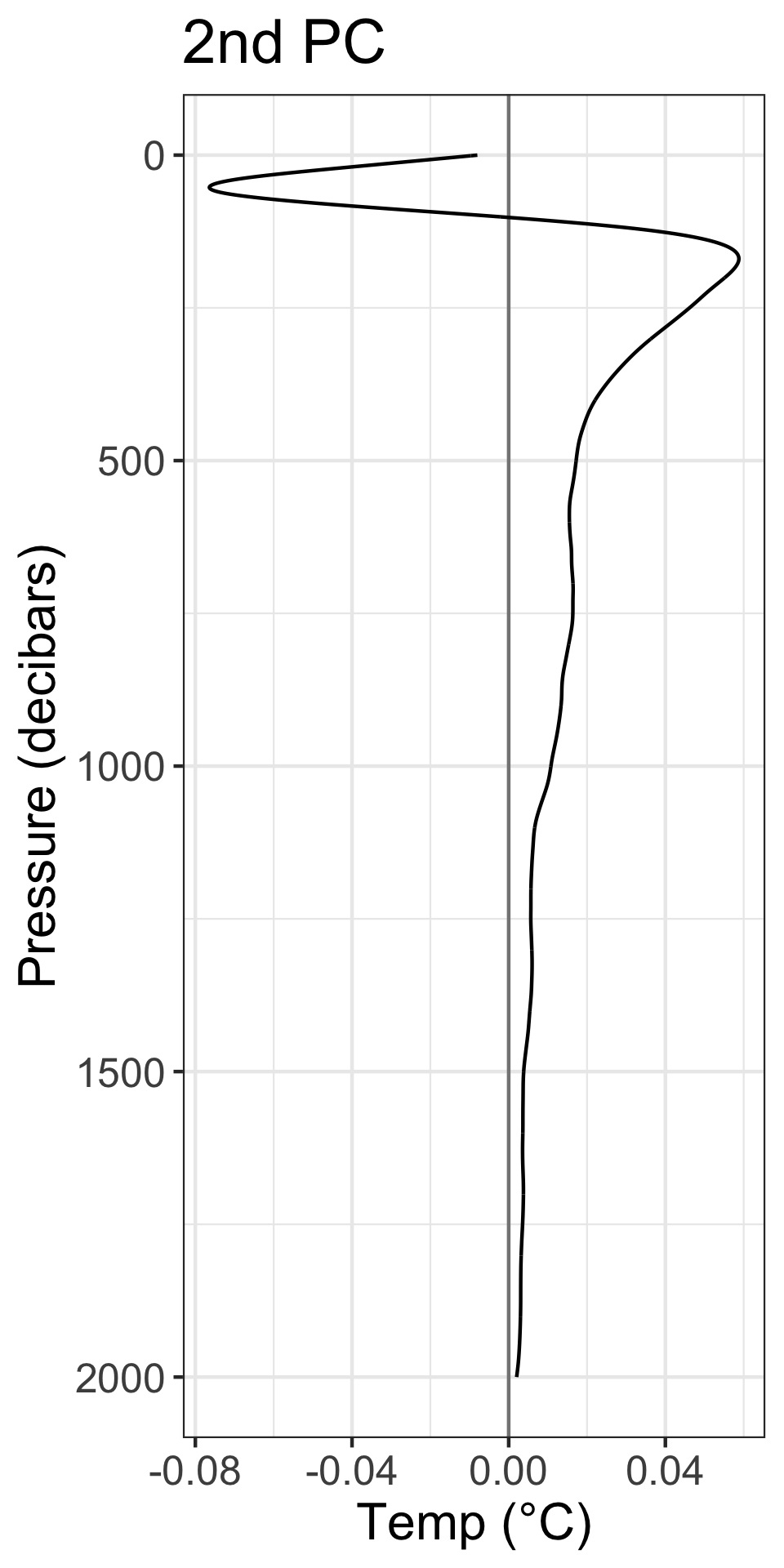}
\includegraphics[scale = .07]{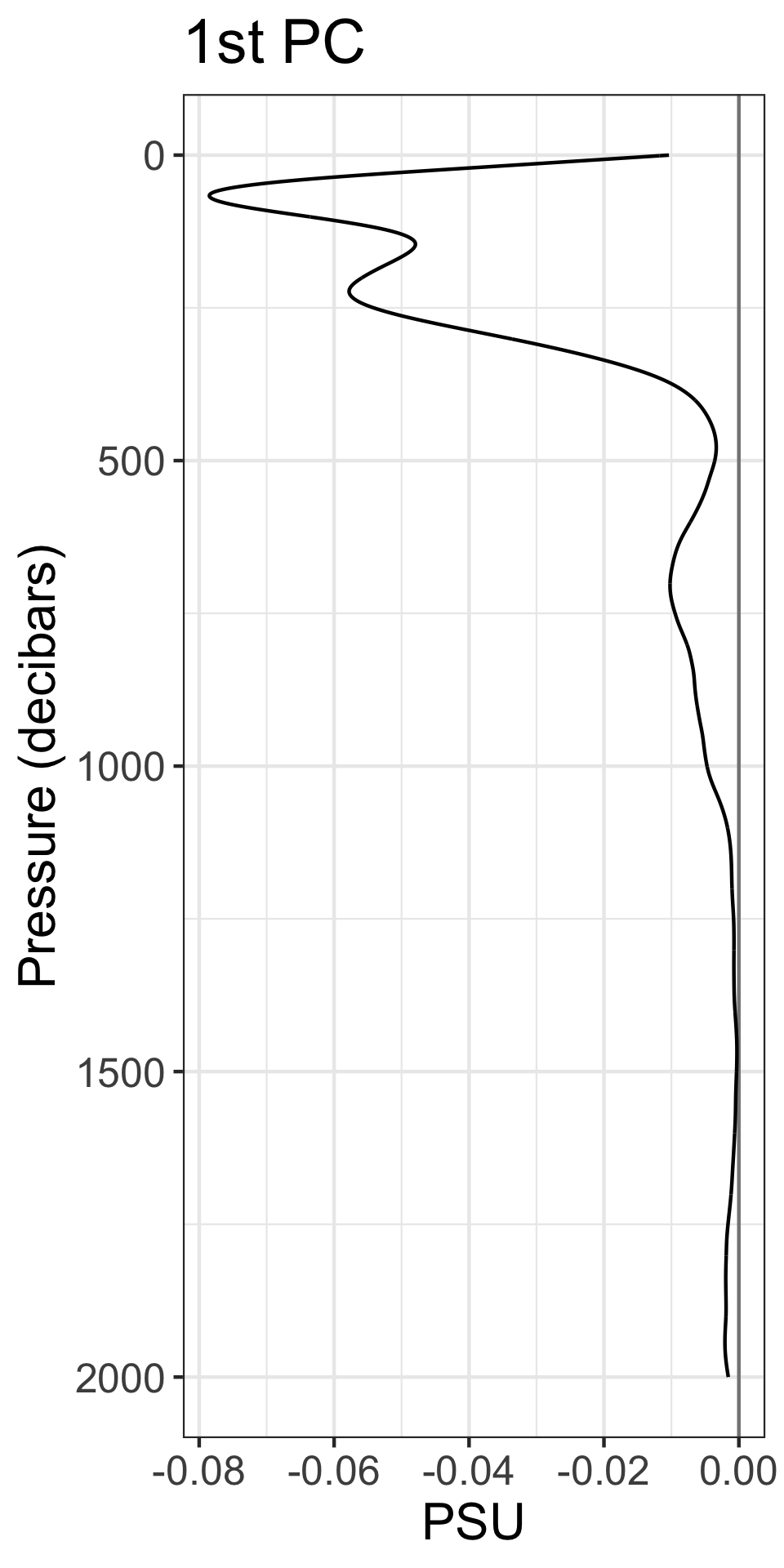}
\includegraphics[scale = .07]{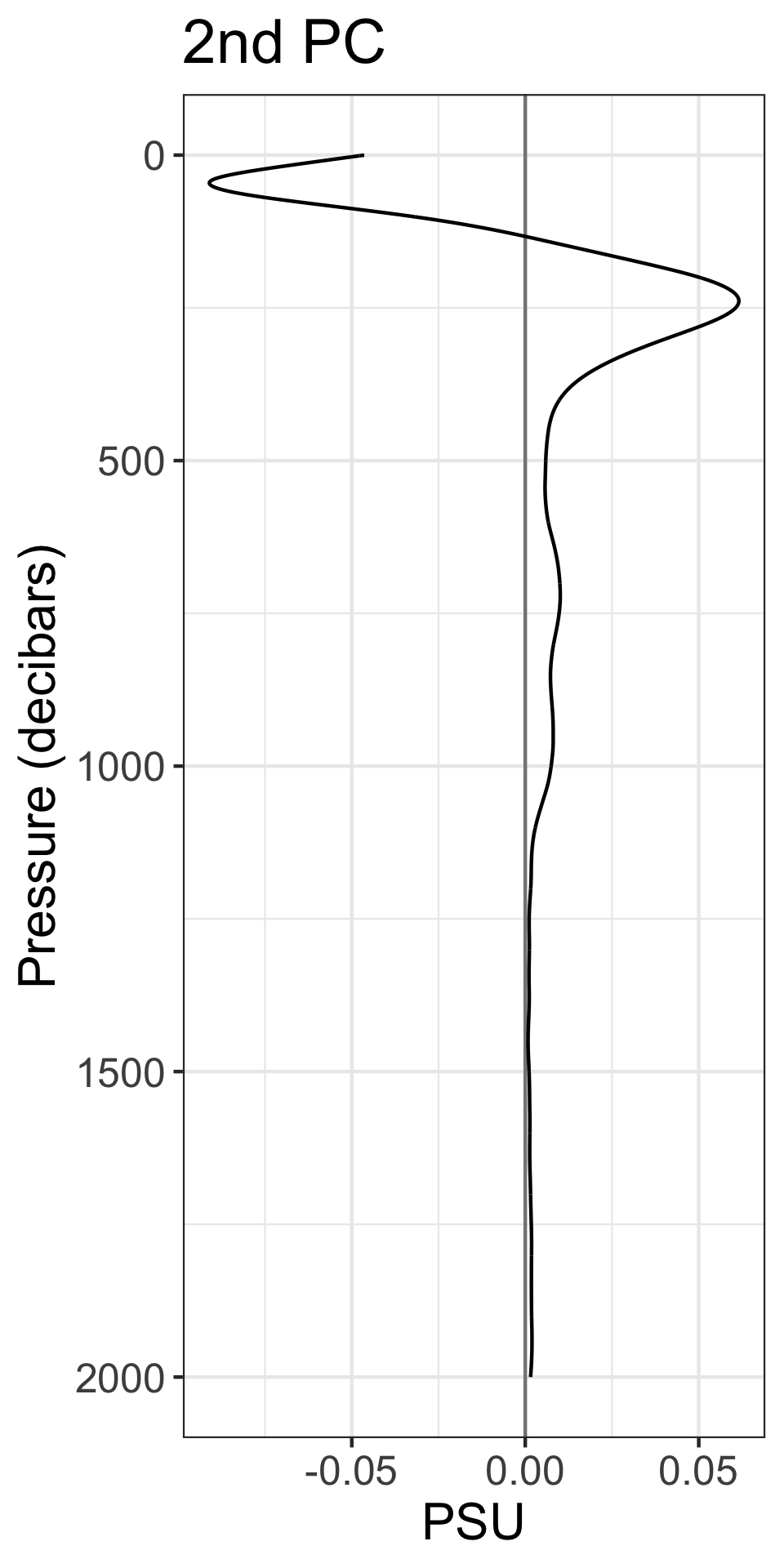}
 \caption{Example of first two estimated functional principal components (\textbf{Left}) temperature (\textbf{Right}) salinity at Long 90.5 W and Lat -10.5 S. The principal components suggest much higher variance near the surface of the ocean, as expected.
 }\label{pcexample}
\end{figure}

An example of the first two functional principal components for a location is shown in Figure \ref{pcexample}. 
Similar plots for other locations can be viewed on an {\tt R} Shiny application \citep{shinyapps}. These principal components can give descriptive information on the variance and dependence of temperature and salinity with respect to pressure.
There is evidence that the covariance and the principal components for temperature and salinity exhibit considerably different structure. 

\begin{remark}
For any fixed location, the principal components are only identifiable up to a sign. For this reason, only one basis $\{\phi_k\}_{k=1}^K$ is used at one time, and the scores are only defined and predicted with respect to this fixed basis. Thus, the products of each score and principal component are invariant to the sign of the principal component. Local regression helps ensure, assuming a sensible smoothing parameter selection, that the estimated covariance varies smoothly as one moves in space. One interesting problem of future research is the estimation of the marginal covariance operator as a function of space and time. This will require a careful registration and alignment of the principal components and scores when moving from location to location.
\end{remark}

\subsection{Space-time modeling of scores}\label{scoremodelling}

In this section, we model the scores for spatio-temporal prediction. 
In standard FDA, the scores are uncorrelated latent variables, and it is not always meaningful to model and predict them. For any two mean-zero, square-integrable random functions $X_i(p) = \sum_{k=1}^\infty Z_{i,k} \phi_k(p)$ for $i=1, 2$, the covariance becomes $\textrm{Cov}(X_1(p_1), X_2(p_2)) = \sum_{k_1=1, k_2 = 1}^\infty \phi_{k_1}(p_1) \phi_{k_2}(p_2) \mathbb{E}(Z_{1,k_1}Z_{2,k_2})$. When $X_1$ and $X_2$ are independent, $\mathbb{E}(Z_{1,k_1}Z_{2,k_2}) = 0$, and one cannot leverage any dependence between the scores. However, for spatially dependent functional data, $\textrm{Cov}(Z_{1,k_1}, Z_{2, k_2})$ may not be $0$ and may depend on the distance between the location of profiles $1$ and $2$. This motivates our approach to model the dependence of the scores and utilize it 
for spatial prediction. 

Focusing on the estimation of the random, mean-zero function in (\ref{pca_approx}), we write \begin{align*}
   T^0(s_i,d_i,y_i,p) = \sum_{k=1}^{K_1}Z_{i,k}\phi_k(p),\ \textrm{and} \ S^0(s_i, d_i, y_i,p) &= \sum_{k=1}^{K_2} W_{i,k}\psi_k(p),
\end{align*}where $T^0$, $Z_{i,k} := Z_{k}(s_i, d_i, y_i)$, and $\phi_k$ denote the respective terms of (\ref{pca_approx}) for temperature, and $S^0$, $W_{i,k}:=W_{k}(s_i, d_i, y_i)$, and $\psi_k$ denote the terms for salinity. For the modeling, we adopt the locally stationary assumption of \cite{kuusela2017locally}. For each location, as described in (\ref{estimate_scores}), we use the $\phi_k$ estimated at that location to compute the temperature scores $Z_{i,k}$ for profiles within some radius of that location, and likewise use the respective terms for salinity, $\psi_k$, to compute $W_{i,k}$. We exclude a small fraction of profiles that do not have sufficient measurements to compute scores. 
The goal of this section is to estimate a predictive distribution for the vector $$\begin{pmatrix} Z_{*,\cdot} \\ W_{*, \cdot}\end{pmatrix} = \left( Z_{*, 1}, \ Z_{*,2}, \ \cdots, \ Z_{*,K_1},\  W_{*,1}, \ W_{*,2}, \ \cdots,\ W_{*,K_2} \right)^\top,$$at an unobserved location to jointly model temperature and salinity. We first introduce our decorrelation step as explained below, which is similar to \cite{bachoc_spatial_2020}. 

For the modeling of the resulting scores, let $\Sigma_{\textrm{scores}}$ be a $(K_1+K_2) \times (K_1+K_2)$ marginal covariance matrix of $\left(Z_{i, \cdot}^\top, W_{i, \cdot}^\top\right)^\top$. 
This matrix $\Sigma_{\textrm{scores}}$ is estimated by \begin{align*}
    \hat{\Sigma}_{\textrm{scores}} &= \frac{1}{|D_{s_0}| - 1}\sum_{i\in D_{s_0}} \begin{pmatrix} Z_{i,\cdot} \\  W_{i,\cdot} \end{pmatrix}\begin{pmatrix} Z_{i,\cdot}^\top ,&  W_{i,\cdot}^\top \end{pmatrix}
\end{align*}where $D_{s_0}$ are the set of nearby delayed mode profiles. Then, consider the standard eigendecomposition $$\hat{\Sigma}_{\textrm{scores}} = V\Gamma V^\top$$where $\Gamma$ is a diagonal matrix, and define \begin{align}\begin{pmatrix} \tilde{Z}_{i,\cdot} \\ \tilde{W}_{i,\cdot} \end{pmatrix} = V^\top \begin{pmatrix} Z_{i,\cdot} \\  W_{i,\cdot} \end{pmatrix}.\label{score_transformation}\end{align}The resulting transformed scores $\left( \tilde{Z}_{i,\cdot}^\top, \ \tilde{W}_{i,\cdot}^\top \right)^\top$ are then approximately decorrelated, with diagonal auto-covariance matrix $\Gamma$.

Let $M(\nu, \Delta)= c_1 \Delta^\nu \mathcal{K}_\nu(\Delta)$ be the Mat\'ern covariance with parameter $\nu$ at distance $\Delta$ with unit variance and scale, studied in, for example, \cite{stein_interpolation_2013}. Here, $c_1$ is a constant so that $M(\nu, 0) = 1$, and $\mathcal{K}_\nu$ is the modified Bessel function of the second kind. The value of $\nu > 0$ governs the smoothness of the field of scores, where larger values give a smoother field. When $\nu = 1/2$, the Mat\'ern model reduces to the exponential function. In our experiments, the choice of $\nu$ had minimal effects on the resulting predictions, and we set it to the common choice $\nu = 1/2$ as in \cite{kuusela2017locally}. For $\tilde{Z}_{i,k}$ and $\tilde{W}_{i,k}$ and each $k$, a Mat\'ern model is fitted for the decorrelated scores of the form $ \mathbb{E}\left(\tilde{Z}_{i,k}\tilde{Z}_{j,k}\right) = C_k(\Delta_{i,j})$ or $\mathbb{E}\left(\tilde{W}_{i,k}\tilde{W}_{j,k}\right) = C_{K_1 + k}(\Delta_{i,j})$ if $y_i = y_j$ with
\begin{align}\begin{split}\label{spatialmodel}C_k(\Delta) &= \gamma_k\cdot M\left(\nu, \sqrt{\left(\frac{\Delta_{s_1}}{\theta_{s_1,k}}\right)^2+\left(\frac{\Delta_{s_2}}{\theta_{s_2,k}}\right)^2+\left(\frac{\Delta_{d}}{\theta_{d,k}}\right)^2}\right)+\sigma^2_k \cdot 1(\Delta = 0),\end{split}\end{align} where $\Delta = (\Delta_{s_1}, \Delta_{s_2}, \Delta_d)$ is a vector of corresponding distances in space and time. The parameters $\theta_{s_1,k}$, $\theta_{s_2,k}$ and $\theta_{d,k}$ are scale parameters that specify the correlation ranges for each of the directions. Lastly, $\gamma_k$ and $\sigma_k^2$ are parameters that describe the variance of the spatial process and the nugget, respectively. This space-time model is considered in \cite{kuusela2017locally} in their fixed pressure level analysis.

In summary, the resulting covariance of temperature and salinity is \begin{align}
\mathbb{E}\left\{\begin{pmatrix} T^0(s_i, d_i, y_i, p_1) \\ S^0(s_i, d_i, y_i, p_1) \end{pmatrix}\begin{pmatrix} T^0(s_j, d_j, y_j, p_2) \\ S^0(s_j, d_j, y_j,p_2) \end{pmatrix}^\top \right\}&= \Xi_{p_2}^\top V C(\Delta_{i,j}) V^\top \Xi_{p_1}\label{full_covariance}
\end{align}if $y_j = y_i$ and $0$ otherwise, where $C(\Delta_{i,j})\in \mathbb{R}^{(K_1+K_2) \times (K_1+K_2)}$ is the diagonal matrix with the $k$-th element $C_k(\Delta_{i,j})$, and $\Xi_p = \begin{pmatrix} \phi(p) & 0 \\ 0 & \psi(p)\end{pmatrix}\in \mathbb{R}^{(K_1+K_2) \times 2}.$ This model, by considering a nugget effect on each of the scores, also results in a kind of ``functional nugget'' 
for the process as described in  \citep{zhang_spatial_fda}. This functional nugget has covariance $$\Xi_p^\top V C_{\sigma} V^\top \Xi_p,$$ where $C_{\sigma} = \textrm{diag}(\sigma_1^2, \sigma_2^2, \dots, \sigma_{K_1+ K_2}^2)$.

We estimate the spatial model for February at each location using data from January, February, and March. For each location, profiles within 1,100 kilometers were used \citep[similar to the size of the moving windows used in][]{kuusela2017locally}. We set $K_1 = K_2 = 10$, which allows the profiles to be well represented by the principal components, though our experiments suggest that using more principal components may slightly improve predictions near the surface. We provide the reasoning of this choice in Section \sref{numpcs}, where we show that 10 components explain a large proportion of the variability in both temperature and salinity. Choosing the number of functional principal components under a smoothly-varying covariance structure in space could be developed based on \cite{li_selecting_2013}. To estimate the parameters $\gamma_k$, $\theta_{s_1,k}$, $\theta_{s_2,k}$, $\theta_{d,k}$, and $\sigma_k^2$ for each $k$, we employ the same approach as \cite{kuusela2017locally} using maximum likelihood summarized below. 
Let $\tilde{Z}_y$ be the scores for one $k$ for year $y$ in each of the above models, and let $\textrm{Var}(\tilde{Z}_y) = \Sigma_y$ be a matrix specified by the parameters in (\ref{spatialmodel}) above. We assume that $\tilde{Z}_y$ are multivariate Gaussian, so that the log likelihood of the data for all $y= 2007, \dots, 2016$ is \begin{align*}
-\frac{1}{2}\left(\sum_{y=2007}^{2016}  \log(\textrm{det}(\Sigma_y) ) + \tilde{Z}_y^\top \Sigma_y^{-1}     \tilde{Z}_y + n_y \log(2\pi) \right). 
\end{align*}where $n_y$ is the number of observations used in year $y$. This likelihood treats data from different years as independent. 
To maximize the likelihood, we use the optimization L-BFGS-B algorithm due to \cite{optim_bfgs} implemented in the \texttt{optim} function in \texttt{R}.

One challenge is that quality control is essential for the salinity data; many Argo profiles that have high-quality temperature data may not have the same quality of salinity data. %
Instead of discarding such profiles, we offer a solution by using an established missing data approach via an expectation-maximization-type (EM) algorithm:

\begin{enumerate}

    \item (E step) Using the temperature, delayed-mode salinity data, and the estimated parameters, form a prediction (the conditional expectation) for the real-time salinity scores. In the first iteration, the prediction of the real-time salinity scores are $0$.

    \item (M step) Using all data (obtained from the E step) as if it were delayed-mode, estimate the model parameters via maximum likelihood.
    
    \item Alternate between the E and M steps and repeat until a convergence criterion is met.  
    
\end{enumerate}

This treats the real-time salinity scores as unobserved, latent variables. The differences between estimated parameters from consecutive steps decrease quickly after the first few steps. Therefore, at each grid point, we decided to perform 6 iterations of the algorithm, and the parameter estimates from the final M step are used. This choice strikes a balance between computation time and statistical accuracy. 

Estimating the joint dependence between temperature and salinity is not considered in \cite{kuusela2017locally} and \cite{roemmich20092004}. While it requires additional computation, accounting for this dependence provides a more comprehensive analysis of the Argo data. In particular, this enables us to predict and provide uncertainty estimates for functionals of temperature and salinity such as potential density and potential temperature.
The estimated parameters can be viewed using an {\tt R} Shiny application \citep{shinyapps}.

\subsection{Predictions, Uncertainties, and Prediction Bands} \label{Predictionsbands}

In this section, we employ the estimated spatial covariance for functional kriging. Under the assumptions of our model, this provides an optimal functional prediction at an unobserved location. 
To detail this approach, let $\Sigma_{y*}$ be covariance matrix of the (true) decorrelated scores $\tilde{Z}_{y*} = (\tilde{Z}_{i,k})_{i=1}^{n_{y_*}}$ for a fixed $k$ in an area around a fixed location for a year $y_*$. 
Notably, using the local stationarity assumption, profiles within 1,100 kilometers are used as in the Mat\'ern estimation step. This provides enough data for prediction while avoiding introducing data that may violate the locally-stationary assumption. The conditional distribution of $\tilde{Z}_{*,k} := \tilde{Z}_{k}(s_*,d_*,y_*)$ at an unobserved location given
$\tilde{Z}_{y*}$ is 
$$
\tilde{Z}_{*,k} \big| \tilde{Z}_{y*}  \sim N\left(\Sigma_{12}^\top (\Sigma_{y*})^{-1}  \tilde{Z}_{y*}, \gamma_k + \sigma^2_k - \Sigma_{12}^\top (\Sigma_{y*})^{-1} \Sigma_{12}\right)
$$
where $\Sigma_{12} = \textrm{Cov}\left(\tilde{Z}_{*,k}, \tilde{Z}_{y*}\right)$. However,
in our prediction problem, $\tilde{Z}_{y*}, \Sigma_{12}$, and $\Sigma_{y*}$ are unknown and are
estimated by the approaches described in Section \ref{scoremodelling}.

We similarly obtain the predictions for the decorrelated salinity scores $\tilde{W}_{*,k}$. From these estimated distributions of the $\tilde{Z}_{*,k}$ and $\tilde{W}_{*,k}$, using the relation that $\begin{pmatrix} Z_{*} \\ W_{*} \end{pmatrix} = V \begin{pmatrix} \tilde{Z}_{*} \\ \tilde{W}_{*} \end{pmatrix}$ described in (\ref{score_transformation}), the conditional distribution of the original scores is $$ \begin{pmatrix} Z_{*} \\ W_{*} \end{pmatrix} \Big| \tilde{Z}_{y*} = V \begin{pmatrix} \tilde{Z}_{*} \\ \tilde{W}_{*} \end{pmatrix} \Big|\tilde{Z}_{y*} \sim N\left( V \mathbb{E}\left\{ \begin{pmatrix} \tilde{Z}_{*} \\ \tilde{W}_{*} \end{pmatrix} \Big| \tilde{Z}_{y*} \right\}, V \textrm{Var}\left\{ \begin{pmatrix} \tilde{Z}_{*} \\ \tilde{W}_{*} \end{pmatrix}\Big| \tilde{Z}_{y*} \right\} V^\top \right).$$The conditional distribution of $T^0(s,d,y, p)$ and $S^0(s,d,y,p)$ can be found using (\ref{full_covariance}) or (\ref{predictivemean}) and (\ref{predictive}), providing a prediction for any pressure. In Section \sref{anomalyexample}, we give an example prediction for one pressure.

We test the uncertainty estimates based on this model in a leave-one-profile-out manner. For each February profile, the profile is left out, and nearby profiles are used to predict at the location and time of the profile. Then, the left-out profile is compared with the predictions. 
For salinity, only delayed-mode profiles are compared.
For each quantity, we use bounds of two standard deviations from the mean, which corresponds to approximately a 95.4 percent prediction interval. For brevity, we develop uncertainties for temperature, and similar bounds are obtained for salinity. 
We consider both pointwise and uniform prediction bounds on the residual curves $Y_{i,j}^0 = \phi(p_{i,j})^\top Z_{*, \cdot} + \epsilon_{i,j}$. 
The pointwise $1-\alpha$ interval for the residual at pressure $p_{i,j}$, based on (\ref{predictivemean}) and (\ref{predictive}), is \begin{align*}
\phi(p_{i,j})^\top \mathbb{E}\left\{Z_{*,\cdot}| \tilde{Z}_{y*}  \right\} \pm q_{1-\alpha/2}\sqrt{\phi(p_{i,j})^\top \Var\left\{Z_{*,\cdot} | \tilde{Z}_{y*} \right\}\phi(p_{i,j}) + \hat{\kappa}(p_{i,j})}.
\end{align*}where $q_{1-\alpha/2}$ is the $1-\alpha/2$ quantile of $N(0,1)$, and $\hat{\kappa}$ is estimated as described in \sref{me_var}.
In addition, we develop simultaneous predictions bands over pressure by using the approach of \cite{choi_geometric_2018} reviewed in Section \sref{bands_appendix}.

In our empirical coverages in Table \ref{coverage} (where $K_1 = K_2 = 10$), the intervals and bands show good coverage for both temperature and salinity. In Table \ref{coverage}, the band coverage refers to the proportion of profiles for which every observation of the left-out profile was covered by the estimated band. The pointwise coverages correspond to the proportion of all measurements covered by the intervals over all pressures. We also summarize the pointwise coverages by pressure in Figure \sref{cv_example}, and the coverage is achieved for most of the pressure dimension, though typically the intervals in the range 20-200 dbar do not meet full coverage due to more complex processes near the surface.

 
  \begin{table}[t]
   \caption{Average pointwise coverages of intervals and bands over all pressures}\label{coverage}
 \begin{tabular}{|c||c|c|c||}
 \hline 
 Quantity & \#  Profiles &  Pointwise  Coverage  &Band  Coverage  \\
 \hline
 Temperature & 76,016 & 96.2 & 95.6 \\
 \hline
 Salinity & 45,188 & 98.2 & 96.6\\
\hline
 Nominal level &  & 95.4 & 95.4 \\
 \hline
 \end{tabular}
 \end{table}
 


\subsection{Validation and comparison}\label{compare_kuusela}

We compare our approach with the Roemmich and Gilson (RG) reference model and Model 5 of \cite{kuusela2017locally} (KS) which provide predictions only at fixed pressure levels. In Section \sref{ks_exact_compare}, we compare the differences between the KS and functional predictions at 10, 300, and 1500 dbar and find them to be generally comparable. Also, we can compare the predictive errors through the cross validation approach described in the previous subsection. Our functional approach enables the prediction of temperature and salinity \textit{without} interpolation onto fixed pressure levels. To provide a comparison with the fixed pressure levels of KS, we compute summaries of the residuals by breaking up the interval $[0,2000]$ using the midpoints of the Roemmich and Gilson pressure levels. For example, the intervals $(6.25, 15]$, $(290,310]$, and $(1456.25,1550]$ correspond to the 10 dbar, 300 dbar, and 1500 dbar levels, respectively.
Not all profiles are included in the comparison. For KS, profiles are removed in boundary seas and where the interpolation fails, that is, where there are no measurements either above or below the relevant pressure level, and we remove them in this comparison as well and only use profiles included in KS at any of 10, 300, and 1500 dbar. The prediction errors are evaluated by the root mean squared error (RMSE) defined as $\sqrt{\frac{1}{n}\sum_{i=1}^n (y_{i,p} - \hat{y}_{i,p})^2 }$ and the $50\%$ (median) and $75\%$ (3rd quartile) quantiles of $|y_{i,p} - \hat{y}_{i,p}|$ where $y_{i,p}$ are the measurements corresponding to pressure level $p$, and $\hat{y}_{i,p}$ are the predictions for that measurement. 

We show the results in Table \ref{KS_compare} and Figure \sref{rg_comparison} and comment on them. Our method outperforms the Roemmich and Gilson-type reference model and has approximately the same the prediction error as KS.
Notably, we suspect that avoiding interpolation onto pressure levels considerably improves our prediction error, especially at greater depths. For example, at 1500 dbar, the RMSE for the functional model outperforms KS, though it trails in the outlier-resistant measures of the median and 3rd quartile. This is due to a small number of profiles that have sparse measurements at greater depths, leading to poor quality of interpolation in pressure. 
At 300 decibars, our functional model improves upon KS for each of the metrics, and at 10 decibars, the functional model is slightly worse. We explain a possible reason for this gap at 10 decibars. 
Mainly, the correlation lengths in space can decrease quickly when moving from 10-20 dbar to 40-50 dbar in some locations. 
Due to this effect, a pointwise approach as in KS can better model the surface pressure levels because the conditions in the small width of the interval near the surface are not easily isolated by the scores based on a limited number of principal components. This motivates future work on a new space-time functional model that allows a scale parameter to change smoothly but quickly as a function of pressure, or, alternatively, an approach to adaptively choose the number of principal components in space (ref. Section \ref{conclusion}). 

In Section \sref{computational_challenges}, the computational costs of our approach and KS are roughly compared. We conclude that, when focusing on temperature, the FDA approach can provide similar predictions for all pressures in roughly the same amount of time it takes to compute a pointwise approach for 13 pressure levels. Thus, our approach can provide approximately a 4 to 5 times speedup when considering the 58 Roemmich and Gilson pressure levels.

   \begin{table}
 \caption{Comparison of KS and Functional Approach prediction errors, temperature. RG residuals and functional residuals refer to the residuals after subtracting the respective mean.}\label{KS_compare}
 \begin{tabular}{|c||c|c|c||c|c|c|c|}
 \hline
 Pressure & Metric &  RG residuals & \begin{tabular}{@{}c@{}} Functional \\ residuals \end{tabular} & \begin{tabular}{@{}c@{}} RG-type  \\ model \end{tabular} & KS & \begin{tabular}{@{}c@{}} Functional  \\ model \end{tabular}  \\
 \hline\hline
 10 & RMSE & 0.8889 &  0.7540 & 0.6135 & 0.5072 & 0.5215 \\ 
 \hline
 10 & Q3 & 0.8670 & 0.6247 & 0.5026 &  0.3735  &0.3940\\ 
  \hline 
 10 & Median & 0.4750 & 0.3193 & 0.2556 & 0.1801  & 0.1961 \\ 
 \hline\hline
  300 & RMSE & 0.8149 &  0.8552 & 0.5782 & 0.5124 &0.4968 \\ 
 \hline
 300 & Q3 & 0.6320 & 0.6845 & 0.4213 &  0.3684  & 0.3644 \\ 
  \hline
 300 & Median & 0.3062 & 0.3494 & 0.1991 & 0.1740  & 0.1720 \\ 
 \hline\hline  
 1500 & RMSE & 0.1337 & 0.1381  & 0.1014 & 0.0883 & 0.0857 \\ 
 \hline
 1500 & Q3 & 0.1043 & 0.1160 & 0.0736 &  0.0641  & 0.0689\\ 
  \hline
 1500 & Median & 0.0530 & 0.0620 & 0.0356 & 0.0311 & 0.0349\\ 
 \hline\hline
 \end{tabular}
 \end{table}

\section{Applications: Ocean Heat Content and potential density estimates}\label{section_functional}

The procedures of Sections \ref{meanest} and \ref{covarianceest} result in estimated functions of temperature and salinity at each location. 
For these functions, derivatives and integrals can be easily calculated. Also, other oceanographic measures of interest, like potential density and conservative temperature, can be derived directly from the estimated temperature and salinity using TEOS-10 \citep[e.g., in R,][]{gsw}. In Section \sref{generalfun} and \sref{intderiv}, we present a general framework for leveraging these estimates for other scientific problems and give specific examples in this section. 

\subsection{Ocean Heat Content}\label{ohc}

The amount of heat contained in the ocean is of great interest for global climate change and has been studied extensively, since the ocean absorbs the majority of the Earth's excess heat. A non-exhaustive list includes \cite{levitus_world_2012}, \cite{roemmich_unabated_2015}, \cite{lyman_estimating_2013}, \cite{roemmich_135_2012}, and \cite{johnson_as_2017}.
While integrating temperature over pressure describes the heat content in the ocean, it is biased since the temperature of two volumes of water with the same amount of heat content at two different pressures is different. For this reason, conservative temperature is more commonly used for heat content estimates \citep{mcdougall_potential_2003}.
Conservative temperature can be calculated using the standard oceanographic toolbox \cite{teos10} which is implemented in {\tt R} from \cite{gsw}. We use the delta method approach described in Section \ref{generalfun} and \ref{ohc_appendix} to estimate its distribution. Following \cite{meyssignac_measuring_2019}, denote conservative temperature as a function of temperature, practical salinity, and pressure at a location as $\Theta(t,s,p)$ and the ocean heat content at a location as $$Q = \int_{0}^{p^*} c_p \rho \Theta(T_{s,d,y}(p), S_{s,d,y}(p), p) dp$$where $c_p$ and $\rho$ are constants (the specific heat capacity and density of seawater, respectively), \newedits{and $T_{s,d,y}$ and $S_{s,d,y}$ are the predicted temperature and salinity functions at location $s$ on day of the year $d$ for year $y$}.

\begin{figure}[t]
\includegraphics[scale = .16 ]{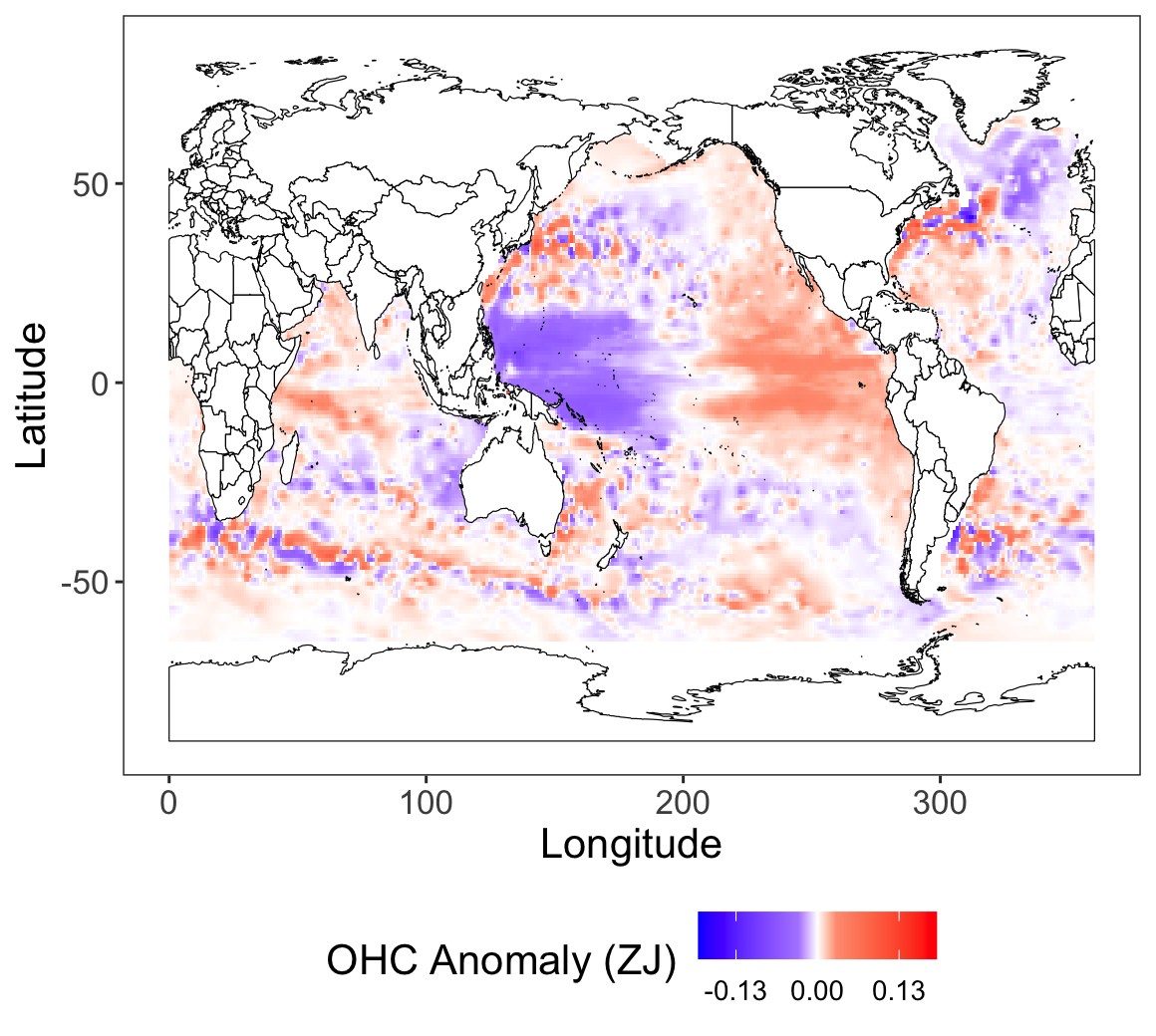}
\includegraphics[scale = .16 ]{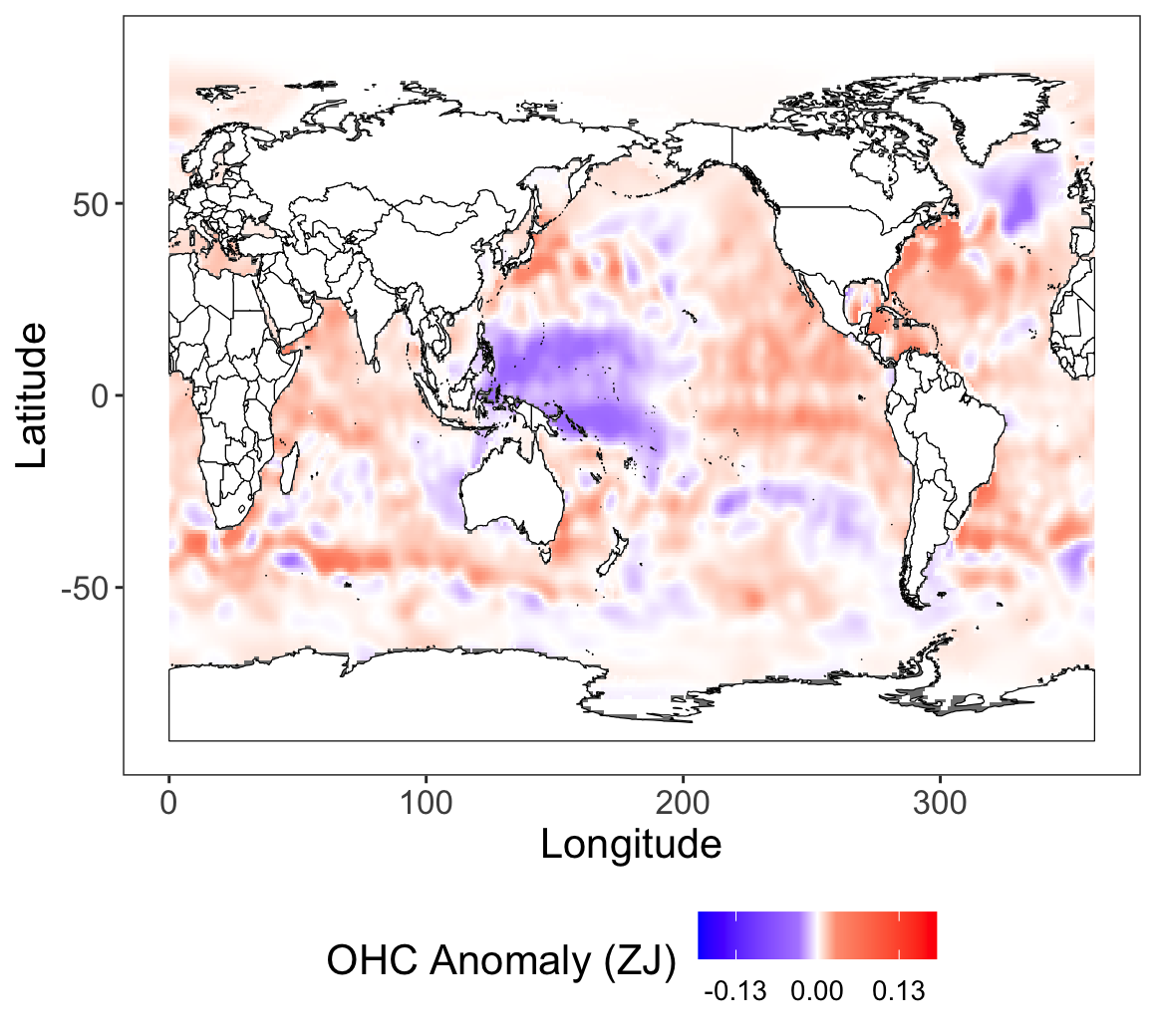}

\includegraphics[scale = .16 ]{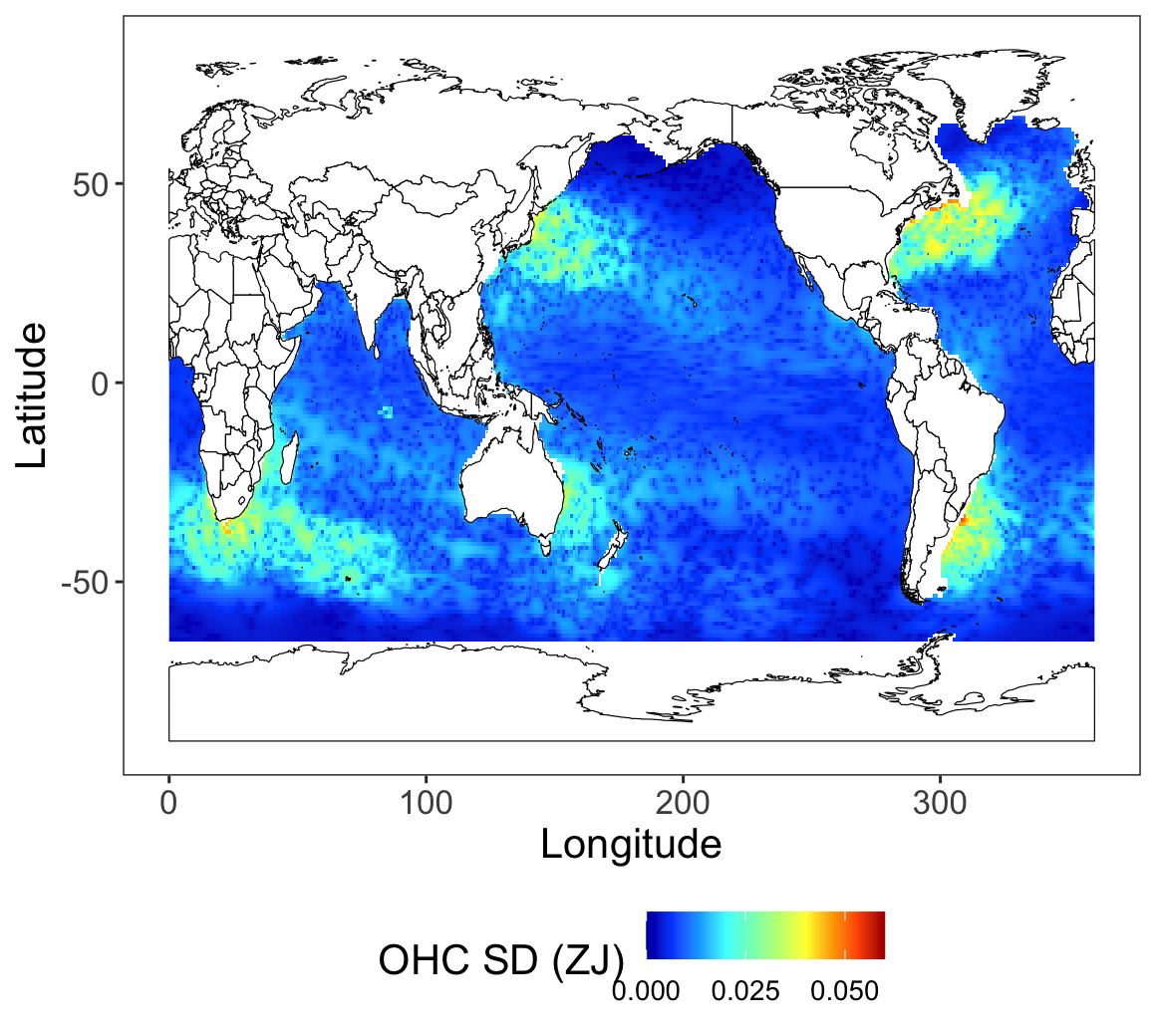}
\includegraphics[scale = .11 ]{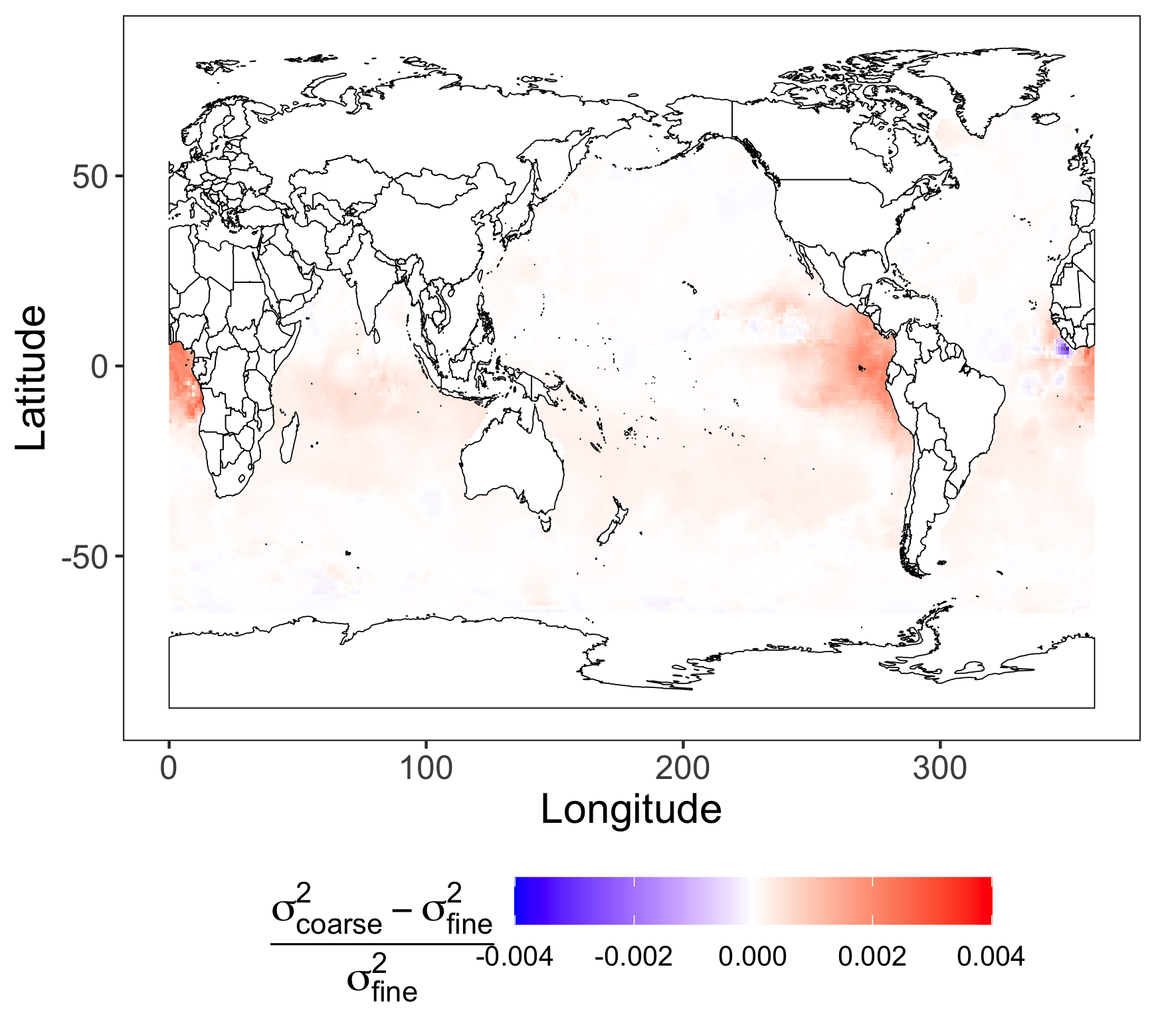}
 \caption{(\textbf{Top Left}) Ocean heat content anomaly estimates from the year-averaged mean, February 2016 functional estimate integrated for 0-700 dbar in zettajoules, (\textbf{Top Right}) NOAA estimate for 0-700 m, January-March 2016 average. (\textbf{Bottom Left}) our estimates of the standard deviation of OHC at each location, February 2016 and (\textbf{Bottom Right}) Comparison of variance estimates on a coarse and fine grid, February 2016. 
 }\label{int_temp}
\end{figure}

\newedits{For a fixed location and day of the year, w}e consider anomalies from the mean as the difference between the ocean heat content (OHC) using a mean averaged over all years and the conditional expectation of OHC for one year. Specifically, anomalies from the mean are computed for each year as $\mathbb{E}\{Q| \tilde{Z}_{y*} \} - \mathbb{E}\{Q\}$, where $\mathbb{E}\{Q\}$ is the ocean heat content given by year-averaged mean $\overline{\beta}(p) = \frac{1}{10} \sum_{y=2007}^{2016} \beta_{0,y}(p)$ from the mean form described in (\ref{functional_form_mean}), and an example of these estimates and standard deviations are shown in Figure \ref{int_temp} for February 2016, while similar plots for other years can be viewed in an R Shiny application \citep{shinyapps}. \newedits{Such estimates are computed for each location and a fixed day of the year in mid-February.} We compare the estimates for 0-700 dbar with the estimates available at \cite{noaa_ohc} that employ the \cite{levitus_world_2012} approach to estimation of ocean heat content. The large-scale features of the fields are similar, though our integrated functions show finer-detail and smaller-scale features as well. \newedits{We hypothesize that much of the difference in smoothness and features is due to different temporal windows. Our field, as a prediction for a fixed day in mid-February, estimates finer-scale activity compared to the NOAA January-to-March average.} 
The FDA approach, by modeling the dependence between different pressures, makes these uncertainty estimates possible between any two pressures in $[0,2000]$ as a natural consequence of our functional kriging approach. 

Our functional data approach can evaluate the level of error for any interpolation scheme when estimating integrals of the ocean properties. \cite{cheng_uncertainties_2014} have evaluated the levels of uncertainty in ocean heat content due to insufficient sampling in pressure; we evaluate here the amount of error one introduces by using a fixed number of pressure levels with respect to the integrated ocean heat content. We compute our estimates for 0-700 dbar on a fine (.5 dbar) grid as well as a coarse (10 dbar) grid that gives similar pressure gaps used in \cite{roemmich20092004} or \cite{li_increasing_2020}. We evaluate our estimates of OHC as $Q=c_p \rho \sum_{m=1}^{M_1} (p_{m+1} - p_m)\Theta(T(p_m), S(p_m), p_m)$ for a grid of pressures. Based on the functional estimates, we derive the mean and variance of the estimates in the Section \sref{ohc_appendix} for two different grids: \begin{align*}
    Q_{fine} &\sim N(\mu_{fine}, \sigma^2_{fine})\\
    Q_{coarse} &\sim N(\mu_{coarse}, \sigma^2_{coarse})
\end{align*}For each location, $(\mu_{coarse} - \mu_{fine})/\mu_{fine}$ is negligible, suggesting there is little bias introduced by using a limited number of pressures. On the other hand, the differences in the estimates of variance $(\sigma_{coarse}^2 - \sigma_{fine}^2)/\sigma_{fine}^2$, plotted in Figure \ref{int_temp}, are larger, and can reach 0.3\% in some areas. This suggests, that if one uses oceanographic products at fixed levels to estimate the ocean heat content, the estimates may be practically unbiased but may be burdened with slightly higher variance. Our functional approach provides this comprehensive estimate of the mean and covariance in pressure which evaluates the consequences of specific discretization approaches in pressure. 

\subsection{Use of potential density estimates for mixed layer depth}\label{mld_section}

Estimating the joint dependence of temperature and salinity for any pressure gives estimates of quantities like potential density that give important information about the vertical structure of the oceans. More dense water sinks below less dense water, and thus potential density helps describe the stratification of the oceans: the larger the potential density gradient is in pressure, the more stratified the water is at this point \citep{talley_chap4}. Potential density can be computed directly from temperature, salinity, pressure, and location using \cite{gsw}. In this section, we use potential density to estimate the depth of the mixed layer (which can be characterized by approximately constant potential density), and in the next section we evaluate the deviations from monotonicity of potential density.

To describe this first application, the mixed layer is a section of the ocean near the surface where the water mixes freely, giving near-uniform properties of temperature, salinity, and density. The mixed layer governs the interaction between the atmosphere and the ocean, and thus its study can reveal information about the carbon uptake and heat content of the ocean, among others features \citep{holte_argo_2017}. During the summer, the temperature at the surface rises considerably, and the mixed layer is more shallow. During the winter, the mixed layer deepens at a lower temperature, resulting in large seasonal changes of its depth. 

Mixed layers are usually estimated using discretely observed profiles; see Sections 4.2 and 7.4 of \cite{talley_chap4} and \cite{holte2009} for algorithms to estimate the mixed layer, and mixed-layer climatologies include \cite{schmidtko_mimoc_2013}, \cite{holte_argo_2017}, and \cite{hosoda_improved_2010}. In comparison to these approaches, our functional approach offers two advantages. First, by basing the mixed layer estimates on entire functions predicted from pooled data, we avoid discretization error in the mixed layer estimates. Second, our estimates produce entire mixed layer distributions even when few profiles have been observed nearby; these are robust to the skewed nature of mixed layers.

These estimates and their variability are assessed using the parametric bootstrap approach described in Section \sref{generalfun}. 
For each location and year, we simulate $B=1{,}000$ times from the distribution of February mixed layer depth based on a year-averaged mean, and $B$ times for each year from the conditional distribution of February mixed layer depth. We present results using the variable density threshold approach described in \cite{holte_argo_2017}, where the mixed layer depth is chosen as the first depth for which potential density decreases an amount corresponding to a temperature decrease of $0.2^{\circ}\textrm{C}$. 


Our modeling framework allows us to examine the within and across year variability of MLD (mixed layer depth) estimates. This is an important step in quantifying significant anomalies and trends in the MLD due perhaps to climate change, which is of fundamental scientific importance. 
Denoting $D_{j}$ as the estimated mixed layer depth for the $j$-th simulation using the year-averaged mean, and $D_{y,j}$ as the estimated mixed layer depth for the $j$-th conditional simulation for year $y$. We define estimates based on summaries of the values: $\overline{D} =\frac{1}{10B} \sum_{y=2007}^{2016}\sum_{j=1}^B D_{y,j}$ estimates the overall mean, the conditional simulation median $\tilde{D}_y$ estimates the median for year $y$, and the year-averaged simulation median $\tilde{D}$ estimates the year-averaged median. Then, consider the estimates of the variation \begin{align*}
    \textrm{MAEY} &= \frac{1}{10 B}\sum_{y=2007}^{2016}\sum_{j=1}^B |D_{y,j}- \tilde{D}_y| \\
    \textrm{MAE} &= \frac{1}{10 B}\sum_{y=2007}^{2016}\sum_{j=1}^B |D_{y,j}- \tilde{D}| 
\end{align*}where $D$ is distributed as the year-averaged mixed layer depth at a given location, and the factor of $10$ adjusts for the ten years. That is, the mean absolute error with yearly estimates (MAEY) and the year-averaged mean absolute error (MAE) gives estimates of the variation that are relatively robust to outliers. The value $p_{year} = \textrm{MAEY}/\textrm{MAE}$ gives an estimate of the relative sizes of errors with year-specific medians versus a single group median, which evaluate yearly variation in the MLD. 

\begin{figure}[t]

\includegraphics[scale = .16]{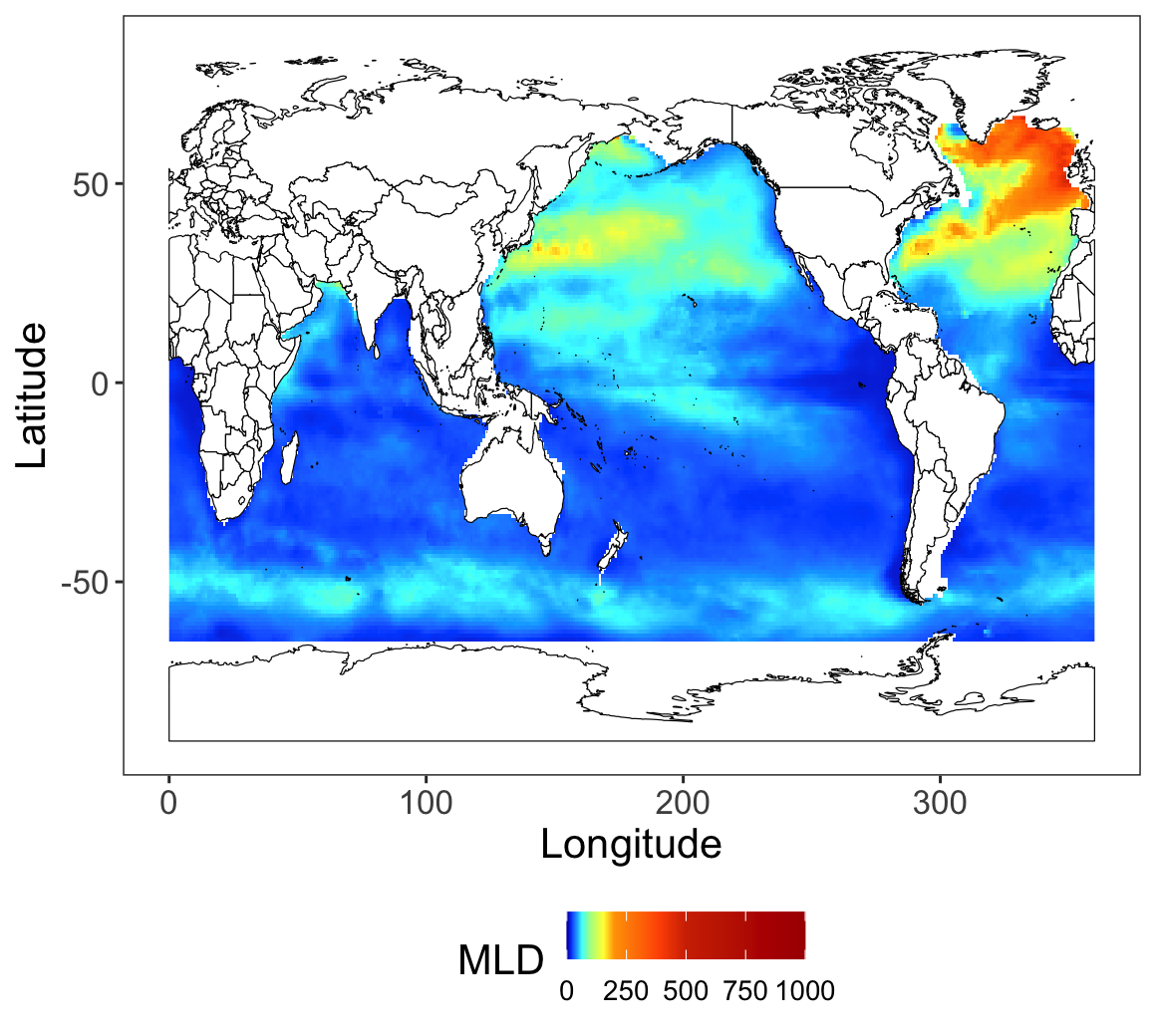}
\includegraphics[scale = .16 ]{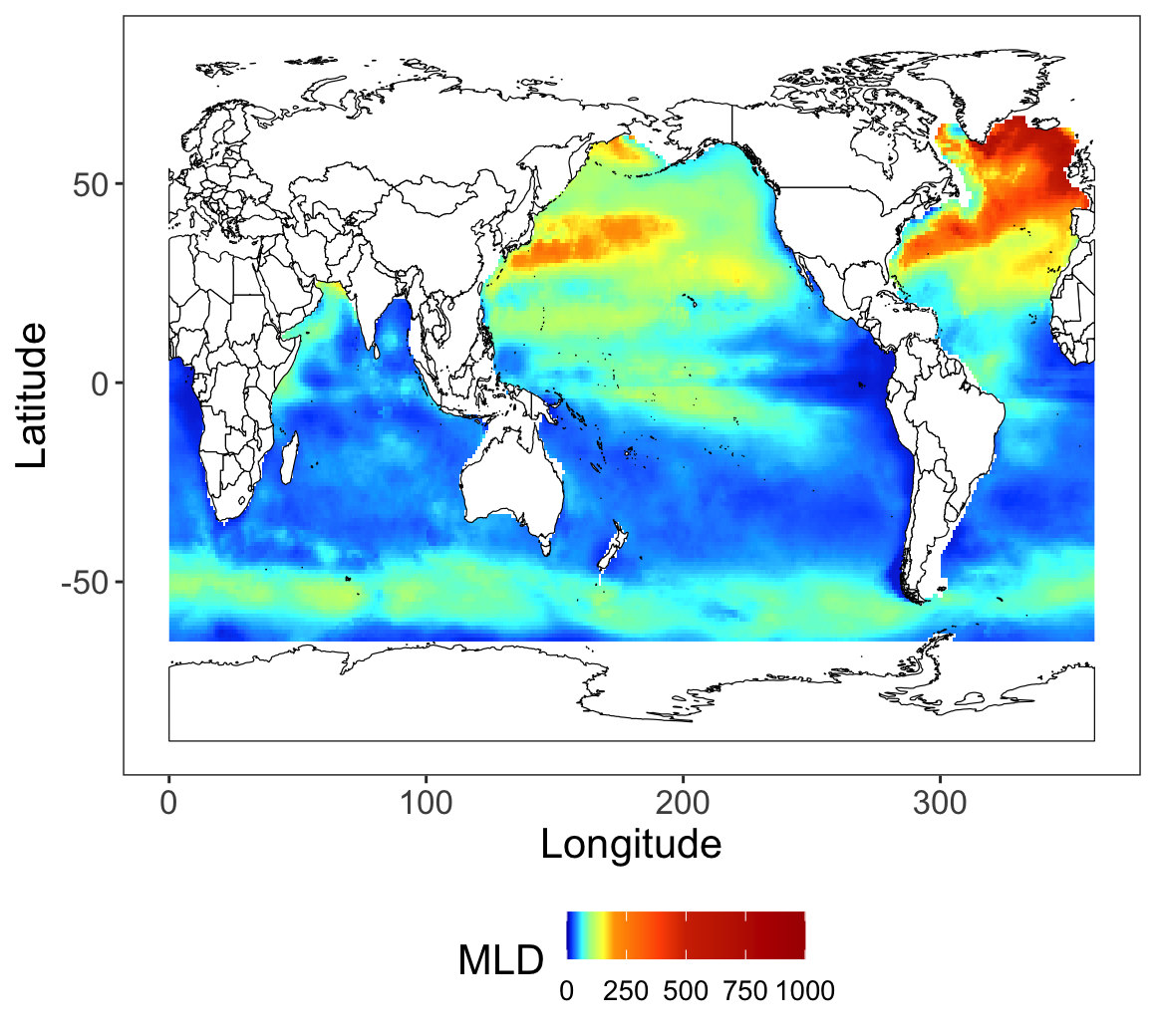}

\includegraphics[scale = .16]{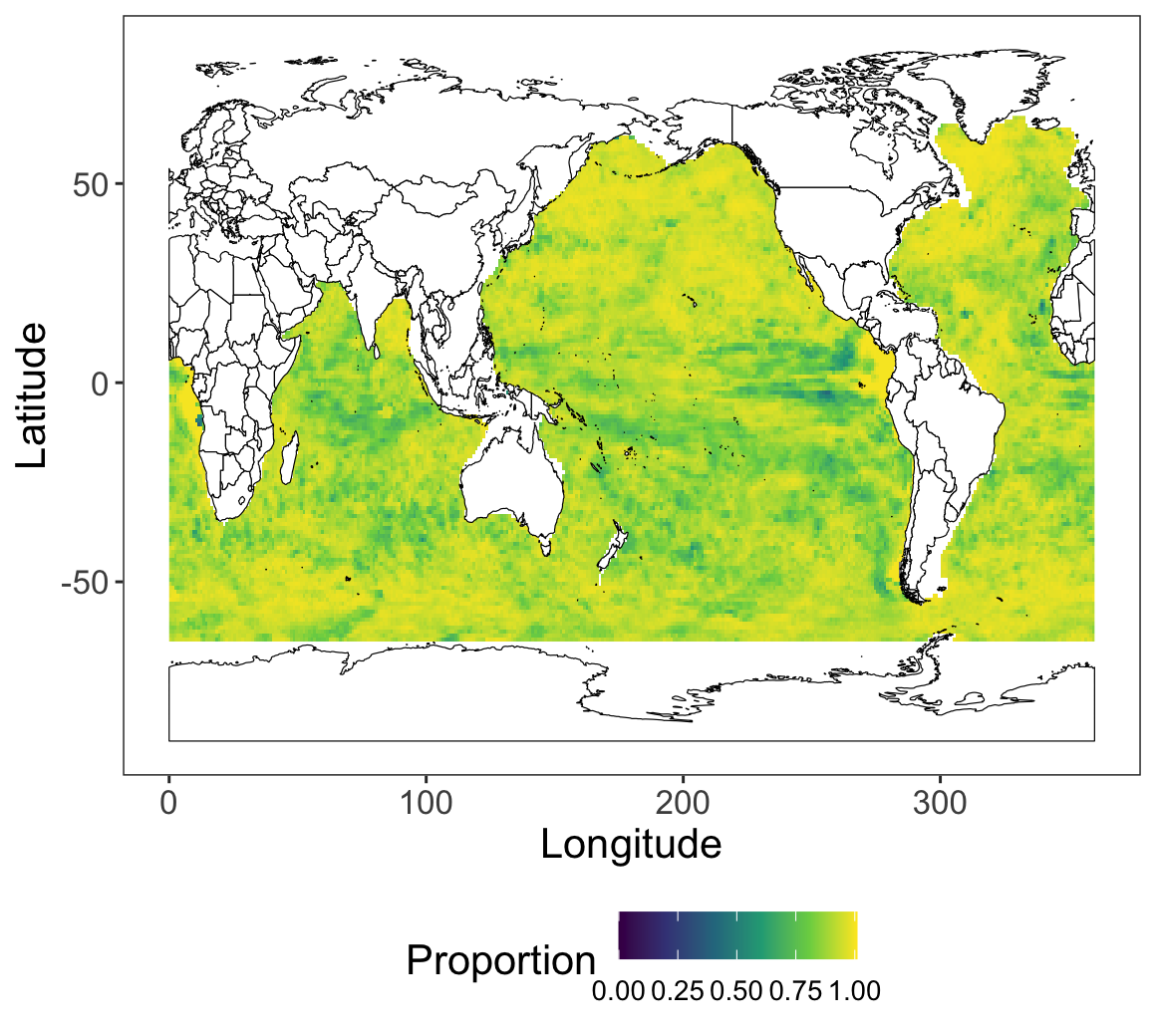}
\includegraphics[scale = .16 ]{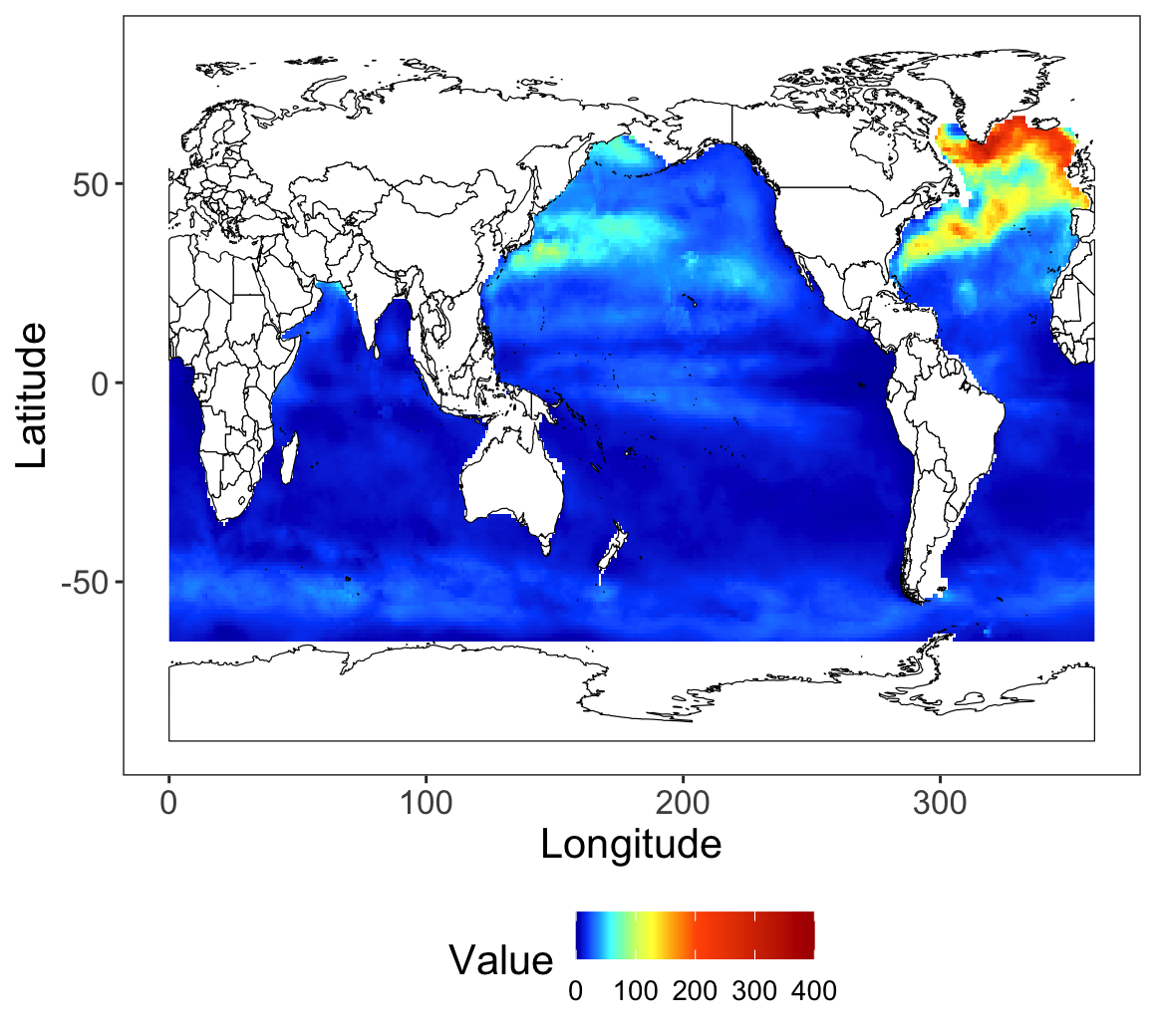}

 \caption{February mixed layer depth in dbar by the variable threshold approach (\textbf{Top Left}) average $\overline{D}$, (\textbf{Top Right}) 80th percentile of the conditional MLD distribution, averaged over the ten years, February mixed layer depth (\textbf{Bottom Left}) $p_{year}$, (\textbf{Bottom Right}) $\textrm{MAEY}$.}
 \label{mld_est}
\end{figure}

In Figure \ref{mld_est}, we show selected summaries from the results. The algorithm picks out both shallow mixed layers during the summer in the Southern Hemisphere as well as deeper mixed layers during the winter for the Northern Hemisphere (top left). Winter mixed layers show more variation than summer mixed layers (bottom right). There are two main reasons for this. First, the distribution of the MLD is truncated near the surface, so distributions of MLD near the surface will show less variation. Also, during the summer, there is more stratification and thus larger differences in ocean properties at the depth of the mixed layer, so the mixed layer is more consistent. On the other hand, in winter the mixed layer depth is less well-defined, and absolute differences of the MLD from its median can be greater than 65 decibars. Finally, the MAE is generally about the same size as MAEY (bottom left), which indicates that there are not large differences in median mixed layer depths between years. Again, this pattern is more evident in the Northern Hemisphere, while, for some locations in the Southern Hemisphere, there is substantial between-year variation. These results indicate that differences in MLD from year-to-year may not be discernible based on the current Argo array alone. 


\subsection{Monotonicity of density}

Potential density generally increases as a function of pressure as water becomes more dense, though this can often be violated for periods of a few hours \citep{talley_chap4}. Here, our estimates are used to evaluate the occurrence of these density ``inversions,'' where potential density becomes non-monotone. \cite{kuusela2017locally} suggest that applying a monotonicity constraint on the density may improve estimation of the mean and covariance structure. \cite{talley_chap4} suggest that these inversions occur only on the order of a few hours, as gravity removes the instability in the density. Though we have not imposed this constraint, we are in a position to evaluate how well this constraint is satisfied based on our estimates. We compare the amount of inversion in our predictions to the amount of density inversion in the raw Argo profile data which we detail in Section \sref{sup:densinv}. Here, we use our functional uncertainty estimates to evaluate how consistent density inversions are at a fixed pressure. 

\begin{figure}
    \centering

\includegraphics[scale = .16]{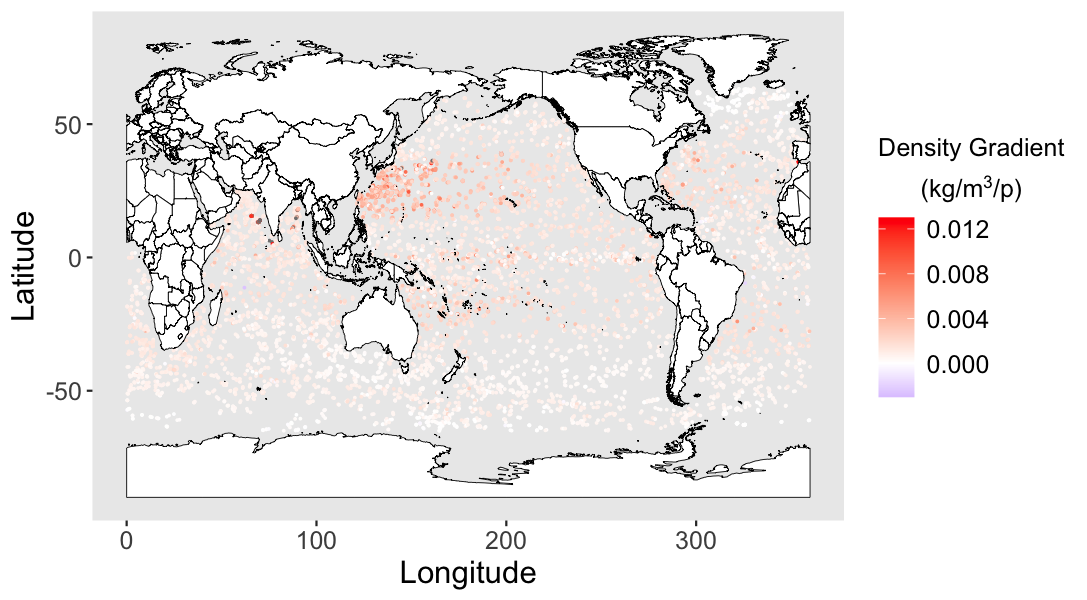}
\includegraphics[scale = .08]{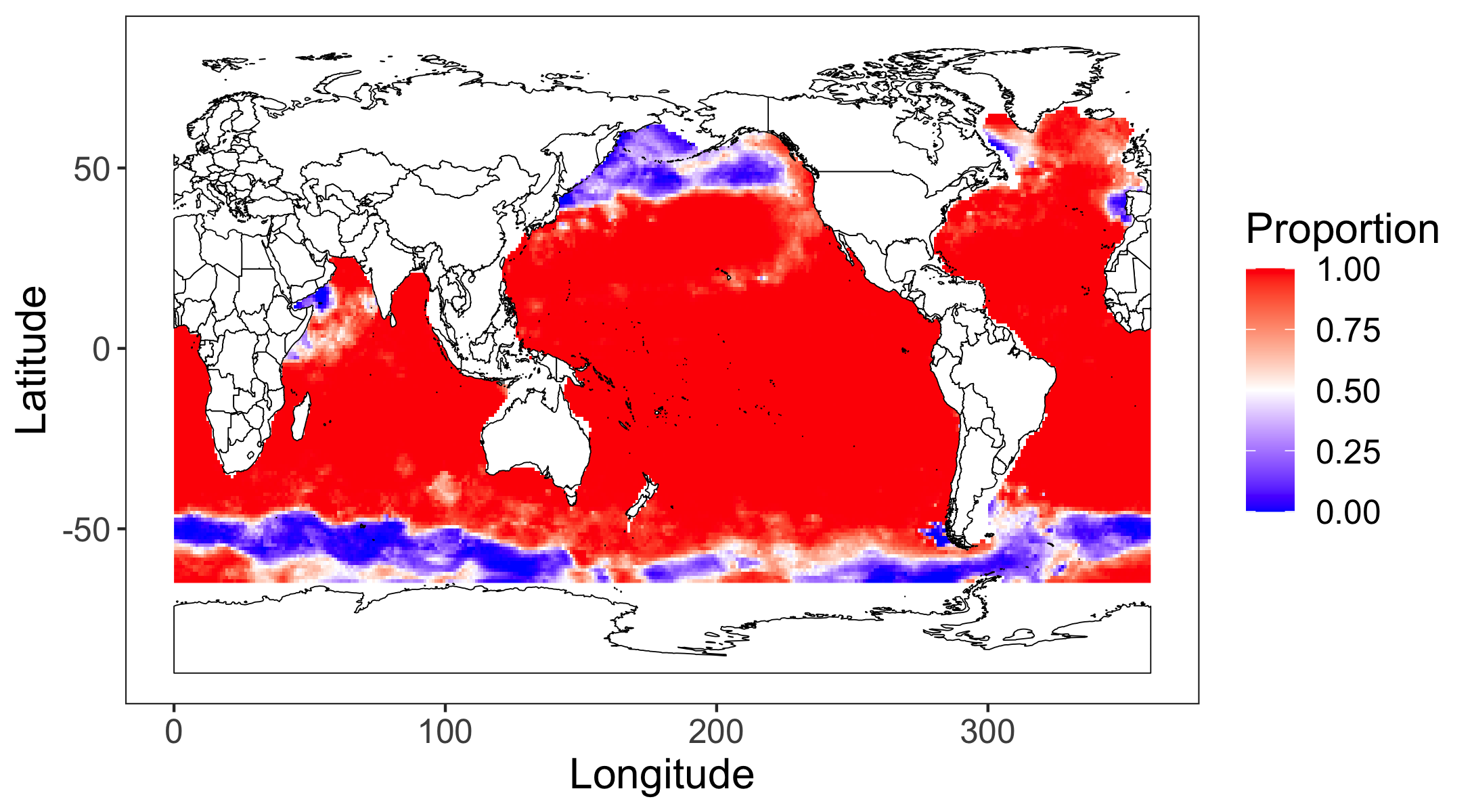}

    \caption{(\textbf{Left}) Estimated gradient using the two closest measurements to 550 decibars for February 2015 profiles.
    (\textbf{Right}) The proportion of conditional simulations that show increasing density at 550 decibars in 2015.}
    \label{fig:dens_inversion}
\end{figure}

To address the salience of the density constraints, we simulate from 1,000 functions using our conditional simulation approach, then compute the proportion of times a density inversion is shown at a particular pressure. In the bottom of Figure \ref{fig:dens_inversion}, we plot this at a pressure of 550 decibars for the year 2015 (right) and compare it to the estimated gradient from raw Argo profiles using finite differences (left). At this pressure, areas of density inversions are consistently shown where marginal seas mix with the open oceans, as well as areas in the Antarctic Circumpolar Current, where there are stronger currents. These areas correspond to areas where negative or low density gradients occur in the profiles. Many of these areas are deep water formation regions, where fresh, cool water sinks due to its high density. On the other hand, in most of the open oceans, there is little evidence of density inversion at this pressure. We conclude that implementing a hard density constraint may not be appropriate, especially in areas of consistent ocean mixing near marginal seas and in the Southern Ocean. Moreover, the conditional simulations show that further study at all 
pressure levels can provide valuable statistical insights to the open scientific question 
on the change of ocean stratification and its effect on thermohaline circulation \citep{li_increasing_2020}.

\section{Conclusion and Future Directions}\label{conclusion}

The Argo data is an exemplary modern dataset that motivates new statistical approaches, development of methodology, and appropriate statistical applications.
In this paper, we have provided the first comprehensive functional-data analysis of the Argo data which addresses methodological and computational challenges for mean estimation, covariance estimation, functional kriging, and estimation of functionals of the estimates. Our approach avoids the simplification of data in pressure via interpolation which limits other methods' ability to provide a comprehensive analysis. The predicted functions give powerful new tools to fully explore important scientific problems. Furthermore, our approach can decrease the computational burden of prediction by sharing information across pressure.
Our estimates match well or outperform existing methodologies that estimate ocean properties at fixed pressure levels. Our analysis also introduces the local estimation of functions and represents a leap forward in the analysis of spatio-temporal functional data. 

The methods we develop could be applied to scientific problems in neuroscience and spatial statistics. Here, we focus on spatial sensor networks. In the context of these applications, our methods are amenable to use time, instead of pressure, as the functional variable. Two specific case studies could include estimating air pollution \citep[as in][]{king_functional_2018} and the Canadian Weather data \citep{fda} studied in \cite{delicado2010} and \cite{koko2019}, among others. Instead of considering the annual cycle of temperature at only 35 locations in the Canadian Weather data, one could provide high-granularity estimates using thousands of weather stations in North America. Notably, our mean estimation approach establishes a new, computationally-efficient, hybrid methodology that combines kernel estimation and smoothing splines. 

Throughout our approach, there are areas for improvement. For mean estimation for Stage 1, one would want to select the amount of nearby data adaptively and allow for elliptical regions in space. This is especially important for areas in the Western boundary currents and other areas where changes in ocean properties are highly directional in space. One approach would be to extend algorithms from local regression that choose the bandwidth to this functional model. Using iteratively reweighted least squares or more careful smoothing parameter selection may also give improvements. 

For our spatial covariance modeling for Stage 2, we employ a relatively simple model that successfully captures key features by jointly modeling temperature and salinity. In general, we are limited by computational challenges, which could be addressed with approximate models, e.g., Vecchia's approximation \citep{guinness_gaussian_2019} or the SPDE approach \citep{lindgren_explicit_2011}.
More complexity should be explored in the models. 
For example, a functional model that allows rapid changes in the scale parameter as a function of depth would likely improve upon our model.
In addition, there is some evidence that the cross-covariance between vectors of principal component scores include non-reversible, i.e. asymmetric, dependence, which is not available in the scalar Mat\'ern-type multivariate models. In addition, one could explore non-Gaussian models, which could provide better coverage for prediction intervals as demonstrated in \cite{kuusela2017locally} and \cite{bolin_multivariate_nodate}. Also, we have only modeled the local spatial dependence, and ideally, one would also like to combine estimates across space with uncertainty, for example, using an approach similar to \cite{wiens_modeling_2020}. This would enable uncertainty estimates for global ocean heat content.

There is a wide variety of statistical research directions using the Argo data, many of which are noted in the conclusion of \citet{kuusela2017locally}. For instance, one would want to consider integrating Argo data with other oceanographic data (e.g. satellite data) as well as using additional biogeochemical variables that a limited set of Argo floats measure. 
Although we have considered many standard approaches in FDA for use on the Argo data, there are more tools that could be applied, including clustering of profiles, functional regression, canonical correlation analysis, hypothesis testing, and data fusion with scalar data like sea surface temperature data. Moreover, the Argo data calls for full methodological and theoretical development of the field of space-time functional data. For example, the large-sample properties of the methodology used in this paper could be explored under functional and spatial dependence.



\section*{Acknowledgements}
We would like to thank the physical oceanography group at the Scripps Institution of Oceanography, including Sarah Gille, Lynne Talley, Matt Mazloff, Isabella Rosso, John Gilson, and Dean Roemmich; in addition, we want to thank
Mikael Kuusela, Alison Gray, and Donata Giglio for helpful comments on our work. We thank Michael Stein for suggesting to look at the observed monotonicity properties of density. We would also like to thank Moritz Korte-Stapff for his work on the covariance estimation. We finally thank the reviewers and an associate editor for important comments and suggestions that substantially improved the quality of the paper.

These data were collected and made freely available by the International Argo Program and the national programs that contribute to it: \url{http://www.argo.ucsd.edu}. The Argo Program is part of the Global Ocean Observing System. This research was supported in part through computational resources and services provided by Advanced Research Computing at the University of Michigan, Ann Arbor. This work used the Extreme Science and Engineering Discovery Environment (XSEDE), which is supported by National Science Foundation grant number ACI-1548562 \citep{xsede}. Code for the analyses presented in this paper are available at \cite{argoFDAcode}.

The authors acknowledge support from grants DMS-1646108 and DGE-1841052 for Drew Yarger and DMS-1916226 for Stilian Stoev and Tailen Hsing.

\begin{supplement}
\stitle{Supplemenary Document PDF}
\sdescription{Description of further modeling choices, computational details, and additional figures.}
\end{supplement}

\bibliographystyle{imsart-nameyear} 
\bibliography{msnew} 

\end{document}




\title{Supplemental Material to\\ ``A functional-data approach to the Argo data''}
\author{}
\date{}
\maketitle





In this document, the sections, figures, and equation references are referred to with a label starting with ``S.'' Labels without references refer to items in the main paper.

\section{Additional Description of Methodologies}
 
\subsection{Orthonomalization of principal components}\label{pc_ortho}
 To find the principal components, the $M \times M$ matrices $B$ with entries $(B)_{k_1, k_2} = \alpha_{k_1, k_2}$ and a Gram matrix $\Omega_0$ with entries $(\Omega_0)_{k_1, k_2} = \int_0^{2000} \chi_{k_1}(p) \chi_{k_2}(p) dp$ are formed. Then, performing standard principal components analysis on the matrix $\Omega_0^{1/2} B \Omega_0^{1/2}$ results in the vectors $\{v_1, v_2, \dots, v_M\}$, and the vectors $\{\Omega_0^{-1/2}v_1$, $\Omega_0^{-1/2}v_2$, $\dots$, $\Omega_0^{-1/2}v_M\}$ give the coefficients for the FPCs in the basis $\{\chi_k\}_{k=1}^M$. The use of $\Omega_0$ rotates the problem into the space with the $\mathbb{L}_2$ inner product, decomposes the principal components in this space, then rotates the vectors back to the space of original coefficients. This ensures that the resulting functions are orthonormal with respect to the $\mathbb{L}_2$ inner product.

\subsection{Estimation of the measurement error variance $\kappa(s,d,p)$}\label{me_var}

In (\ref{dgp}), we have outlined a model involving the additional term $\epsilon_{i,j}$, modeled as a measurement error with some variance $\kappa=\kappa(s_0,d_0,p)$. To give full uncertainties for our predictions, we estimate $\kappa$ as follows \citep[see also][]{li_uniform_2010}. At each location, a smoothing spline is fit with the observations $$R_{i,j} = \log\left( \left(Y_{i,j}^0 - \sum_{k=1}^{K_1}  Z_{i,k} \phi_k(p_{i,j})\right)^2\right)$$by solving the minimization problem \begin{align*}
    \frac{1}{n}\sum_{i=1}^n \frac{K_{h_s, h_d}(s_i - s_0, d_i - d_0)}{m_i} \sum_{j=1}^{m_i}\left(R_{i,j} - \beta(p_{i,j})\right)^2 + \lambda \left\lVert \beta^{(2)} \right\rVert_{\mathbb{L}_2}^2,
\end{align*}an approach similar to the mean estimation in Section \ref{meanest} with cross validation. The logarithm ensures that the variances are nonnegative, and the transformed estimate $\hat{\kappa}(p) = 2\cdot\textrm{exp}(\beta(p) + \varphi(1))$, where $\varphi(x)$ is the digamma function, gives a bias-corrected estimate of the remaining variance not captured in the modeling of the first $K_1$ scores.  
We compute this for temperature and salinity separately.

\subsection{Leveraging estimated functions of temperature and salinity}\label{generalfun}

Denote the estimated functions of temperature and salinity at a fixed location $s_0$ and time $d_0$ as \begin{align}
   T(p) = \mu_T(p) +  \sum_{k=1}^{K_1} Z_k \phi_k(p), \ \textrm{and}\
    S(p) = \mu_S(p)+  \sum_{k=1}^{K_2} W_k \psi_k(p)
    \label{TandS}
\end{align}where $\mu_T$ and $\mu_S$ are the respective mean functions and $$\begin{pmatrix} Z \\ W \end{pmatrix}\Big| \tilde{Z}_{y*}  \sim N\left(\begin{pmatrix}\theta_1 \\ \theta_2\end{pmatrix}, \begin{pmatrix} \Sigma_{11} & \Sigma_{12} \\ \Sigma_{21} & \Sigma_{22}\end{pmatrix}\right).$$

We present two general methods for estimating the distribution of some function of $T(p)$ and $S(p)$, which we denote $g(T(p), S(p))$. The first is a conditional simulation or parametric bootstrap approach: simulate $T_b(p)$ and $S_b(p)$ according to their estimated distributions and compute $g(T_b(p), S_b(p))$, repeating this process a large number of times $b=1, \dots, B$. The average and variability of the bootstrapped values $g(T_b(p), S_b(p))$ give estimates for the distribution of $g(T(p), S(p))$.

The second approach is due to the Delta method, a commonly-used tool used in statistics. Suppose that $g(T(p), S(p))$ has continuous first partial derivatives $$\nabla g(p) = \begin{pmatrix}
\frac{\partial g(T(p),S(p))}{\partial t}, &
\frac{\partial g(T(p),S(p))}{\partial s}\end{pmatrix}^\top$$ at $\mathbb{E}\{T(p) | \tilde{Z}_{y*} \} =\mu_T(p) +  \phi(p)^\top \theta_1$ and $\mathbb{E}\{S(p) | \tilde{Z}_{y*} \} = \mu_S(p) + \psi(p)^\top \theta_2$.
By the multivariate Delta method, $g(T(p), S(p))$ converges to a normal distributed random variable in distribution with mean \begin{align*}
    \mu_g(p) &= g(\mathbb{E}\{T(p) | \tilde{Z}_{y*} \} ,\mathbb{E}\{S(p) | \tilde{Z}_{y*} \})
\end{align*}as the matrix $\begin{pmatrix} \Sigma_{11} & \Sigma_{12} \\ \Sigma_{21} & \Sigma_{22}\end{pmatrix}$ decreases to the $0$ matrix. Furthermore, letting $$ C(p_1, p_2) = \begin{pmatrix} \phi(p_1)^\top  & 0\\ 0 & \psi(p_1)^\top \end{pmatrix} \begin{pmatrix} \Sigma_{11} & \Sigma_{12} \\ \Sigma_{21} & \Sigma_{22}\end{pmatrix}\begin{pmatrix} \phi(p_2)  & 0\\ 0 & \psi(p_2) \end{pmatrix} \in \mathbb{R}^{2\times 2},$$ the limiting variance of $g(T(p), S(p))$ is $
    \Sigma_g(p) 
    \approx \nabla g(p)^\top C(p, p) \nabla g(p).$
This gives a natural way to estimate differentiable functions of temperature and salinity using a linear approximation to the function $g$. 

\subsection{Integrals and Derivatives}\label{intderiv}

Here, we demonstrate how, for a fixed location, the distributions of integrals and derivatives can be estimated naturally from our estimates. Using the notation for temperature, for any $p^* \in (0, 2000]$, the integral $I_{p^*} = \int_0^{p^*} T(p) dp$ is normally distributed with mean and variance \begin{align*}
    \mathbb{E}\left(I_{p^*} \bigg| \tilde{Z}_{y*}  \right) &=  \int_0^{p^*} \mathbb{E}(T(p) | \tilde{Z}_{y*} ) dp = \int_0^{p^*} \mu_T(p) dp + \int_0^{p^*} \phi(p)^\top dp  \theta_1 \\
    \textrm{Var}\left(I_{p^*} \bigg| \tilde{Z}_{y*}  \right)
    &=\left(\int_0^{p^*}\phi(p)^\top  dp \right)\Sigma_{11}\left( \int_0^{p^*}\phi(q)dq\right)
\end{align*}by exchanging the order of integration and using (\ref{predictivemean}), interpreting $\int_0^{p^*} \phi(p)^\top dp$ as the integral applied element-wise to $\phi(p)^\top$. The $\ell$th derivative of the prediction is also normally distributed, with mean and variance
\begin{align*}
    \mathbb{E}\left( T^{(\ell)}(p) \bigg| \tilde{Z}_{y*}  \right) &= \mu_T^{(\ell)}(p)  +   \phi^{(\ell)}(p)^\top  \theta_1, \ \textrm{and} \ 
    \textrm{Var}\left(T^{(\ell)}(p) \bigg| \tilde{Z}_{y*}  \right)
    =\phi^{(\ell)}(p)^\top \Sigma_{11} \phi^{(\ell)}(p)
\end{align*}for $\ell=1,2$.
The integral $\int_0^{p^*} \phi_k(p)dp$ for each $p^*$ and $k$ and the derivatives $\mu_T^{(\ell)}$ and $\phi_k^{(\ell)}(p)$ can be computed quickly using the \texttt{inprod} and \texttt{eval.basis} functions, respectively, in the \texttt{fda} package in R \citep{fda}. The integrals are easily extended to arbitrary bounds within $[0,2000]$.

\subsection{Delta Method for Ocean Heat Content} \label{ohc_appendix}

The Delta method estimate of the conditional mean of $Q$ is $$\mathbb{E}\{Q | \tilde{Z}_{y*} \} = c_p\rho\int_0^{p^*}\Theta\left(\mathbb{E}\{T(p) | \tilde{Z}_{y*}  \}, \mathbb{E}\{S(p)| \tilde{Z}_{y*}  \}, p\right) dp$$where $c_p$ and $\rho$ are the specific heat capacity and density of seawater, respectively, treated as constants. 
Also, following the steps and using the definitions in section \ref{generalfun}, $$\textrm{Var}\{Q|\tilde{Z}_{y*} \}\approx c_p^2\rho^2 \int_0^{p^*}\int_0^{p^*} \nabla\Theta(p_1)^\top C(p_1, p_2) \nabla\Theta(p_2)dp_1 dp_2$$where $
\nabla\Theta(p)$ are the derivatives of $\Theta$ with respect to temperature and salinity at pressure $p$. The closed form expressions of the partial derivatives of $\Theta$ with respect to temperature and absolute salinity are available through TEOS-10 \citep{teos10}. 
We use these as the expression for $\nabla\Theta(p)$, suggesting that the derivative for absolute salinity approximates the derivative for practical salinity well on the interval 0 to 2000 dbar. Since these are very similar quantities and the salinity plays a small role in conservative temperature compared to temperature, this approximation is justified. 
Since conservative temperature is a nonlinear function of temperature and salinity, exact expressions for these integrals are not available, but the integral can be approximated on an arbitrarily fine grid of pressure. 

For a fixed grid, $\{p_j\}_{j=1}^M$, the estimate of ocean heat content is given by \begin{align*}
    Q_{grid} = c_p \rho \sum_{m=1}^{M} (p_{m+1} - p_m)\Theta(T(p_m), S(p_m), p_m)
\end{align*}with conditional expectation and variance \begin{align*}
    \mathbb{E}\{Q_{grid} | \tilde{Z}_{y*} \} &= c_p \rho \sum_{m=1}^{M} (p_{m+1} - p_m)\mathbb{E}\{\Theta(T(p_m), S(p_m), p_m) | \tilde{Z}_{y*} \}\\
    \textrm{Var}\{Q_{grid} | \tilde{Z}_{y*} \}&= c_p^2 \rho^2 \sum_{m_1=1}^{M}\sum_{m_2=1}^{M} (p_{m_1+1} - p_{m_1})(p_{m_2+1} - p_{m_2})\\
    & \ \ \ \ \ \ \ \ \ \ \ \ \ \ \ \textrm{Cov}\{\Theta(T(p_{m_1}), S(p_{m_1}), p_{m_1}), \Theta(T(p_{m_2}), S(p_{m_2}), p_{m_2}) | \tilde{Z}_{y*} \}.
\end{align*}The Delta method, as described above, is used to estimate the expectations and covariances.

\subsection{Simultaneous Prediction Bands}\label{bands_appendix}

To develop simultaneous prediction bands, we extend the approach of \cite{choi_geometric_2018} to our model. They focus on the statistic $$W_\theta = \sum_{k=1}^\infty \frac{\lambda_k}{c_k^2} \left(\frac{ Z_{*,k} }{\sqrt{\lambda_k}}\right)^2$$for $\lambda_k = \textrm{Var}\{Z_{*,k} | \tilde{Z}_{y*} \}$ and some chosen constants $c_k$. 
In practice, the sum in $W_\theta$ must be truncated to a finite number $K_1$.
Since $W_\theta$ can be written as the quadratic form of a random Gaussian vector, the \texttt{imhof} function in \texttt{R} is used to compute its quantiles $\{q_{\alpha} | P(W_\theta > q_\alpha) = \alpha\}$.
The band is of the form $$B = \left\{h \in \mathbb{W}_2, \left|h(p) - \sum_{k=1}^\infty \phi_k(p)\mathbb{E}\{ Z_{*,k} | Y\} \right| \leq r(p) + u(p)\right\}$$where $r(p) = \sqrt{\xi \sum_{k=1}^\infty c_k^2 \phi_k(p)^2}$ and $\xi$ is the $1-\alpha_1$ quantile of $W_\theta$. Here, we have $u(p) =  q_{\alpha_2/(2 m_i)}\sqrt{\kappa(p) }$ where $\alpha = \alpha_1 + \alpha_2$. Using $\alpha_1$ and $\alpha_2$ gives a Bonferroni correction to turn the confidence interval into a prediction interval for the profile $i$ that was left out. Here, $\alpha_1 = \alpha_2 = .02275$, which corresponds to $\alpha = .04550$ or a 95.4\% prediction band.
We use their choice of $c_k^2 = \sqrt{\textrm{Var}\{Z_{*,k} | \tilde{Z}_{y*}  \}} = \sqrt{\lambda_k}$. 
Although this creates bands that may not have favorable theoretical coverage properties, as discussed in Section 3 of \cite{choi_geometric_2018}, in practice they work well for both temperature and salinity. This is likely due to the extra measurement error term $\kappa(p)$, which \cite{choi_geometric_2018} did not consider.

\subsection{Discussion of Separability Assumption}\label{discussseparability}

One assumption for the covariance used in \cite{gromenko_evaluation_2017} and \cite{koko2019} is separability between the spatial and functional dimensions or that the spatial dependence is measured only through an integral over pressure. In our context, the separability assumption of the covariance would amount to having $C(s_1,t_1,p_1; s_2, t_2,p_2) = C_{s,t}(s_1, t_1; s_2, t_2)C_p(p_1, p_2)$. Such an assumption is not appropriate for the Argo data, since the spatial dependence varies considerably by pressure. In particular, the spatial range and variances parameters are much larger near the ocean surface, as shown in \cite{kuusela2017locally} and supported by our analysis. 


Our approach to modeling the spatial dependence via the scores gives a more flexible and non-separable space-time-pressure covariance model than those assuming separability. Specifically, we propose a model of the form $$C(s_1,t_1,p_1; s_2, t_2,p_2) =  \phi(p_1)^\top C_{s,t}(s_1, t_1;s_2, t_2)\phi(p_2),$$ where $\phi(p)$ is a vector of principal component functions evaluated at $p$, and $C_{s,t}(s_1, t_1;s_2, t_2)$ is a $K\times K$
matrix specified by a space-time covariance of the principal component scores. 
The principal components give a locally optimal representation of profiles in $K$ dimensions.
This model assumes separability only on a principal component by principal component basis, and captures the nonstationary features of the pressure dimension through the principal components. 

In addition, we do not assume that the matrix $C_{s,t}(s_1, t_1;s_2,t_2)$ is diagonal (i.e. that the scores for different principal components are independent). This weakly-separable \citep{lynch_test_2018} assumption may not be satisfied for the Argo data. In particular, one set of principal components may not best represent all profiles within a spatial area, and there may be correlation between scores of different components at different locations. Also, to jointly model temperature and salinity, we must take into account their dependence. These two challenges motivate an initial decorrelation step, after which the transformed scores are modeled independently. To our knowledge, applying this step to the scores before spatial modeling is novel in spatial FDA and is similar in spirit to spatial blind source separation \citep{bachoc_spatial_2020} or the Linear Model of Coregionalization \citep{Cressie:2015uu}.

\section{Computational Details}
\subsection{Details of Leverage Score Computation}\label{leverage_score_computation}

In this section, we detail how one can take advantage of the local nature of the B-spline basis in computations while specifying a working correlation structure for the minimization problems of the type used in the mean estimation. 
There are algorithms for the efficient computation of the leverage scores when the matrices $\Sigma_i^{-1}$ are diagonal \citep{hutchinson1985smoothing}. 
However, using a working correlation matrix results in a matrix that is no longer banded, but still very sparse. This is a simple extension from past work, but this approach has not been considered for spline estimation before. Let  $\{\chi_\ell\}_{\ell=1}^m$ be the B-spline basis used. Let $\Phi$ be the $(\sum_{i=1}^n m_i) \times M$ matrix, where $\sum_{i=1}^n m_i$ is the total number of observations and $M$ is the number of basis functions, that gives the basis functions evaluated at each pressure. Specifically, the first row is given by $\{\chi_\ell(p_{1,1})\}_{\ell = 1}^M$, and subsequent rows are evaluations for different $p_{i,j}$. Letting $\Sigma^{-1}$ be the block diagonal matrix of $\frac{1}{n m_i}K_{h_s, h_t}(s_i - s_0, d_i - d_0)\Sigma_i^{-1}$, where $\Sigma_i$ is the working covariance matrix such that $\Sigma_i^{-1}$ is tridiagonal. Furthermore, let $\Omega$ be the penalty matrix of $\int_{0}^{2000} \chi_i^{(2)}(p) \chi_j^{(2)}(p) dp$ where $\{\chi_k\}_{k=1}^M$ is the B-spline basis used. Then the coefficients for the solution to (\ref{meanoptim}) are $(\Phi^\top \Sigma^{-1} \Phi + \lambda\Omega)^{-1} \Phi^\top \Sigma^{-1} Y$. The leverage scores are defined as the diagonal elements of the matrix $$A(\lambda) = \Phi (\Phi^\top \Sigma^{-1} \Phi + \lambda\Omega)^{-1} \Phi^\top \Sigma^{-1} \in \mathbb{R}^{n\times n}.$$ Let $B = \Phi^\top \Sigma^{-1} \Phi + \lambda\Omega$ and note that if $\Sigma^{-1}$ is diagonal, $B$ is also banded due to the sparsity pattern of $\Phi$ and $\Omega$. Thus, one does not need to invert the entire matrix $B$, and since $\Phi$ is sparse, the diagonal of $A(\lambda)$ only relies on certain elements of $B^{-1}$. In this case, (\cite{hutchinson1985smoothing}) give an algorithm to compute the leverage scores.

Using a non-diagonal matrix $\Sigma^{-1}$ means that $B$ will no longer be banded, but may have additional non-zero entries depending on the basis used and the amount of separation between measurements in the same profile. Thus, the approach in \cite{hutchinson1985smoothing} no longer applies. 
However, for a more general sparse matrix, one can compute the leverage scores as given in \cite{Erisman1975}. We detail the algorithm below. First, we obtain the Cholesky decomposition of $B = U^\top D U,$ where $U$ is upper triangular with unit diagonal entries, and $D$ is diagonal. The Takahashi equation for the inverse is given by \begin{align}B^{-1} = D^{-1}U^{-\top} + (I - U) B^{-1},\label{Takahashi_test}\end{align}where we use the shorthand of notation $U^{-\top}$ to signify the transpose of $U^{-1}$.

One can simplify the computation of leverage scores as follows. Since $B^{-1}$ is symmetric, one only needs to compute the upper triangular part of $B^{-1}$. Thus, since $U^{-\top}$ is lower triangular, we need only compute the diagonal entries of $D^{-1} U^{-\top}$, or, equivalently, the diagonal entries of $D^{-1}$. For a sparse matrix $I - U$, the relation (\ref{Takahashi_test}) gives a recurrence relation for certain elements of $B^{-1}$. Each of these elements can be computed using elements previously computed below and to the right of the right of the element. 
Theorem 1 in \cite{Erisman1975} implies that one can compute $(B^{-1})_{ij}$ for all $i$ and $j$ such that $U_{ij} \neq 0$ using $U$, $D$, and the values of $(B^{-1})_{kl}$ such that $U_{kl} \neq 0$ and $k > i$, $l > j$. 
That is, \begin{align*}
(B^{-1})_{ij} &= \frac{1}{d_{ii}} \delta_{ij} -\sum_{k=i+1}^p U_{ik} (B^{-1})_{kj}
\end{align*}where $\delta_{ij}$ is the Kronecker delta function. Thus, one can fill in certain elements of $B^{-1}$ beginning in the bottom right and working one's way up the matrix.

In our context, these elements of $B^{-1}$ are a superset of those needed for computing the diagonal of $A(\lambda) = \Phi B^{-1}\Phi^\top \Sigma^{-1}$. Furthermore, we need only the trace of $A(\lambda)$, so we compute $
    \textrm{tr}\left(\Phi B^{-1} \Phi^\top \Sigma^{-1}\right) = \textrm{tr}\left(B^{-1} \Phi^\top \Sigma^{-1} \Phi \right)$
    using the circulant property of the trace and only using the elements of $B^{-1}$ computed. We have the correct entries to compute because $B$ is as or less sparse compared to $\Phi^\top \Sigma^{-1} \Phi$. 
These are implemented as a Matlab module and made available in {\tt R} \citep{sparseinvmatlab,sparseinv}. 

It is not clear how to include a correlation structure in the estimation of the covariance operator while maintaining the sparse structure of $B$. Although potentially very computationally intensive, this could offer some improvements in the estimates of the principal component functions.

\subsection{Computational Complexity}\label{computational_challenges}

In Table \ref{comp_challenges}, we give the computational run times and information on each of the stages of our analysis. We have developed efficient code for the B-spline evaluation matrices with speed and flexibility that are not currently available in {\tt R}. These functions are based on the {\tt smooth.spline} function that quickly computes a univariate spline and cannot include covariates. This function is functionally identical to the {\tt eval.basis} function from the {\tt fda} package for B-splines, but has better speed.
Of each of the stages, the maximum likelihood estimation has the largest computational cost since it includes numerical optimization. Furthermore, it must be maximized for 20 scores, for six separate EM iterations. After the first iteration, we start the optimization at previous iteration parameters to speed things up and find that each successive EM iteration takes slightly less time. However, if one were to consider 10 scores for temperature, this would speed up the computations by more than half since the EM algorithm (designed to handle missing salinity measurements) is not necessary. A smaller radius would also decrease the computation time, since this would reduce the size of the dense covariance matrices, for which computing the Cholesky decomposition is the largest computational cost in computing the likelihood. 
However, note that once the parameters are estimated, one can quickly predict as demonstrated by the cross validation column.

For the marginal covariance estimation, we reduced the number of observations used for each profile so that the number of cross products did not become too large. Without reducing the number of observations used for each profile, the median number of observations would be in the tens of millions, which is not realistic for computation with respect to both memory and time. Though sparse matrices lessen the challenge, even forming the basis evaluation matrices can be prohibitive.

Based on the computational overview, multiplying the mean running time by the number of grid points, the sections represented takes approximately 7,100 core-hours to compute. However, given a new year of data, one can use the previously-estimated mean and covariance parameters at each grid point in approximately 50 core-hours, a tiny fraction of the overall total. For the computation, we use Michigan's ARC-TS resources for these computations and to host the Shiny Applications. Also, some computations were run using XSEDE Bridges cluster.

 \begin{table}[h]
  \caption{Overview of computations. We give the number of grid points for each step and the median amount of running time in seconds, profiles, observations, cross validation evaluations, and the approximate computational complexity. High-performance computing and parallelization allow for substantial but manageable computation for all grid points. Note that the number of observations for the marginal covariance estimation refers to the number of cross products used. Here, $N = \sum_{i=1}^n m_i$, $N_2 = \sum_{i=1}^n m_i^2$, $\tilde{n}_y = \max_y n_y$, and approximate computational complexity refers to the naive complexity with fixed bandwidth for each stage. }\label{comp_challenges}
 \begin{tabular}{|c||c|c|c|c|c|c|c|c|c||}
 \hline
 Metric &\begin{tabular}{@{}c@{}} Mean \\ (T) \end{tabular}& \begin{tabular}{@{}c@{}} Mean \\ (S) \end{tabular} & \begin{tabular}{@{}c@{}} Marginal \\ Cov (T) \end{tabular}  & \begin{tabular}{@{}c@{}} Marginal \\ Cov (S) \end{tabular} & \begin{tabular}{@{}c@{}} ML 20 scores  \\ (6 EM iter) \end{tabular} & Nugget & \begin{tabular}{@{}c@{}} Cross \\ Validation \end{tabular}\\
 \hline\hline
 \# Grid Points & 47,938 & 46,023 & 43,646 & 41,896  & 41,888 & 41,271 & 76,050 \\ 
 \hline
 Med \# sec & 41.9 & 31.1 & 96.2 & 64.2 & 251.1  & 3.8 &  3.3\\ 
  \hline
   Mean \# sec & 42.4 & 32.7 &113.7 & 86.9 & 322.4  & 4.0 & 3.9\\ 
  \hline
 Med \# Profiles & 1,362 & 931 & 646 & 450 & 2,137 & 543 & 283 \\ 
 \hline
  Med \# Obs & 302,790  & 177,573 & 82,460,735  & 48,036,594& -- & -- & --  \\
  \hline
  Med \# CV evals & 10  & 10 & 17 & 17 & -- & -- & -- \\
  \hline
  Comp. Complex& $O(Nm)$  & $O(Nm)$ & $O(N_2m^2)$ & $O(N_2m^2)$ & $O(\tilde{n}_y^2)$ & $O(Nm)$ & $O(\tilde{n}_y^2)$
  \\ 
 \hline
 \hline
 \end{tabular}
 \end{table}
 
We briefly and roughly compare the computational cost compared to a pointwise approach like \cite{kuusela2017locally}. If we focus on temperature and use 10 scores with 40,782 grid points and a radius of 1250 kilometers for the maximum likelihood and prediction, we get similar cross-validated prediction errors (results are not presented here), and the mean time to compute the likelihood maximum over all grid points is approximately 169 seconds. By considering only the steps for temperature, the computations take approximately 3,000 core-hours. The naive comparison is 10 Mat\'ern likelihoods to maximize compared to 58 for 58 separate pressure levels. One must also take into account the extra step of marginal covariance estimation, estimating the nugget variance, and estimating the scores by least squares for each grid point. Since estimating the scores and nugget typically takes less than 5 seconds, we estimate that the time for these extra steps is less than 40 seconds per grid point, which is approximately equivalent to the time it takes to compute the maximum of 3 likelihoods. Thus, our approach can get comparable predictions for each pressure level and the entire pressure dimension in approximately one-fifth to one-fourth (13/58) of the time of a pointwise approach for 58 levels.

\section{Modeling Choices and Checks}

\subsection{Comparison of Estimated and Working Correlation} \label{comparecor}

\begin{figure}[h]
\centering
\includegraphics[scale = .07]{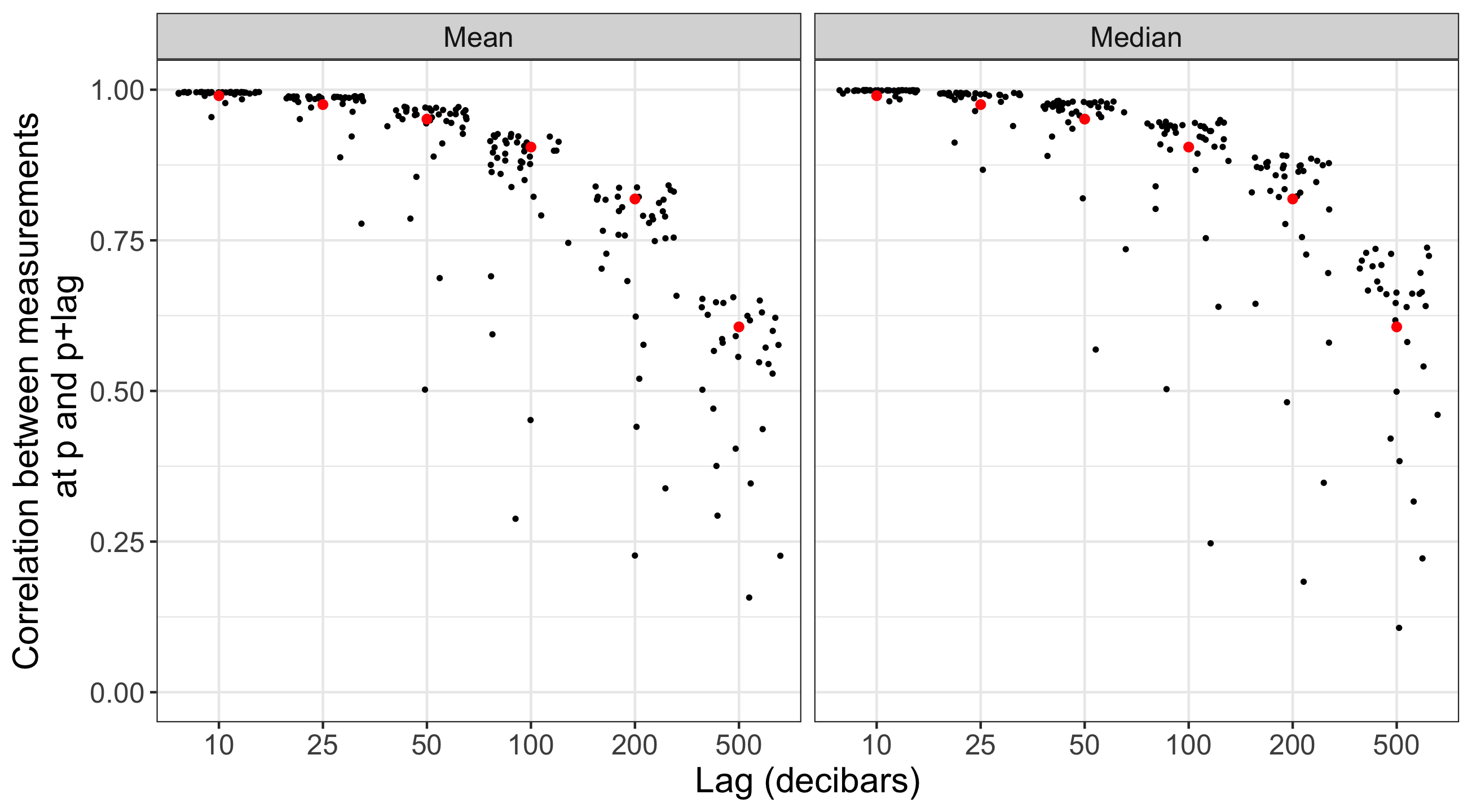}
\includegraphics[scale = .07]{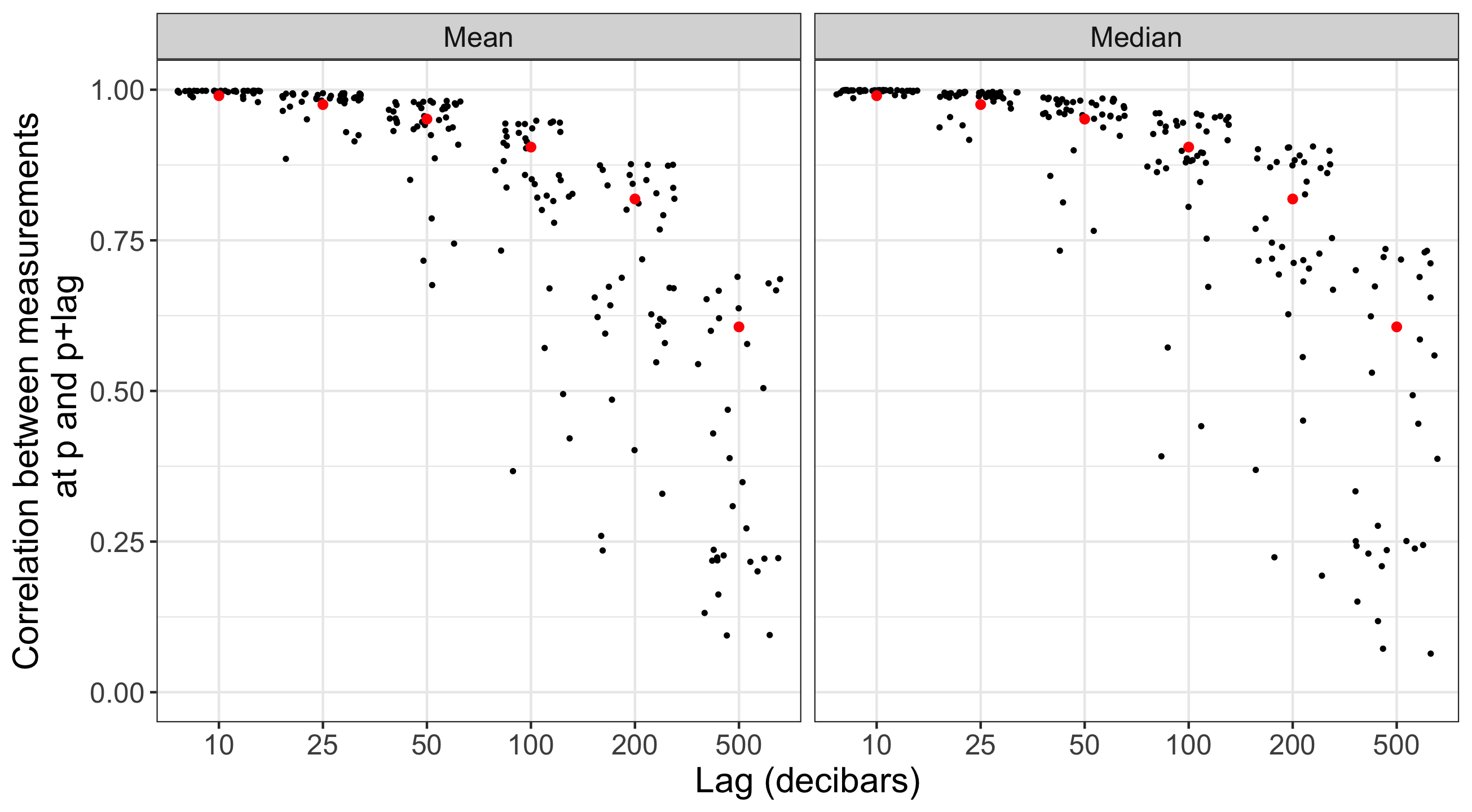}

 \caption{Estimated correlation versus working correlation used (\textbf{Left}) temperature (\textbf{Right}) salinity for measurements different lags apart. Each black point represents an average over locations for one $p_1$ and $p_1 + p_{\textrm{lag}}$. The black dots within the same lag describe the correlation at different baseline pressure $p_1$. The points have been randomly jittered horizontally for visualization. The working correlation is in red.}\label{working_cor}
\end{figure}

For each location, we compute the estimated marginal correlation in pressure $$\rho(p_1, p_2) = \frac{C(p_1, p_2)}{\sqrt{C(p_1, p_1)C(p_2, p_2)}},$$ where $C(p_1, p_2)$ is our estimated covariance, at points $p_1=p$ and $p_2 = p+p_{\textrm{lag}}$ for $p = 0$, $50$, $100$, $\dots$, $1900$, $1950$ for both temperature and salinity and lags $p_{\textrm{lag}} = 10$, $25$, $50$, $100$, $200$, $500$. Next, we consider the average and median over locations. The results of this are plotted in Figure \ref{working_cor}, with comparison to the working covariance used in the mean estimation. The working covariance provides an adequate approximation of the covariance at each lag and considerably improves compared to no specification of the correlation structure.

\subsection{Comparison with \cite{roemmich20092004}}\label{RG_comparison}

In this section, we compare and contrast our mean estimation approach with the excellent oceanographic standard for the Argo data due to \cite{roemmich20092004}. While \cite{roemmich20092004} treats years equally and uses data from the entire year simultaneously, we use data from only the first three months of the year and model the variation across years.
Accounting for the changes in ocean properties from year to year changes the structure of the residuals, especially near the ocean surface. This mean estimation approach differs from the standard approach in oceanography, where the mean field often is an average over all years, and the deviation from this mean is studied separately and referred to as the ``anomaly." Estimating these anomalies gives a natural way to compare conditions from different years. From a statistical perspective, this approach leads to a challenge, since the yearly variation can overwhelm the spatio-temporal variability in the residuals. For example, in \cite{kuusela2017locally}, the scale parameters for longitude and latitude in the covariance models near the ocean surface are often on the scale of thousands of kilometers. Such large-scale covariance models are likely artefacts of the presence of anomalies, which we claim can be better captured in the mean rather than the covariance component of the model. Modeling the year-to-year variation in the mean decreases the range of dependence for the resulting residuals. This makes our functional covariance modeling easier and more flexible by ensuring that the resulting residuals have zero mean for each local fit for each year. After modeling the covariance, the predictions are more closely compared with \cite{kuusela2017locally} in Section \ref{compare_kuusela}. 

\begin{figure}[t]
\centering
\includegraphics[scale = .095]{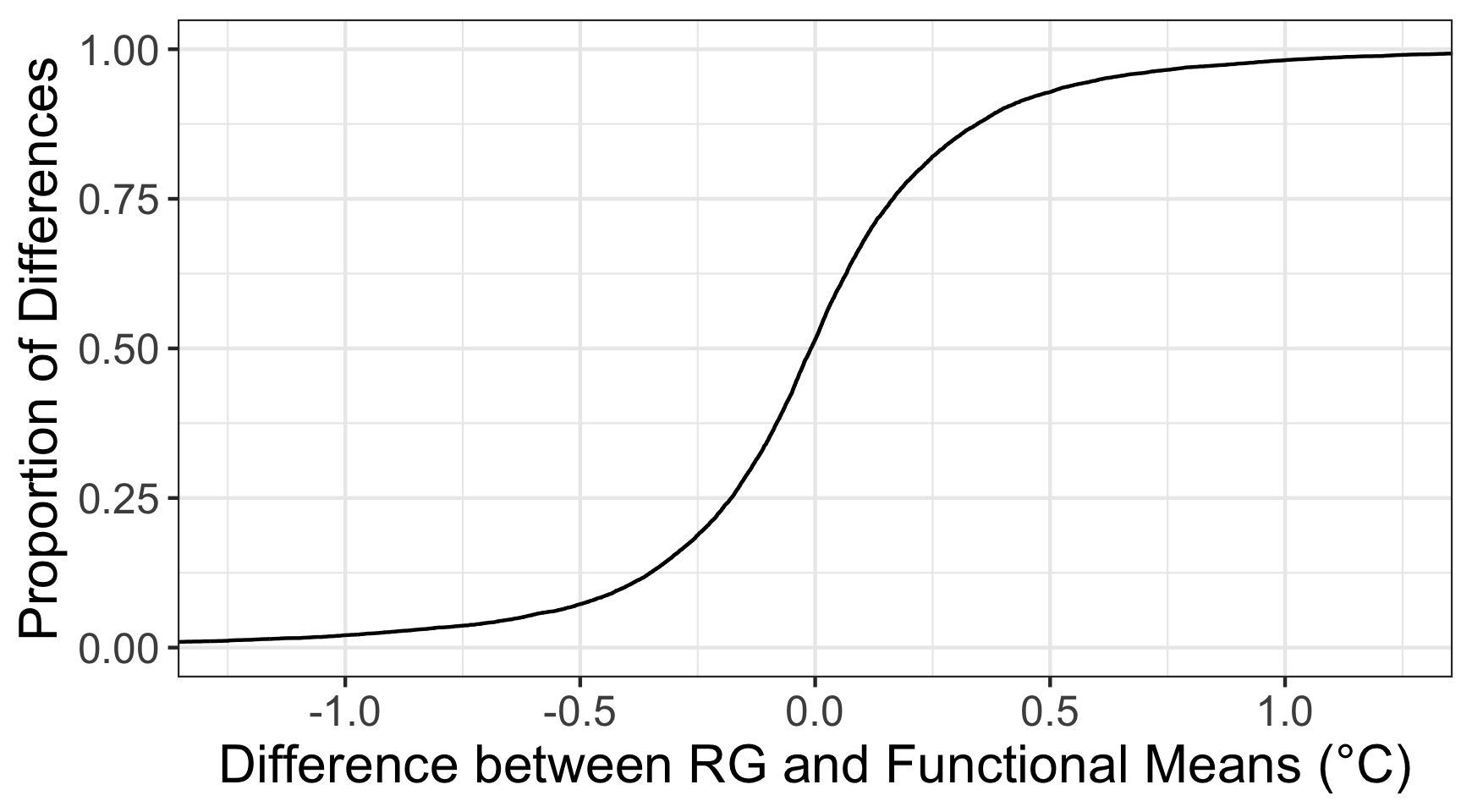}
\includegraphics[scale = .095]{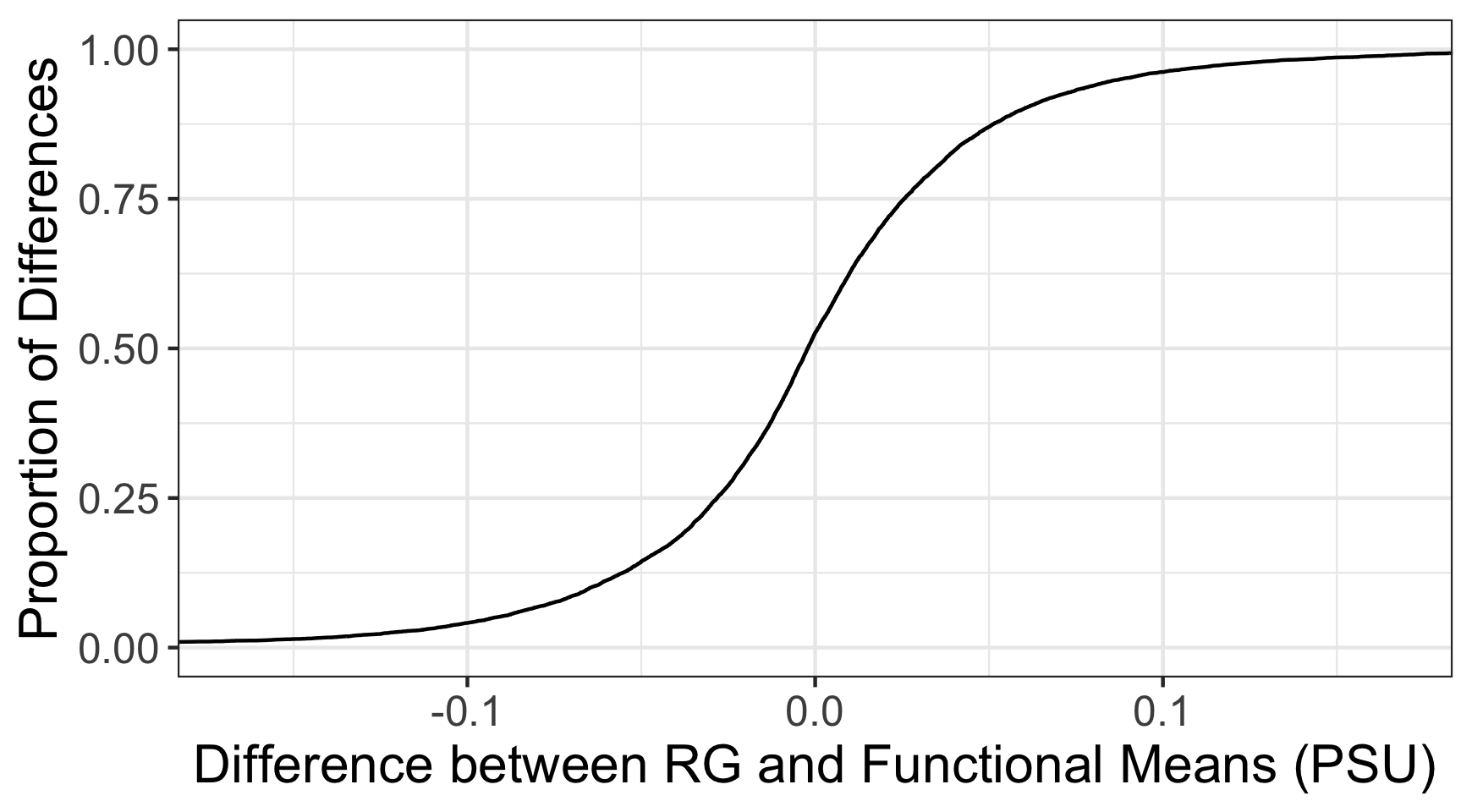}
\caption{Empirical cumulative distribution functions of Roemmich-Gilson mean estimates minus functional mean estimates at 300 dbar for (\textbf{Left}) temperature and (\textbf{Right}) salinity over each grid point. Both means refer to mid-February, and most differences are less than $0.5$ degrees Celsius and $0.1$ PSU.}\label{rg_compare_ecdf}
\end{figure}

There are additional differences between the two approaches for mean estimation, mainly in the data used. The Roemmich and Gilson mean uses additional data from 2004 to 2006 and less stringent data quality checks that require more care. In addition, we use a fixed kernel bandwidth rather than the nearest 100 profiles from each of the 12 months utilized in Roemmich and Gilson product. Our year-averaged mean for mid-February is compared with the Roemmich and Gilson mean in mid-February in Figure \ref{rg_compare_ecdf} for one pressure level. As seen, the two estimates match very closely.  Detailed comparisons at additional pressure levels are available using an {\tt R} Shiny application \citep{shinyapps}.
The mean fields were kindly provided to us by John Gilson. In general, there are two notable spatial differences in the estimates when comparing the two fields. In the Western boundary currents, the fields differ likely due to the distance metrics used; since Roemmich and Gilson account for differences in the depth of the ocean in the distance metric, they employ more data along the coasts compared to our estimates. Our mean is more smooth in space since we choose a large bandwidth, and more of the spatial variability is modeled through the covariance.

\subsection{Choice of Number of Principal Components}
\label{numpcs}

\begin{figure}[t]
    \centering
    \includegraphics[scale = .11]{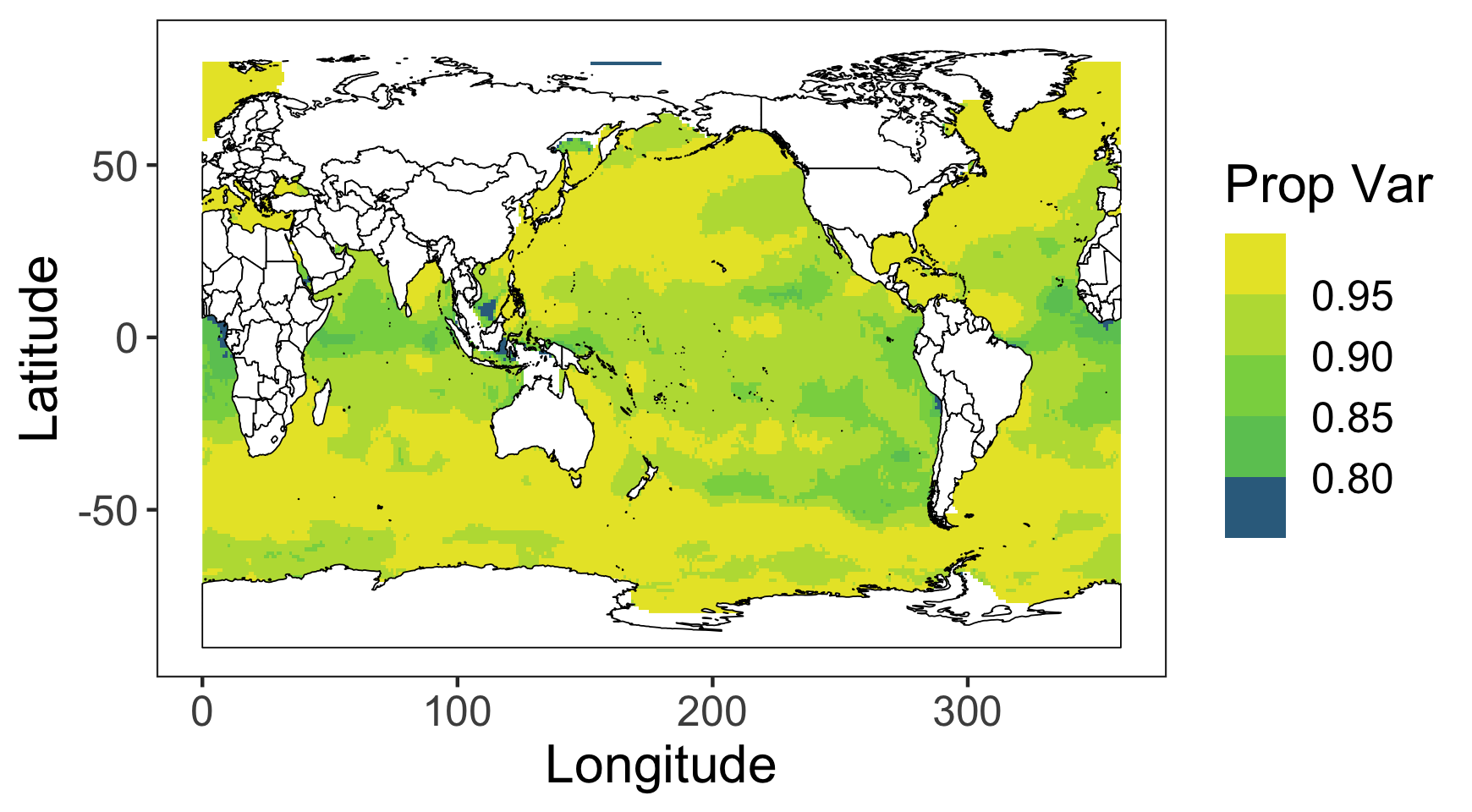}
\includegraphics[scale = .11]{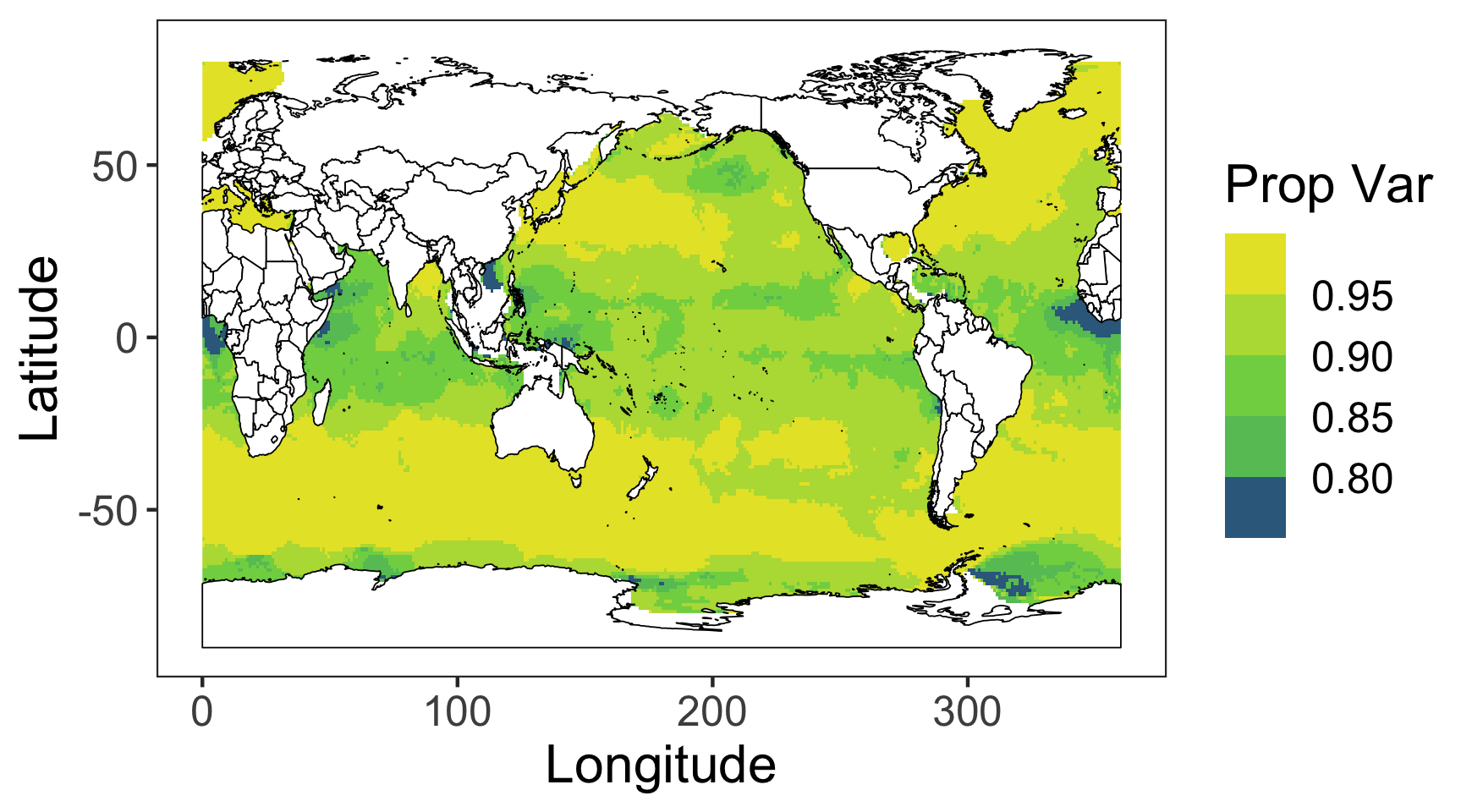}

    \caption{Proportion of variance explained by the first 10 principal components for temperature (\textbf{Left}) and
    salinity (\textbf{Right}).}
    \label{numpcsplot}
\end{figure}

Here, we describe our approach for choosing the number of principal components in the spatial modelling. To roughly choose the number of principal components, our general approach is to increase the number of principal components until adequate spatial modeling could be done and so that profiles could be well represented by this choice. To further explore this choice, we plot the percent of variability explained by location, that is, computed \begin{align*}
 \frac{\sum_{i=1}^n\sum_{j=1}^{m_i}(Y_{i,j} - \sum_{k=1}^K Z_{i,k}\phi_k(p_{i,j}))^2}{\sum_{i=1}^n\sum_{j=1}^{m_i}Y_{i,j}^2}
\end{align*}for profiles within 500 kilometers of each grid point. In Figure \ref{numpcsplot} the respective plots for temperature and salinity are given for $K_1 = K_2 = 10$, which shows that the first 10 principal components represent a large proportion of the variance of the residuals. We want to point out that this is a global measure rather than a local measure. In particular, the first 10 principal components explain less of the variance near the surface in \% in comparison to pressure $> 1000$.  
In addition, we have explored using more principal components 
($K_1 = 15$), but this resulted only in negligible improvement over the choice $K_1 = 10$ at the expense of increasing the computational cost considerably.

For this problem for iid functional data, \cite{li_selecting_2013} propose AIC and BIC based estimates for the number of principal components. This is complicated by the spatial nature of our data. Most notably, in order to apply the AIC criterion, the covariance structure of the scores must be modeled for different numbers of principal components. Since this is the most computationally challenging part of our computations, this is not a feasible approach in practice, though future work could adapt the BIC criterion to spatial data. 

\subsection{Marginal covariance estimation plots}

Our approach estimates the covariance for two pressures, which we show an example of this for pressures 10 and 800. Such estimates are available at any combination of pressures for $p \in (0, 2000)$.

\begin{figure}[t]
\centering
    \includegraphics[scale = .11]{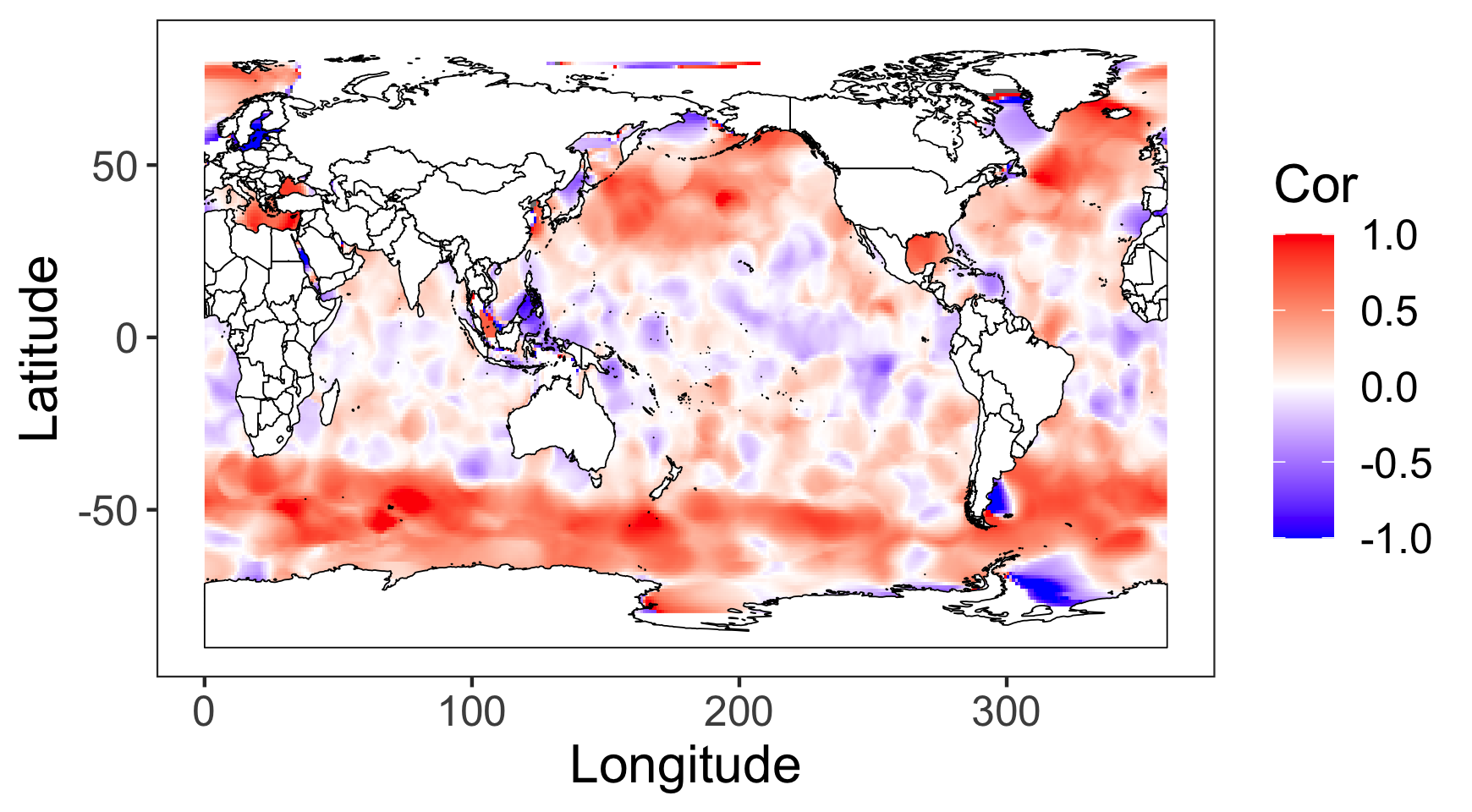}
    \includegraphics[scale = .11]{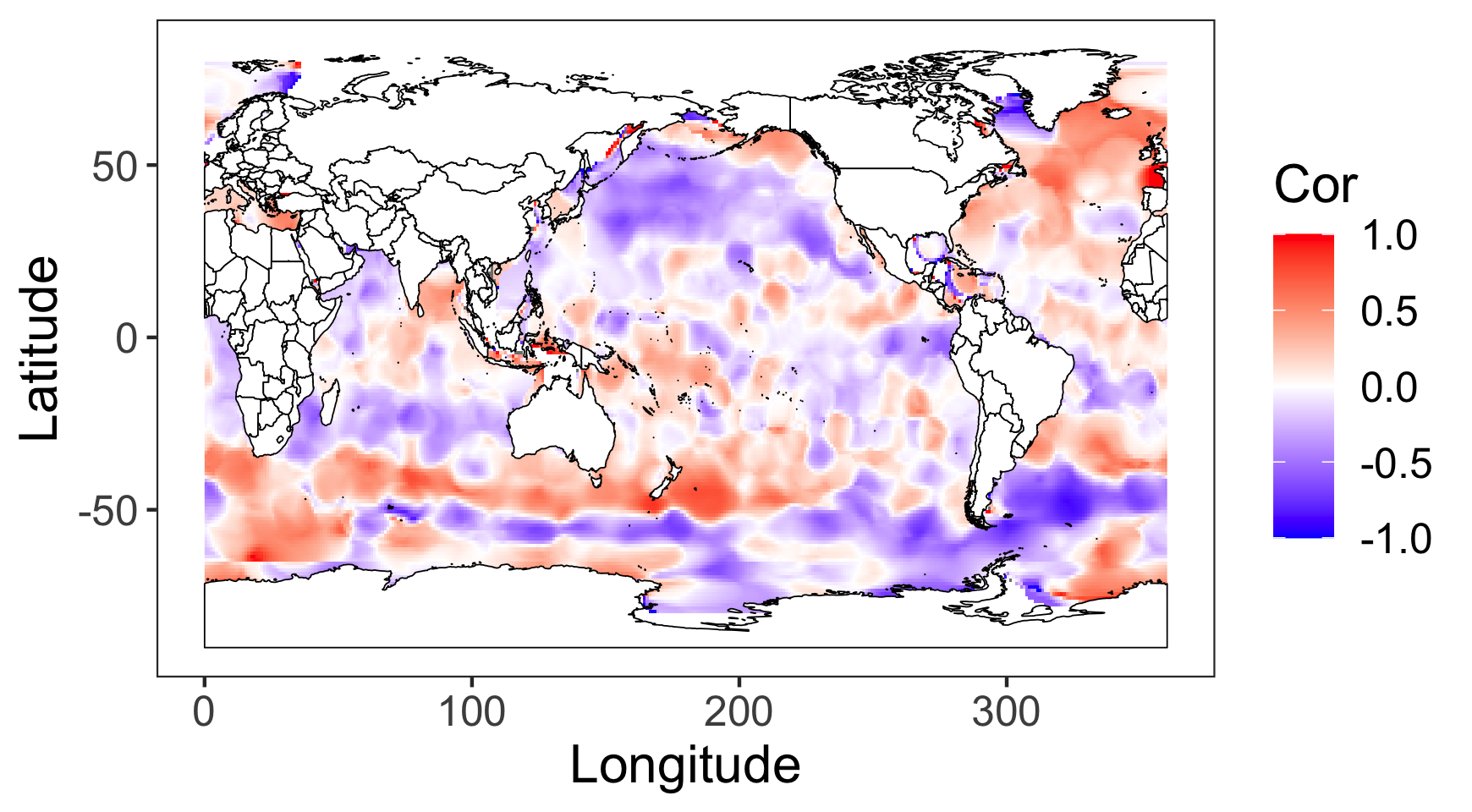}

\caption{Estimated correlation between observations in the same profile at pressures of 10 and 800 dbar
 (\textbf{Left}) temperature (\textbf{Right}) salinity.}\label{covcomp}
\end{figure}

\subsection{Empirical and Model-Based Variance Estimates}\label{variance_est_comparison}

\begin{figure}[h]
\centering
\includegraphics[scale = .2]{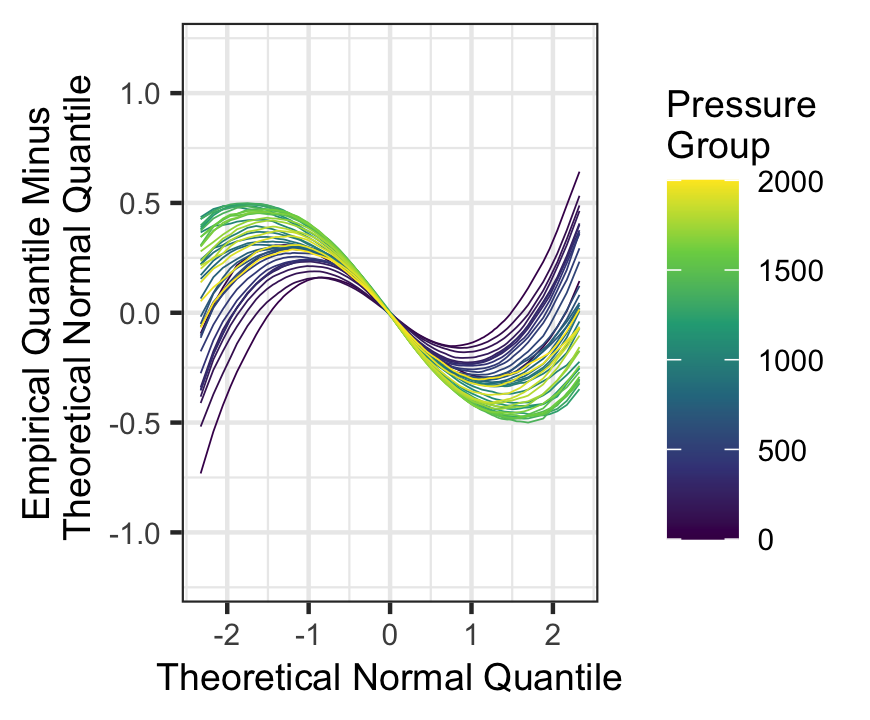}
\includegraphics[scale = .2]{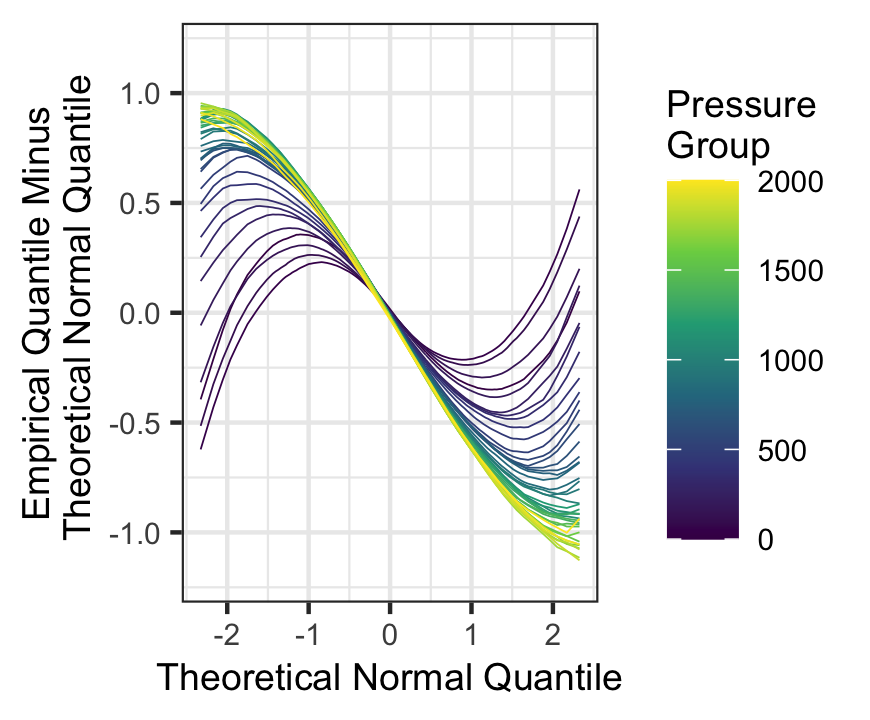}
 \caption{Quantile plots for predicted residuals, grouped by 50 dbar pressure increments (\textbf{Left}) temperature and (\textbf{Right}) salinity. }\label{qq_plots}
\end{figure}

We compare the empirical and theoretical variance estimates for the residuals. After breaking pressure into 50 decibar increments, we plot the difference in quantiles for the two variances against the theoretical quantile. In Figure \ref{qq_plots}, exact normality would have a horizontal line at $0$. There appear to be some differences from normality, especially for salinity. For quantiles near the median, the variance estimates tend to be conservative, with the opposite effect near the extremes. A similar relationship is seen in \cite{kuusela2017locally}. The variance estimates at deeper pressures 
($ > 1000$ dbar) are generally more conservative than those in the first $500$ dbar.

\subsection{Example Predictions at 300 dbar}\label{anomalyexample}

\begin{figure}[t]
\centering
\includegraphics[scale = .17]{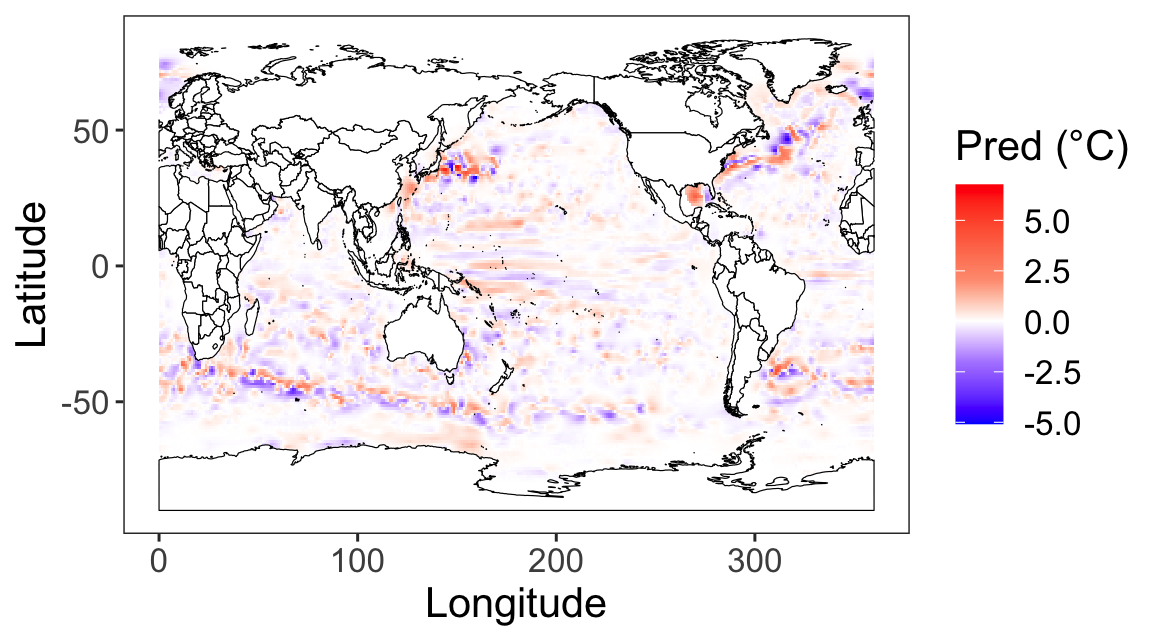}
\includegraphics[scale = .17]{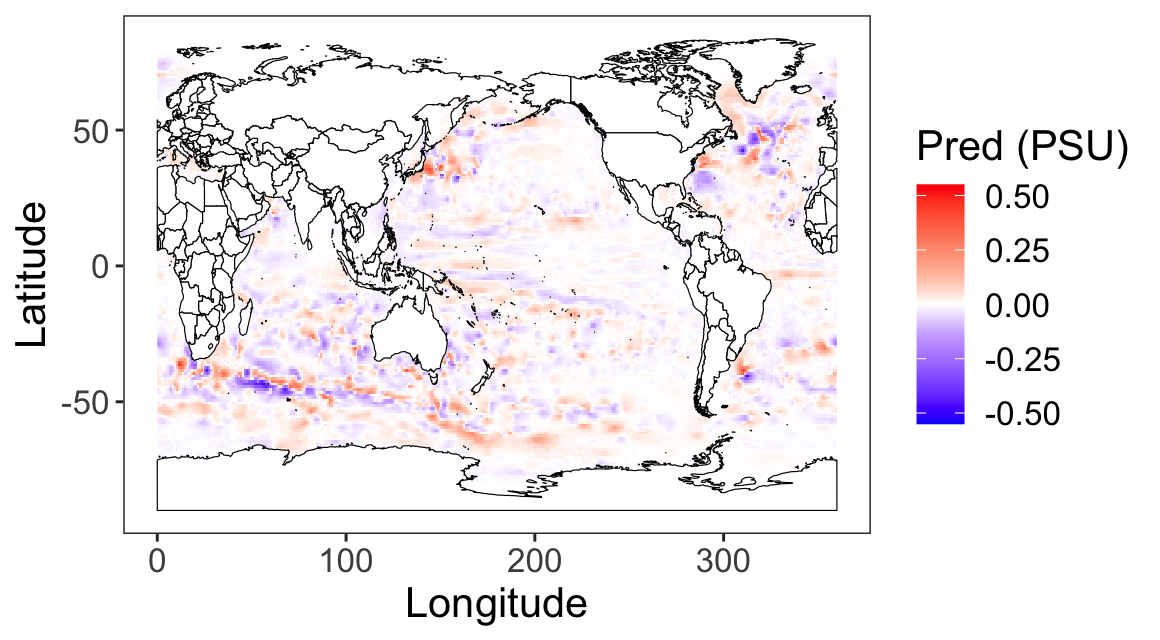}
\caption{February 2012 predictions for temperature (\textbf{Left}) and salinity (\textbf{Right}) residuals, 300 dbar. 
 }\label{anomaly300}
\end{figure}

In Figure \ref{anomaly300}, we give the predictions for a fixed pressure of 300 dbar on the 1 degree by 1 degree grid. 

\subsection{Cross Validation Plots}

\begin{figure}[t]
\centering
\includegraphics[scale = .18]{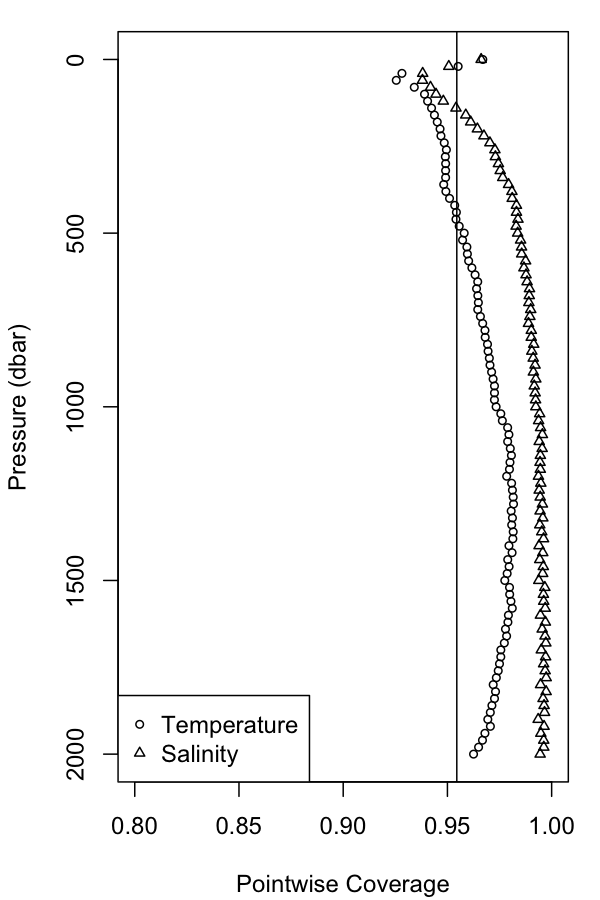}
\includegraphics[scale = .14]{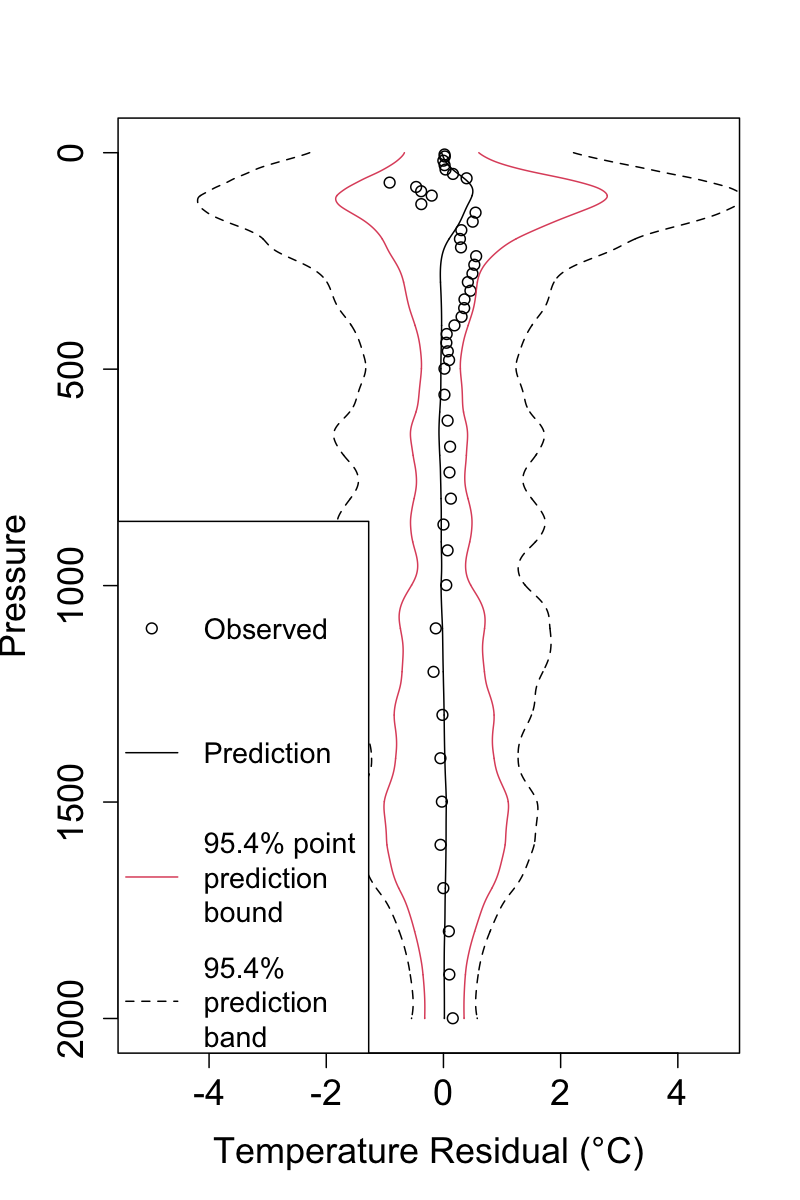}
\includegraphics[scale = .14]{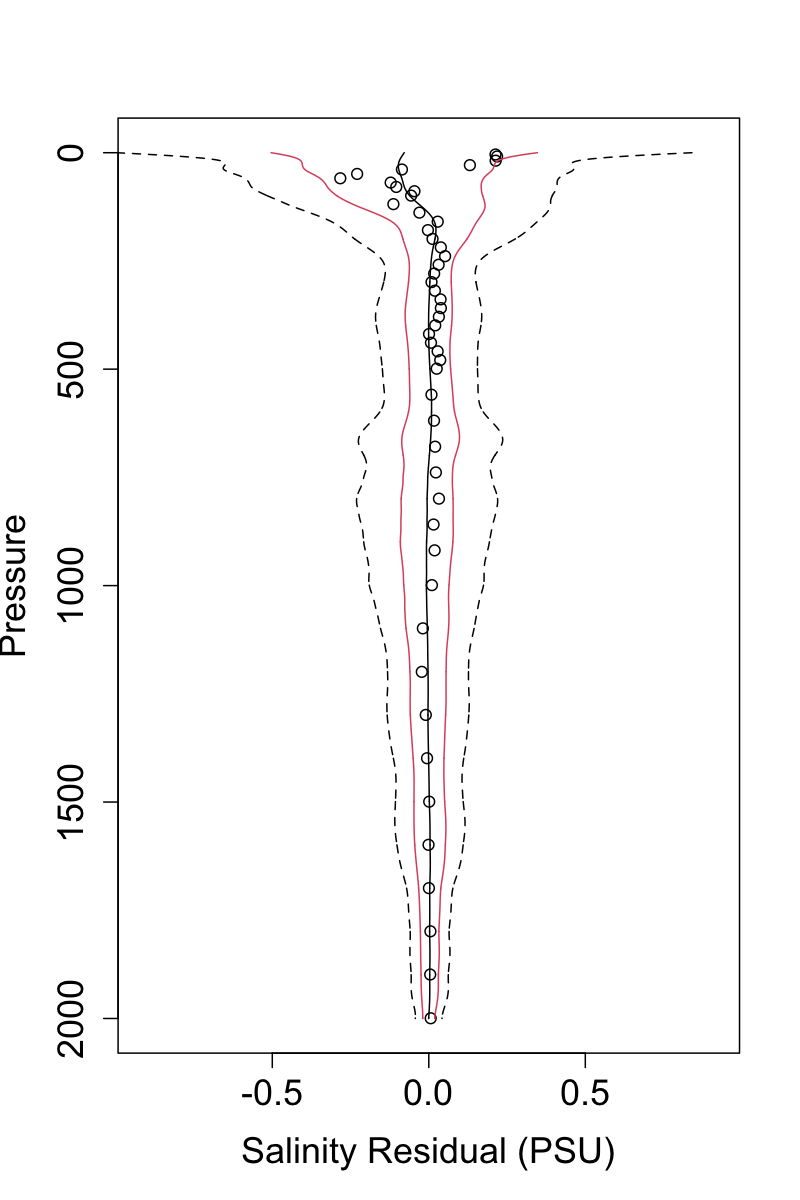}
\caption{Average pointwise coverage of $95.4\%$ confidence intervals summarized in 20 dbar increments for temperature and salinity (\textbf{Left}). The horizontal line indicates the nominal level. The intervals reach coverage for most pressures. Example of cross validation for (\textbf{Middle}) temperature and (\textbf{Right}) salinity. The shown profile is float 2900881, cycle 15, observed in the Northern Indian Ocean (64.5 degrees East, 6.8 degrees North, 2011). A functional prediction over all pressure is provided, and the intervals capture the increased variability near the ocean surface. The simultaneous band gives considerably larger width than the pointwise bound. }\label{cv_example}
\end{figure}

We also summarize the pointwise coverages by pressure in Figure \ref{cv_example}, and such the coverage is achieved for most of the pressure dimension, though typically the intervals in the range 20-200 dbar do not meet full coverage. The conservatism of the uncertainty estimates for deeper pressures may be due to a lack of empirical independence between $Z_k(s,d,y)$ and $\epsilon_{i,j}$. In Figure \ref{cv_example}, we show the functional prediction, interval, and band for one profile. Closer to the surface, there is more variability and complex processes, and as a result more principal components may be needed to reach full coverage for these pressures.

\subsection{Comparison of Predictions with \cite{kuusela2017locally}}\label{ks_exact_compare}

\begin{figure}[h]
\centering
\includegraphics[scale = .155]{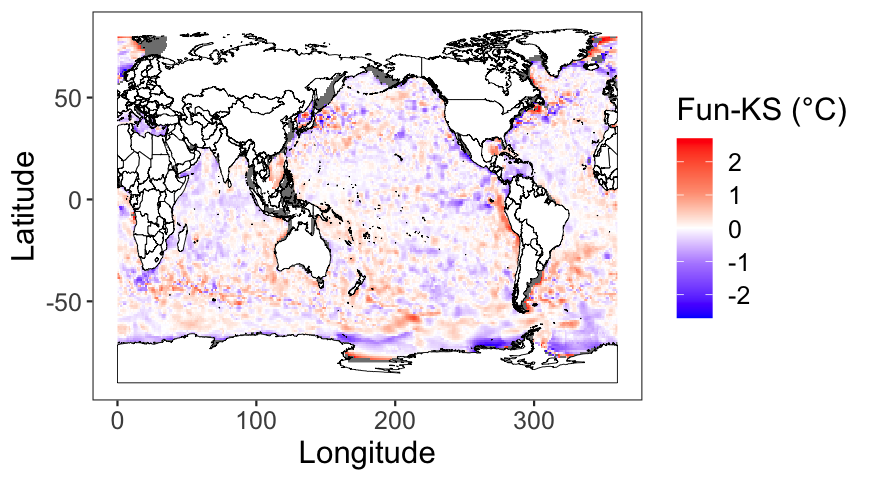}
\includegraphics[scale = .155]{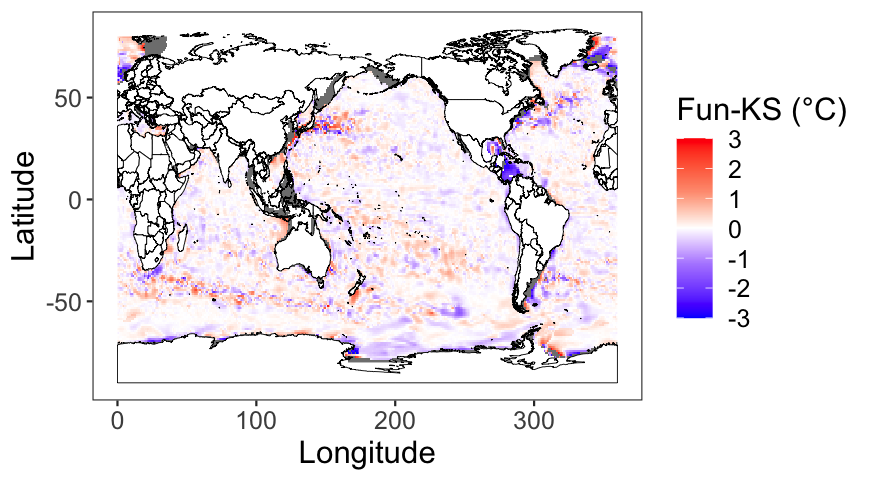}
\includegraphics[scale = .155]{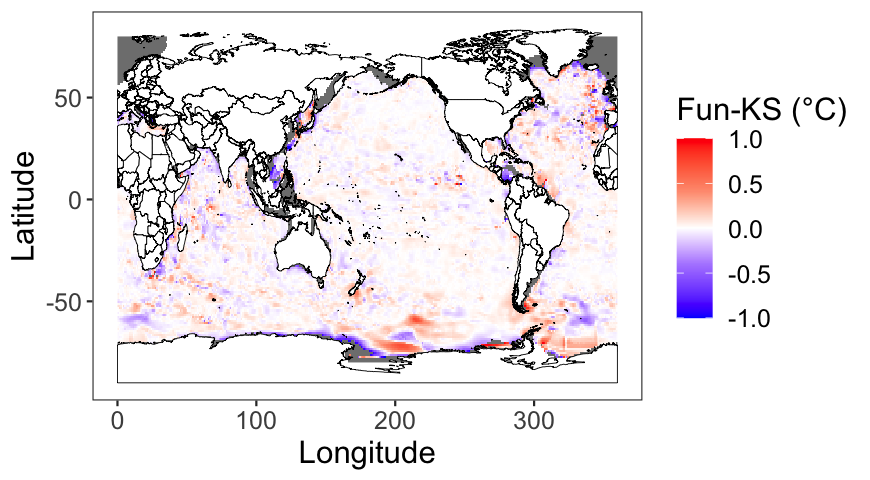}

\label{KS_comparison_at_levels}

 \caption{Comparison between functional predictions and Kuusela and Stein at fixed pressure levels for 2012, (\textbf{Left}) 10 dbar, (\textbf{Middle}) 300 dbar,(\textbf{Right}) 1500 dbar.}
\end{figure}

In Figure \ref{KS_comparison_at_levels}, we evaluate our functions at fixed pressure values and compare our predictions (including the mean and conditional prediction) to those of \cite{kuusela2017locally} for 2012. In general, the differences between the predictions are relatively small. Near Antarctica, there are more differences, possibly because of the differences in the mean functions used. At 1500 dbar, there appear to be more differences in the North Atlantic. 

\begin{figure}[t]
\centering
\includegraphics[scale = .17]{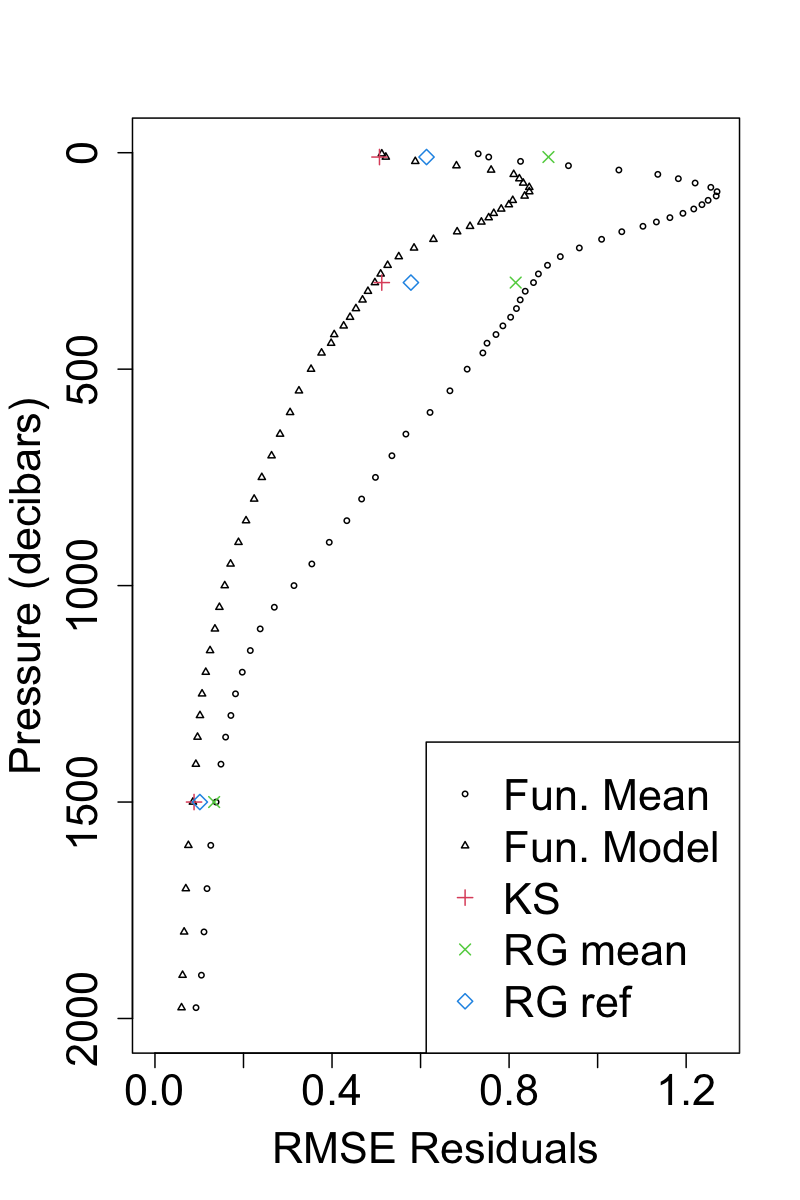}
\includegraphics[scale = .17]{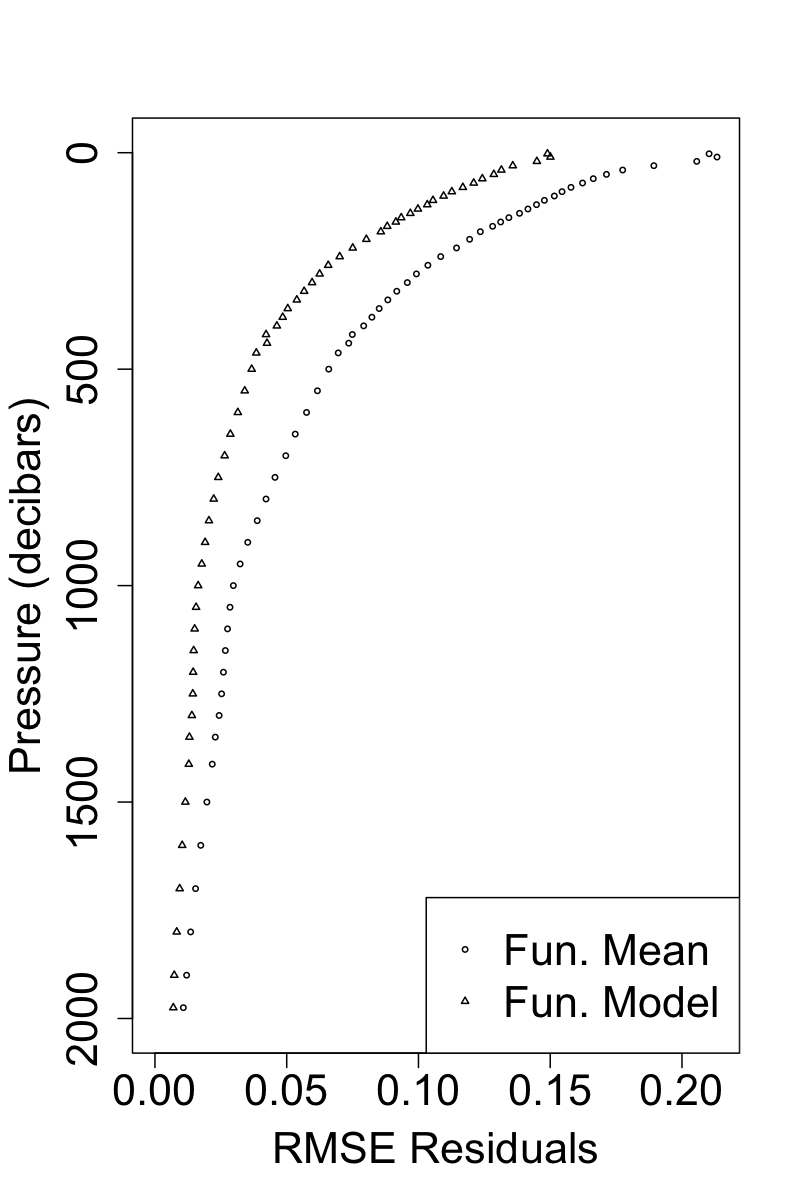}
 \caption{Comparison of RMSE by RG pressure level for temperature (\textbf{Left}) and salinity (\textbf{Right}). The KS, RG mean, and RG ref numbers are included from \cite{kuusela2017locally}, where KS refers to the space-time Gaussian model, the RG mean refers to the Roemmich and Gilson mean, and RG ref refers to the implemented Roemmich and Gilson-like covariance. The variability is largest for salinity near the surface, while for temperature the largest variability lies near the typical thermocline area below the surface. The functional mean results in less variable residuals near the surface compared to the RG mean because yearly effects are included in the functional mean.}
\label{rg_comparison}
\end{figure}

\subsection{MLD Plots}
 
\begin{figure}[t]
\centering
\includegraphics[scale = .16]{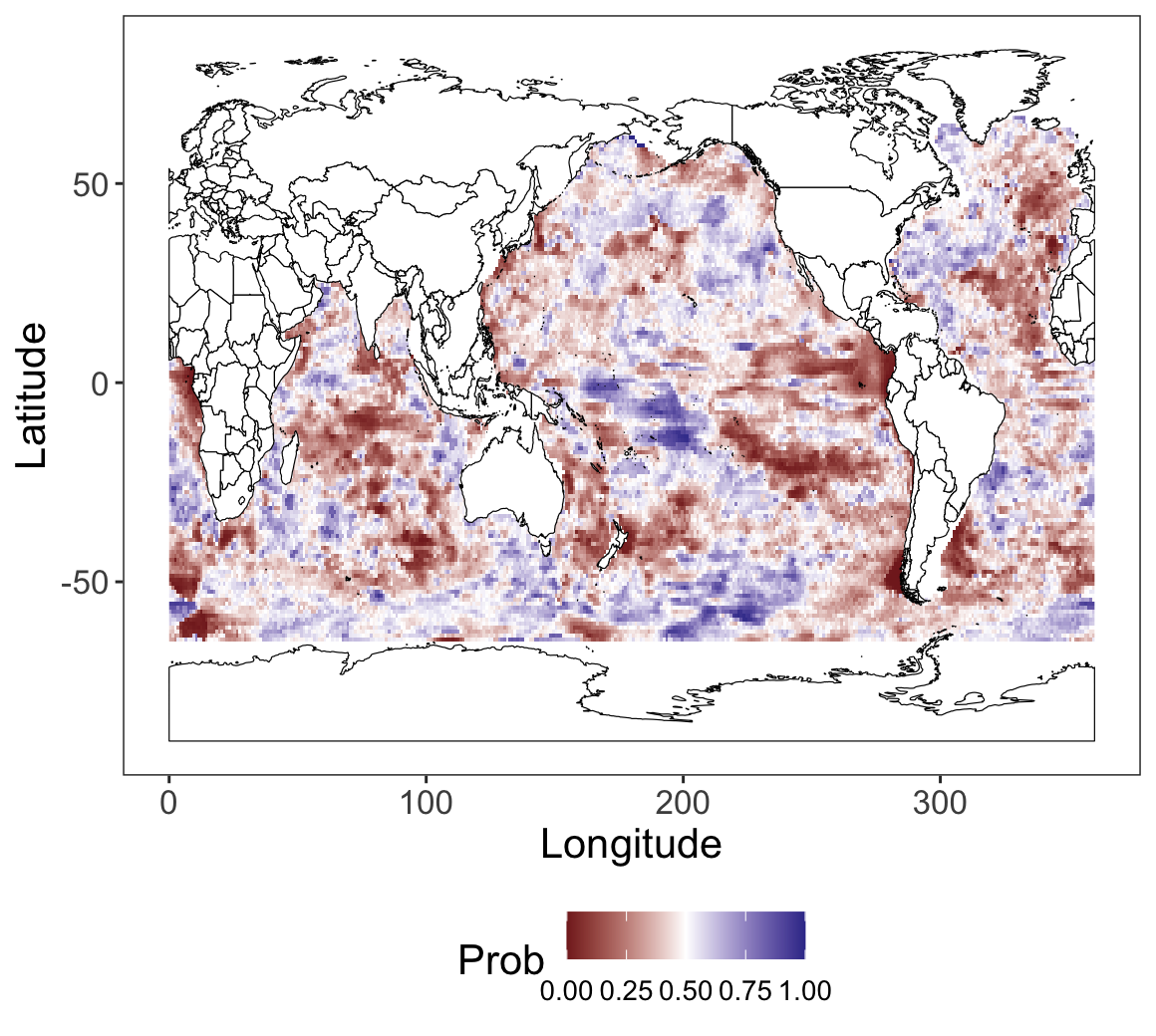}
\includegraphics[scale = .16]{mld/mld_prop_var.png}

\includegraphics[scale = .16 ]{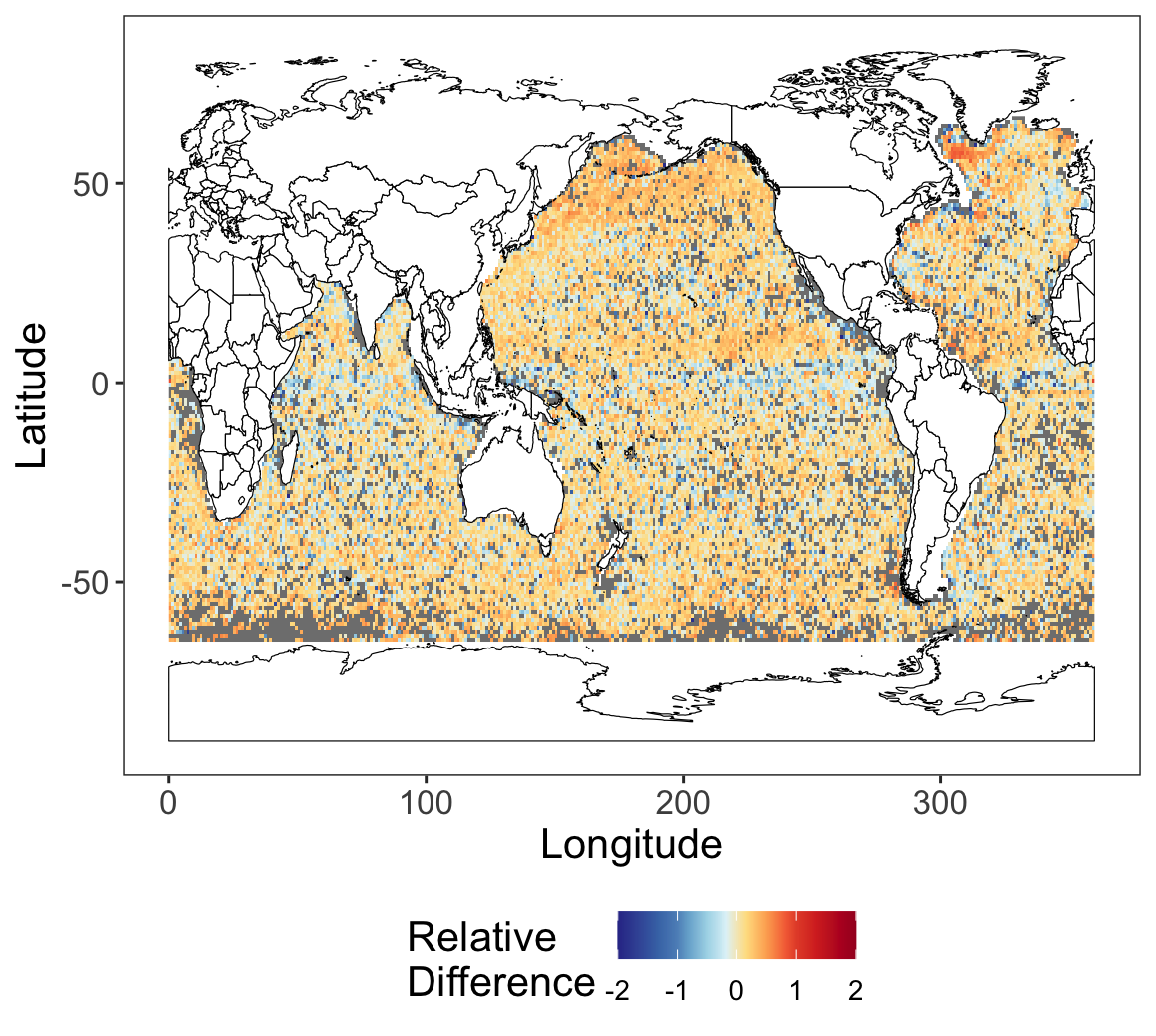}
\includegraphics[scale = .16]{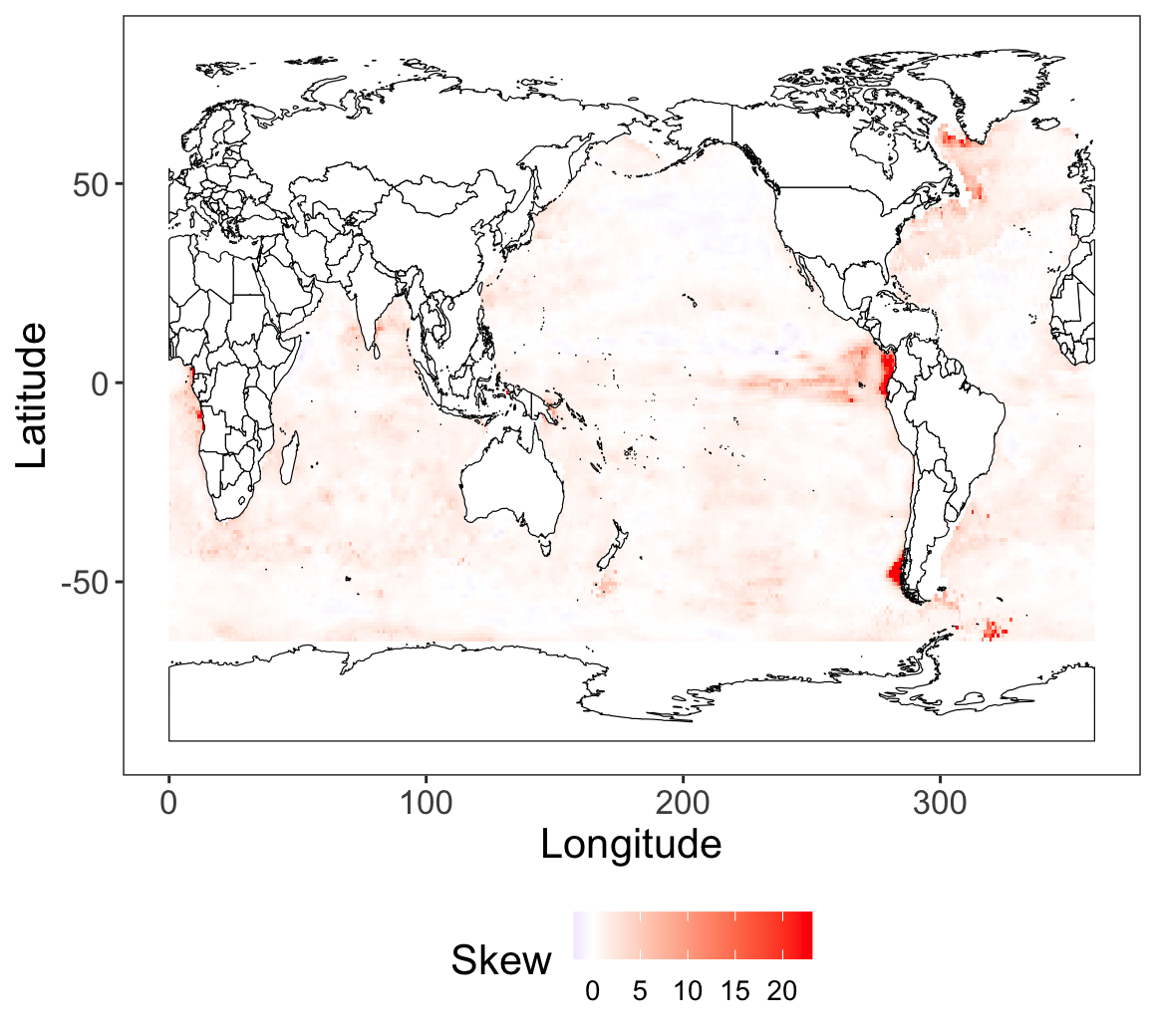}

\caption{(\textbf{Top Left}) $P(D > \tilde{D}_{2014})$, (\textbf{Top Right}) relative difference of mean functional approach and \cite{holte_argo_2017} threshold method (i.e. (HT - FUN)/HT), (\textbf{Bottom}) Measure of skewness $(1/B) \sum_{j=1}^B ((D_{y,j}- \overline{D}_y)/\hat{\sigma}_y)^3$ for $y=2012$ where $\overline{D}_y$ and $\hat{\sigma}_y$ are the mean and standard deviation, respectively, for year $y$.}
\label{mld_est_sd}
\end{figure}

In Figure \ref{mld_est_sd}, we give a few more plots relating to mixed layer depth. In addition to the quantities referenced in the paper, we also refer to 
$$    P(D > \tilde{D}_y) \approx \hat{P}(D > \tilde{D}_y)= \frac{1}{B}\sum_{j=1}^B 1(D_j > \tilde{D}_y),$$the year-averaged distribution of mixed-layer depth to give a probability that the year-averaged mixed layer depth is greater than the observed median for a specific year. Compared to \cite{holte_argo_2017} (on the bottom left), the functional mixed layer depths are somewhat more shallow, which is likely due to the increased vertical resolution of the functional estimates. Finally, the mixed layer depth shows strong right-skew (bottom), which suggests that directly mapping MLD values may be challenged by this skewness. Our functional approach overcomes this by modeling temperature and salinity first.

\subsection{Density comparison over all pressure}\label{sup:densinv}

\begin{figure}
    \centering
    \includegraphics[scale = .174]{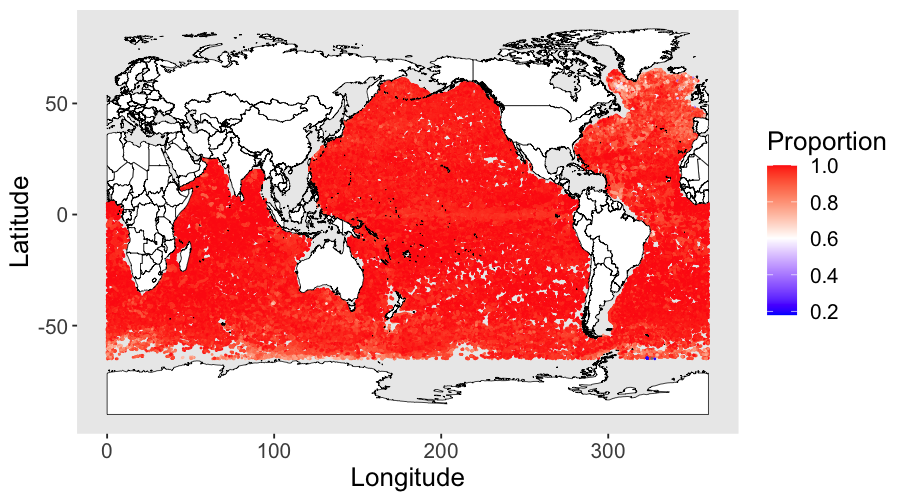}
\includegraphics[scale = .0725]{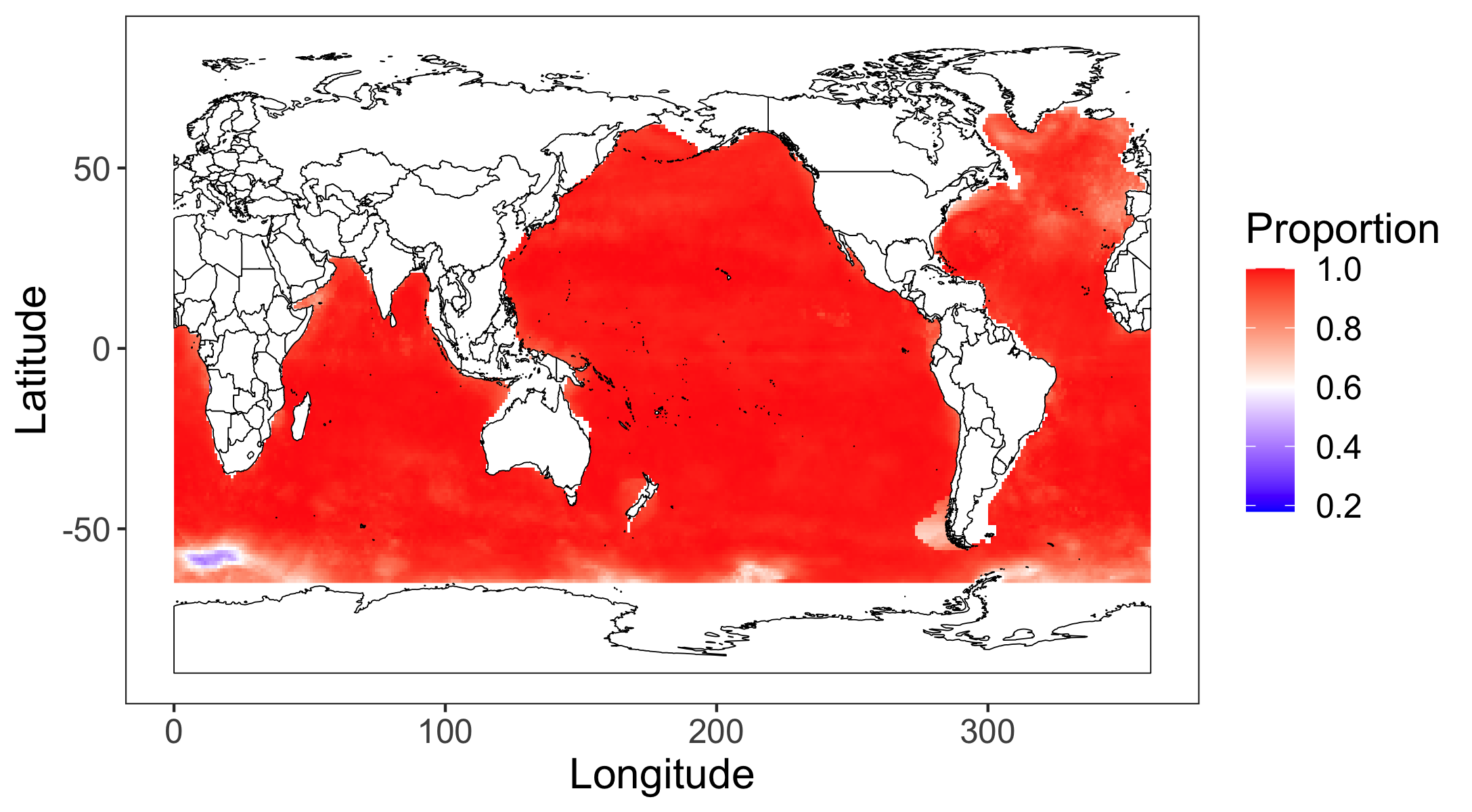}

    \caption{(\textbf{Left}) Proportion of pressure dimension with increasing potential density, evaluated for each January-March Argo profile. (\textbf{Right}) Proportion of pressure dimension with increasing potential density, calculated using functional estimates.}
    \label{fig:dens_inversion2}
\end{figure}

First, we find that our estimates of density inversions generally match with those seen in Argo profiles. In the top panel of Figure \ref{fig:dens_inversion2}, we plot the density inversions for each Argo profile January to March in the following manner. Let $\rho_{i,j}$ be the $j$-th density measurement from the $i$-th Argo profile. We then compute $\Delta\rho_{i,j} = 1(\rho_{i,j+1} - \rho_{i,j} > 0)$ and summarise $\Delta \rho_i = \sum_{j=1}^{m_i - 1} (p_{i,j+1} - p_{i,j})\Delta\rho_{i,j}/\sum_{j=1}^{m_i-1}(p_{i,j+1} - p_{i,j}) $, that is, the proportion of measurement lags with increasing density, weighted by the length of the gap between the measurements. This gives an approximation of the proportion of $[0,2000]$ that exhibits a density inversion. We compare this with a summary of density inversions of our estimates (right). For each of our predicted functions for each of the 10 years, we calculate the density differences on a fine grid, compute the proportion of the pressure dimension with inversions, then average over the 10 years. Overall, the proportion of the pressure dimension with increasing density are similar between the two estimates, with the most notable areas are in the North Atlantic, where deep mixed layers exhibit little stratification, and in the Southern Ocean, where interactions with ice may destabilize the stratification. Otherwise, there is not considerable portions of the oceans of Argo profiles with large amounts of depth with density inversions. The above analysis gives a fixed snapshot of the density inversions, but it does not say much about the persistence of the inversions. For example, if the winter mixed layers in the North Atlantic have approximately constant potential density, then small and insignificant density inversions may be detected.

\bibliographystyle{abbrvnat}
\bibliography{msnew}